\documentclass[]{aa} 

\usepackage{graphicx}   
\usepackage{amsmath}    
\usepackage{amssymb}    
\usepackage{comment}
\usepackage{soul}
\usepackage{color}
\usepackage{xcolor}
\usepackage{hyperref}
\usepackage{placeins}
\usepackage{ulem}
\usepackage[utf8]{inputenc}
\newcommand{\bl}[1]{\mbox{\boldmath$ #1 $}}

\begin{document}

\title{Hydrodynamics modeling of the water snow line in young protoplanetary disks with dust-size-dependent opacities}

   \author{ Eduard I. Vorobyov\inst{1,2}, Anastasiia Topchieva\inst{3}, Aleksandr Skliarevskii\inst{2}, Yaroslav Pavlyuchenkov\inst{3}, and Konstanze Zwintz\inst{1} }
    \institute{  
    Institut für Astro- und Teilchenphysik, Universität Innsbruck, Technikerstraße 25, 6020 Innsbruck, Austria
    \and
       Research Institute of Physics, Southern Federal University, Rostov-on-Don 344090, Russia
  \and
  Institute of Astronomy, Russian Academy of Sciences, Pyatnistkaya str., Moscow, 119017, Russia
  }

  \date{}

   \titlerunning{Water snow line in young protoplanetary disks}
   \authorrunning{Vorobyov et al.}

  \abstract
  {The water snow line in protoplanetary disks plays an important role in the planet formation process. However, most studies are focused on its properties in the advanced stages of disk evolution.}
  {We investigated the properties of the water snow line during the early stages of disk evolution, paying particular attention to the effects of gravitational instability and dust growth on the snow line's shape and position.}
  {We used the FEOSAD numerical hydrodynamics code to simulate the disk formation and evolution in the thin-disk limit. The simulations incorporate the coevolution of gas, dust, and volatiles, including dust growth, volatile phase transitions, and dust-size-dependent opacities.}
  {The position of the water snow line is highly nonsteady during the considered disk evolution period, first moving outward during the disk build-up and then retreating back as the disk cools. Its form in the disk midplane deviates strongly from a circular shape in the early gravitationally unstable phase of disk evolution. An increase in the amounts of grown dust and water ice as well as in the maximum dust size just beyond the snow line, as is readily observed in one-dimensional viscous disk evolution models, in our hydrodynamic models occurs only after gravitational instability diminishes. Dust-growth-induced opacity changes have a profound effect on the position of the water snow line, shifting it closer to the star by almost a factor of two compared to models that do not take this effect into account.}
  {The shape, position, and properties of the water snow line in young, gravitationally unstable disks 
  differ from those of older, axisymmetric disks.
  Our results highlight the importance of taking into account the dependence of opacity on dust size when studying disk evolution.}


   \maketitle

\section{Introduction}
The growth of dust particles via mutual collisions is an important initial step toward planet formation.  However, this process is halted at the dust growth barriers \citep{Birnstiel2024}, which prevent dust from growing above a few meters in size. Alternative mechanisms are needed to circumvent this dust growth barrier problem and form planetesimals, the initial building blocks of planetary cores. One such mechanism is the streaming instability, which causes dust particles to pack together until their self-gravity takes over, leading to the formation of macroscopic bodies \citep{Youdin2005,Yang2017,Lim2025,Lyra2026}.  Therefore, knowing where in the disk the streaming instability can operate is of great importance for the planet formation theories.

\begin{figure*}   
\includegraphics[width=2\columnwidth]{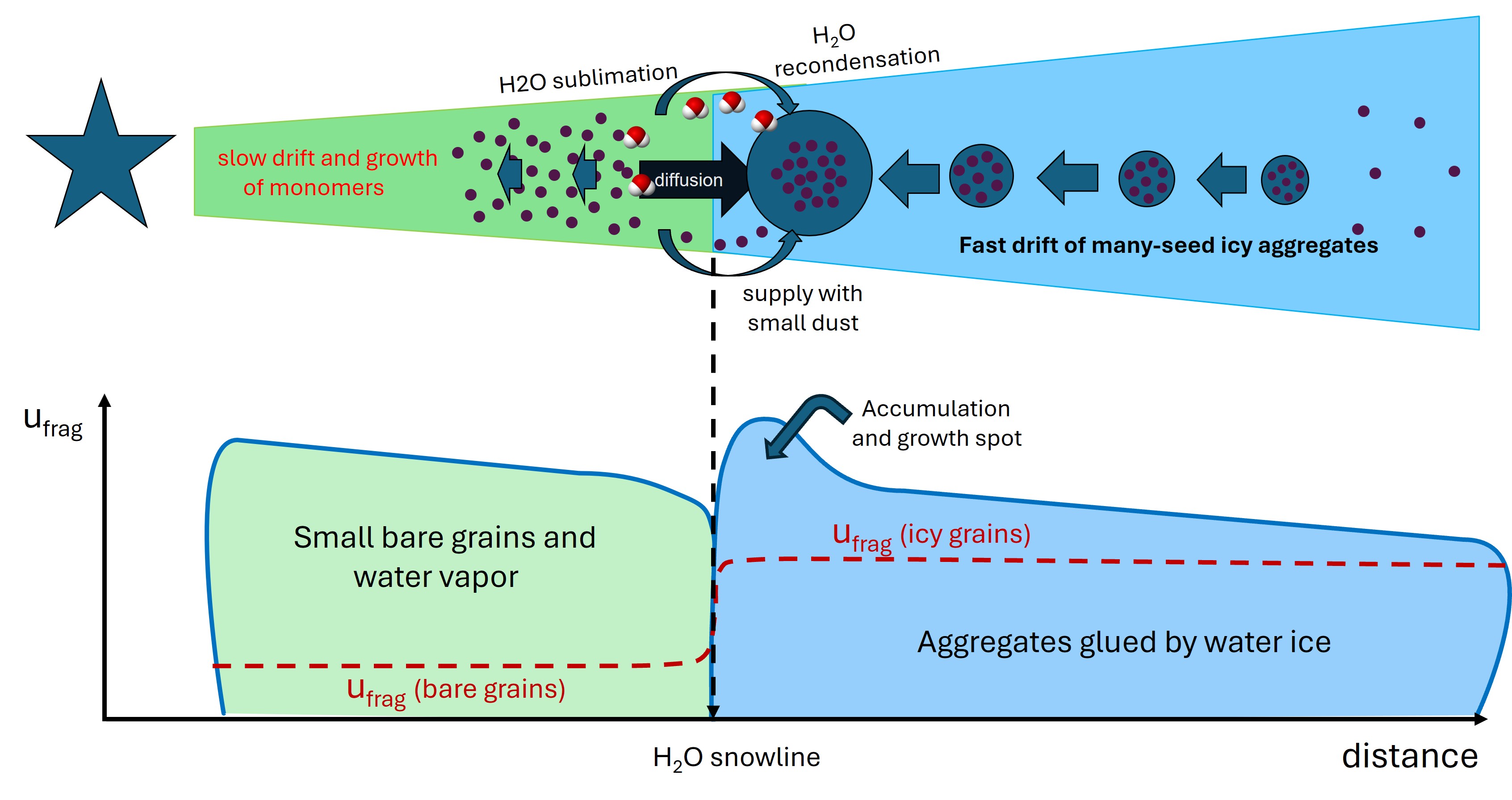}
\caption{Mechanisms that contribute to dust growth just beyond the snow line. Many-seed icy aggregates grow in the outer disk's regions due to the high sticking efficiency of water ice and disintegrate as they cross the water snow line (green-to-blue interface) into silicate monomers and water vapor. Both diffuse back across the snow line due to turbulence, facilitating further accumulation of icy water and the growth of icy aggregates just behind the snow line, effects found in our numerical model as well (see Figs.~\ref{fig:radial-zoom} and \ref{fig:radial-zoom-add}). }
\label{fig:scheme}
\end{figure*}

One such location in the disk is the water snow line, which is the distance in a protoplanetary disk where H$_2$O vapors begin to freeze out onto dust grains.  Inward-drifting icy dust aggregates that cross the water snow line can break apart while losing water ice that glues together their micron-sized silicate constituents \citep{Saito2011}. These small particles, together with water vapor, can diffuse back beyond the snow line, locally enhancing the dust-to-gas mass ratio and promoting dust growth just exterior to it \citep{Cuzzi2004,Schoonenberg2017}. The enrichment mechanism by dust and water ice just behind the water snow line is schematically depicted in Fig.~\ref{fig:scheme}.
This process can provide conditions favorable for the onset of the streaming instability and the formation of water-rich planetesimals \citep{2017Drazkowska,Ryodo2021,Cridland2022,2022LauDrazkowska}. If other mechanisms, such as dust settling to the midplane followed by direct gravitational collapse or dust clumping within spiral arms, can operate interior to the snow line \citep{Goldreich1973,2022Baehr,Rice2025}, then its position would serve as a dividing line between the original location of water-deficient and water-rich planetesimals.

The position of the water snow line is also important in the context of primordial planetary atmospheres. Pebbles exterior to the water snow line are rich in water and oxygen-bearing molecules \citep{Topchieva2024}. Their accretion onto planetary cores in the framework of the pebble accretion model \citep{2012Lambrechts} may enhance primordial atmospheres with the chemical compounds that are required for planetary habitability and may leave observable signatures that will allow us to interpret the composition of exoplanet atmospheres \citep{Oberg-snow2011,Booth2019}.

The radial position of the snow line is determined by the rates of adsorption and desorption onto and from dust grain surfaces, both of which depend on disk temperature \citep{Molyarova2021}.  Numerical studies of snow lines typically assume the underlying disk temperature profile to either be fixed \citep{Oberg-snow2011,2017Drazkowska} or determined from the balance of viscous heating in the midplane and stellar heating through the disk surface against dust radiative cooling \citep{Booth2019,Molyarova2021}. However, the disk temperature in the optically thick limit is sensitive to the disk optical depth, which is the product of dust column density and dust opacity. The latter depends on the spectrum of the dust particle sizes in the disk, which may change during the evolution of the disk. Furthermore, the temperature structure in dynamical disks can be affected by nonstationary processes such as adiabatic heating.

Many numerical studies of protoplanetary disks use dust opacities that are typical of protostellar cores with a maximum dust size of a few microns, such as those of \citet{Semenov2003}. However, dust is known to already rapidly grow in the very early stages of disk formation \citep{Lebreuilly2020, Bate2022, VorobyovKulikov2024}. This may change the thermodynamics of the inner optically thick disk regions where the water snow line may be located \citep{Bitsch2020}. It is therefore important to develop theoretical and numerical models that take into account opacity changes due to dust growth \citep{Yamamuro2023, Nayakshin2025}.

In this work, we modified the Formation and Evolution Of Stars And Disks (FEOSAD) code \citep{2018VorobyovAkimkin,Molyarova2021} to include the dust-size-dependent opacities calculated using the OpTool \citep{OpTool2021}  and gas opacity calculated by \citet{Malygin2014}. We then modeled the coevolution of gas, dust, and volatiles in a protoplanetary disk, focusing on the first 0.6~Myr of its evolution. A monodisperse dust growth model implemented in FEOSAD together with phase transitions and transport of H$_2$O allowed us to explore complex feedback loops between dust growth, dust opacity changes, and the evolution of the water snow line in a young protoplanetary disk. 

The paper is organized as follows. In Sect.~\ref{Sect:model} a  description of the FEOSAD numerical hydrodynamics model is provided, and the dust-size-dependent opacities are presented and analyzed. In Sect.~\ref{Sect:results} we present the main results, and their robustness is discussed in Sect.~\ref{Sect:Discuss}. Section~\ref{Sect:conclusions} presents a summary our conclusions.

\section{Model description}
\label{Sect:model}
Numerical simulations were carried out using the FEOSAD numerical hydrodynamics code, which computes the coevolution of gas, dust, and several volatile species, starting from the cloud core collapse and proceeding through the disk formation and early evolution stages. The equations of hydrodynamics were solved in the thin-disk limit \citep{VorobyovBasu2009}. In this study, we stop the simulations after 0.5~Myr. The detailed description of the code is presented in \citet{2018VorobyovAkimkin}, with modifications to include the phase transitions and advection of volatile species in \citet{Molyarova2021}. Below, we provide a concise description of the main constituents of the code, paying specific attention to modifications related to this study. More details can be found in the appendix.

\subsection{Gas dynamics}
\label{sec:gaseous}
The system of equations for the disk's gaseous component consists of the continuity equation, equations describing the gas dynamics in the disk midplane, and the energy balance equation. The dynamics of gas is determined by gravity (both central source and disk self-gravity), viscosity, and friction between gas and dust. 
The disk's thermal energy is controlled by viscous heating, radiative heating (including the radiation of a nascent star and background radiation), dust radiative cooling, and adiabatic work. The corresponding equations in the thin-disk limit are as follows:

\begin{equation}
\label{eq:cont}
\frac{{\partial \Sigma_{\rm g} }}{{\partial t}}   + {\bl \nabla}  \cdot 
\left( \Sigma_{\rm g} {\bl v} \right) = 0,  
\end{equation}
\begin{eqnarray}
\label{eq:mom}
\frac{\partial}{\partial t} \left( \Sigma_{\rm g} {\bl v} \right) +  {\bl \nabla} \cdot \left( \Sigma_{\rm
g} {\bl v} \otimes {\bl v} \right) & =&   - {\bl \nabla} {\cal P}  + \Sigma_{\rm g} \, {\bl g} + \nonumber
\\ 
&+& {\bl \nabla} \cdot \mathbf{\Pi}  - \Sigma_{\rm d,gr} {\bl f},
\end{eqnarray}
\begin{equation}
\frac{\partial e}{\partial t} + {\bl \nabla} \cdot \left( e {\bl v} \right) = -{\cal P} 
({\bl \nabla} \cdot {\bl v}) -\Lambda +\Gamma + 
\left(\nabla {\bl v}\right):\bl{\Pi}, 
\label{eq:energ}
\end{equation}
where  $\Sigma_{g}$, $\Sigma{\rm d,gr}$, and $e$ are the gas surface density, grown dust surface density, and the internal energy per surface area, respectively; ${\bl v}=v_r\hat{{\bl r}}+v_\phi \hat{{\bl \phi}}$ is the gas velocity in the disk plane; $\cal{P}$ is the pressure, integrated in the vertical direction using the ideal equation of state ${\cal P}=(\gamma-1) e$ with $\gamma=7/5$;  and ${\bl f}$  is the drag force per unit mass between gas and dust. The gravitational acceleration in the disk plane ${\bl g}$ takes into account the gravity of the central star and the gas and dust self-gravity in the disk, which is found by solving the integral form for the potential using the convolution method as laid out in \citet{BT1987} and, in application to protoplanetary disks, in \citet{Vorobyov2024}. Information on the calculations of the viscous stress tensor $\bl \Pi$, cooling $\Lambda$ and heating $\Gamma$ rates is provided in Appendix~\ref{App:visc-cool-heat}, while the expression for the symmetrized velocity gradient tensor $({\bl \nabla} {\bl v})$ is given in \citet{SN1992}.
We note that the magnetohydrodynamics effects, not considered in this work, may be important for the evolution of the water snow line \citep{Mori2021}. The magnetic disk winds and the associated effects are under development in FEOSAD (Redkin et al. 2026).

\subsection{Dust dynamics}
\label{sect:dustgrowth}

In FEOSAD the dust component is divided into two populations: (i) small dust, which are grains with a size\footnote{Here and further in the text by the size of dust grains we mean its radius.} between $a_{\rm min}=5\times 10^{-3} \ \mu \rm m$  and $a_{*} = 1 \ \mu \rm m$ and (ii) grown dust ranging in size from $a_{*}$ to a maximum $a_{\rm max}$, the value of which is variable in space and time, depending on the efficiency of dust growth, drift, and fragmentation.  It is assumed that dust in both populations is distributed over size according to a simple power law: $N(a) = C \cdot a^{ - {\rm p}}$, where $N(a)$ is the number of dust particles per unit dust size, $C$ is a normalization constant, and ${\rm p} = 3.5$ is the power index, which is kept constant during the considered disk evolution period. 
If a discontinuity between the dust size distributions occurs at $a_\ast$ due to, for example, differential drift of small and grown dust populations, we smoothed it out to prevent strong depletion in $\Sigma_{\rm d,sm}$ \citep[see][for details]{Vorobyov2023a}.

We solved the continuity equations separately for the grown and small dust ensembles. Because small dust is assumed to be dynamically linked to the gas due to its small size, the momentum equation was solved only for the grown dust.  The system of hydrodynamics equations for dust in the zero-pressure limit is written as follows:
\begin{equation}
\label{eq:contDsmall}
\frac{{\partial \Sigma_{\rm d,sm} }}{{\partial t}}  + \bl{\nabla} \cdot 
\left( \Sigma_{\rm d,sm} {\bl v} \right) = - S(a_{\rm max}), 
\end{equation}
\begin{equation}
\label{eq:contDlarge}
\frac{{\partial \Sigma_{\rm d,gr} }}
{{\partial t}}  + \bl{\nabla}  \cdot 
\left( \Sigma_{\rm d,gr} {\bl u} \right) =  {\bl \nabla} \cdot \left[ D \Sigma_{\rm g} {\bl \nabla} \left( {\Sigma_{\rm d,gr} \over \Sigma_{\rm g}} \right)  \right] +
S(a_{\rm max}),
\end{equation}
\begin{eqnarray}
\label{eq:momDlarge}
\frac{\partial}{\partial t} \left( \Sigma_{\rm d,gr} {\bl u} \right) +  \bl{\nabla} \cdot \left( \Sigma_{\rm d,gr} {\bl u} \otimes {\bl u} \right)  &=&   \Sigma_{\rm d,gr} \, {\bl g} + \nonumber \\
 + \Sigma_{\rm d,gr} \, {\bl f} + S(a_{\rm max}) {\bl v},
\end{eqnarray}
where $\Sigma_{\rm d,sm}$ is the surface density of small dust, and ${\bl u}$ are the planar components of the grown dust velocity. Details on the calculations of the dust diffusion coefficient $D$, drag force $\bl{f}$, and small-to-grown dust conversion rate $S(a_{\rm max})$ are provided in Appendix~\ref{App:diff-fric}.

Dust growth in our model, as described in more detail in Appendix~\ref{App:dust-growth}, is limited by collisional fragmentation and drift. The drift barrier is accounted for self-consistently via the computation of the grown dust dynamics. The fragmentation barrier is accounted for by calculating the characteristic fragmentation size as \citep{2016Birnstiel}
\begin{equation}
    a_{\rm frag}=\frac{2\Sigma_{\rm g} \mathit{u}_{\rm frag}^2}{3\pi\rho_{\rm s} \alpha_{\rm visc} c_{\rm s}^2},
\label{eq:afrag}
\end{equation}
where $\mathit{u}_{\rm frag}$ is the fragmentation velocity, namely, a threshold value of the relative velocity of dust particles at which collisions result in fragmentation rather than coagulation,  $\rho_{\rm s} = 2.24$~g~cm$^{-3}$ is the dust material  density \citep{Draine2001}, $\alpha_{\rm visc}$ is the coefficient of turbulent viscosity set equal to $10^{-3}$ following \citet{2023Rosotti}, and $c_{\rm s}$ is the sound speed.
Following \citet{Okuzumi2016} we adopt $\mathit{u}_{\rm frag} = 5$~m~s$^{-1}$ for dust grains covered with ice and $\mathit{u}_{\rm frag} = 0.5$~m~s$^{-1}$ for bare grains (see Appendix~\ref{App:volatiles}). 
If $a_{\rm max}$ becomes greater than $a_{\rm frag}$, we suspend the growth of dust and set $a_{\rm max} = a_{\rm frag}$.

We note that the local disk conditions in which a dust particle finds itself as it drifts inward can change such that the value of fragmentation barrier $a_{\rm frag}$ decreases below the local value of $a_{\rm max}$. This occurs, for instance, when inward-drifting dust grains cross the water snow line,  the fragmentation velocity drops by an order of magnitude, and the fragmentation barrier drops accordingly. If this occurs, we reduce $a_{\rm max}$ to adjust it to the new value of $a_{\rm frag}$. This results in fast conversion of large amounts of grown dust into small dust. Therefore, this procedure mimics the disintegration of icy aggregates (glued together by water ice) into smaller silicate grains when crossing the snow line in the "many-seeds" model of \citet{Saito2011,Aumatell2011}. This is a complex physical phenomenon and it is simulated in our model by means of a sharp drop in $u_{\rm frag}$.

\subsection{Volatiles and the position of the snow lines}
\label{Sect:volatiles}

We compute the dynamics and phase transitions of four volatile species: H$_2$O, CO$_2$, CO and CH$_4$, but only water is presented in this study by the reason explained in Appendix~\ref{App:volatiles}. Each of these species can be present in three states: in the gas, in the ice on the surface of small dust, and in the ice on the surface of grown dust. Each species $s$ is described by its surface density in the gas $\Sigma_{s}^{\rm gas}$, on small dust $\Sigma_{s}^{\rm sm}$, and on grown dust $\Sigma_{s}^{\rm gr}$. These surface densities can change through three main processes: advection together with the corresponding component (gas, small dust, or grown dust), exchange of mantles between small and grown dust populations due to grain growth, and phase transitions due to adsorption from gas to dust, and also thermal and photo-desorption. Because of high computational costs of multidimensional hydrodynamics simulations, no other chemical processes, either gas-phase or surface reactions are included.
The corresponding equations describing the evolution of surface densities of volatile species are provided in Appendix~\ref{App:volatiles}.

\begin{table}
\centering
\caption{Dust composition as a function of temperature.}
\label{tab:abundances}
\begin{tabular}{lcccc}
\hline\noalign{\smallskip}
Species & $T_{[10-150]}$ & $T_{[150-425]}$ & $T_{[425-680]}$ & $T_{[680-1500]}$ \\
 & [K] & [K] & [K] & [K] \\
\hline\noalign{\smallskip}
MgFeSiO$_{4}$ & 40\% & 40\% & 80\% & 100\% \\
FeS           & 10\% & 10\% & 0\% & 0\% \\
CHON          & 50\% & 50\% & 20\% & 0\% \\
H$_{2}$O ice  & 0-90\% & 0\% & 0\% & 0\% \\
\noalign{\smallskip}\hline
\end{tabular}
\end{table}

\subsection{Dust and gas opacities}
\label{Sect:opacity}

\begin{figure*}  
    \centering
    \includegraphics[width=1\columnwidth]{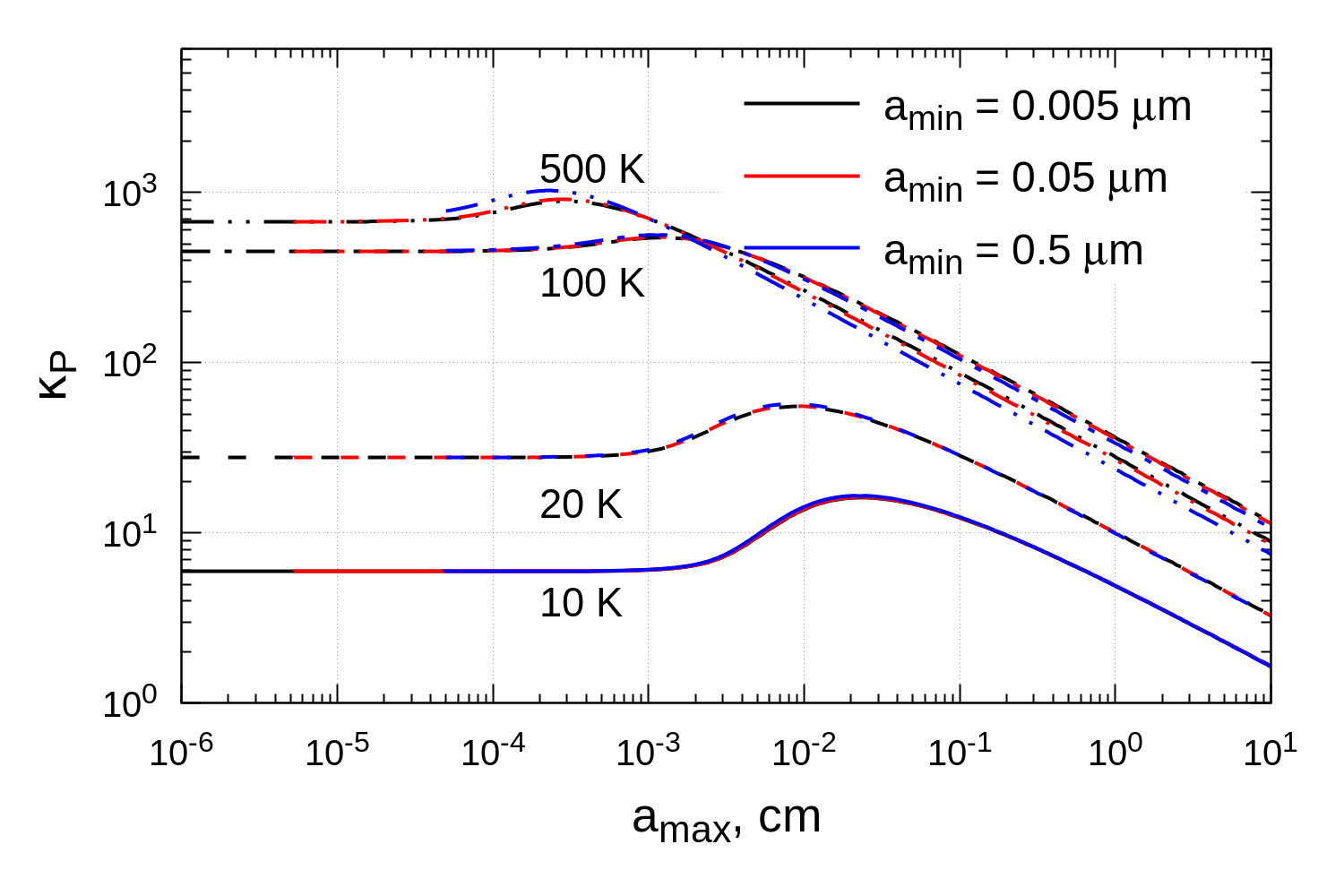}
    \includegraphics[width=1\columnwidth]{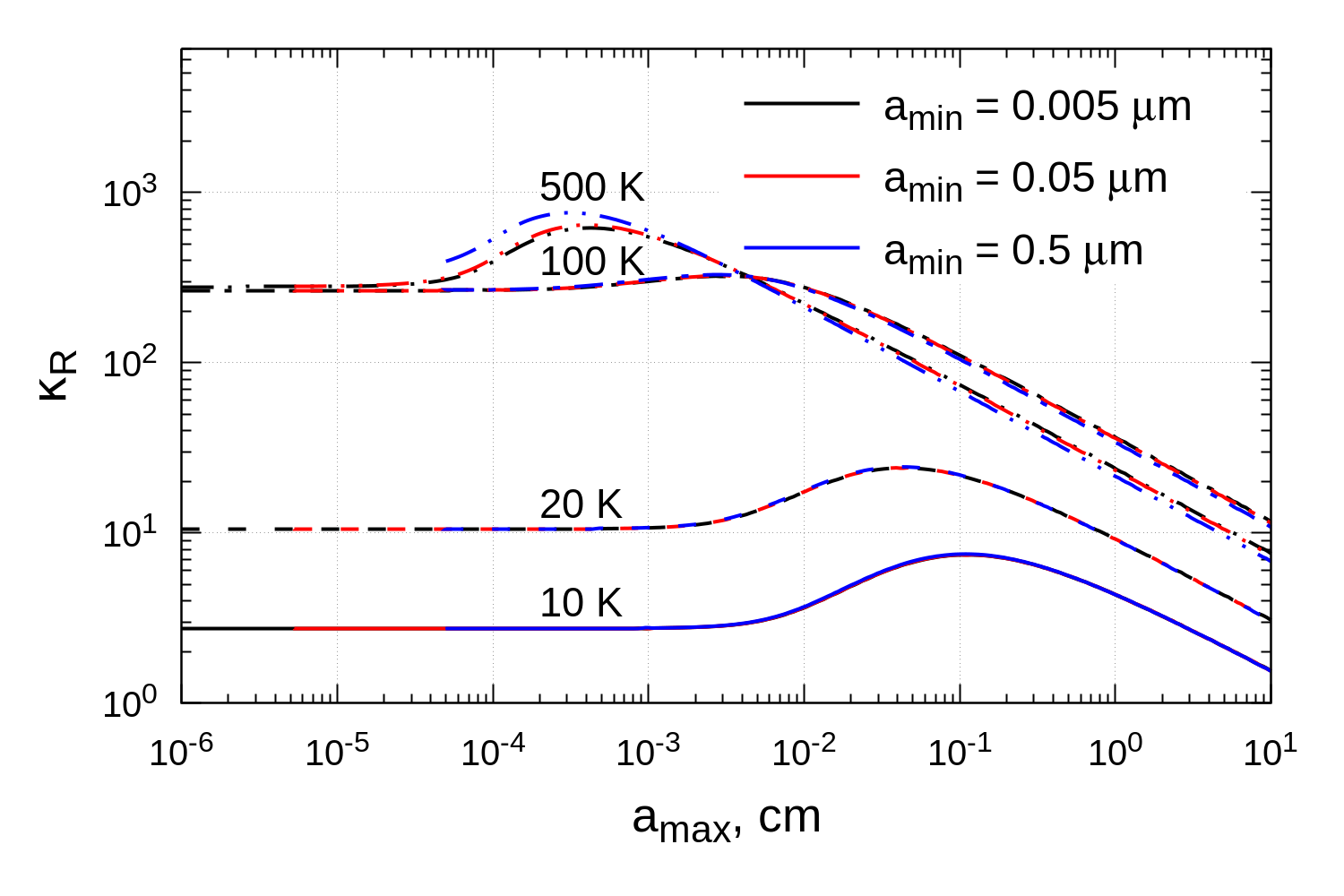}
    \caption{Dependence of the Rosseland (\textit{right}) and Planck mean opacity (\textit{left}) on the maximum dust grain size $a_{\rm max}$ for several minimum dust sizes $a_{\rm min}$ at fixed temperatures of 10, 20, 100, and 500~K. Note that the lines merge at low temperatures. }
    \label{fig:kappavs_size_Temp}
\end{figure*}

\begin{figure}
    \centering
    \includegraphics[width=1\columnwidth]{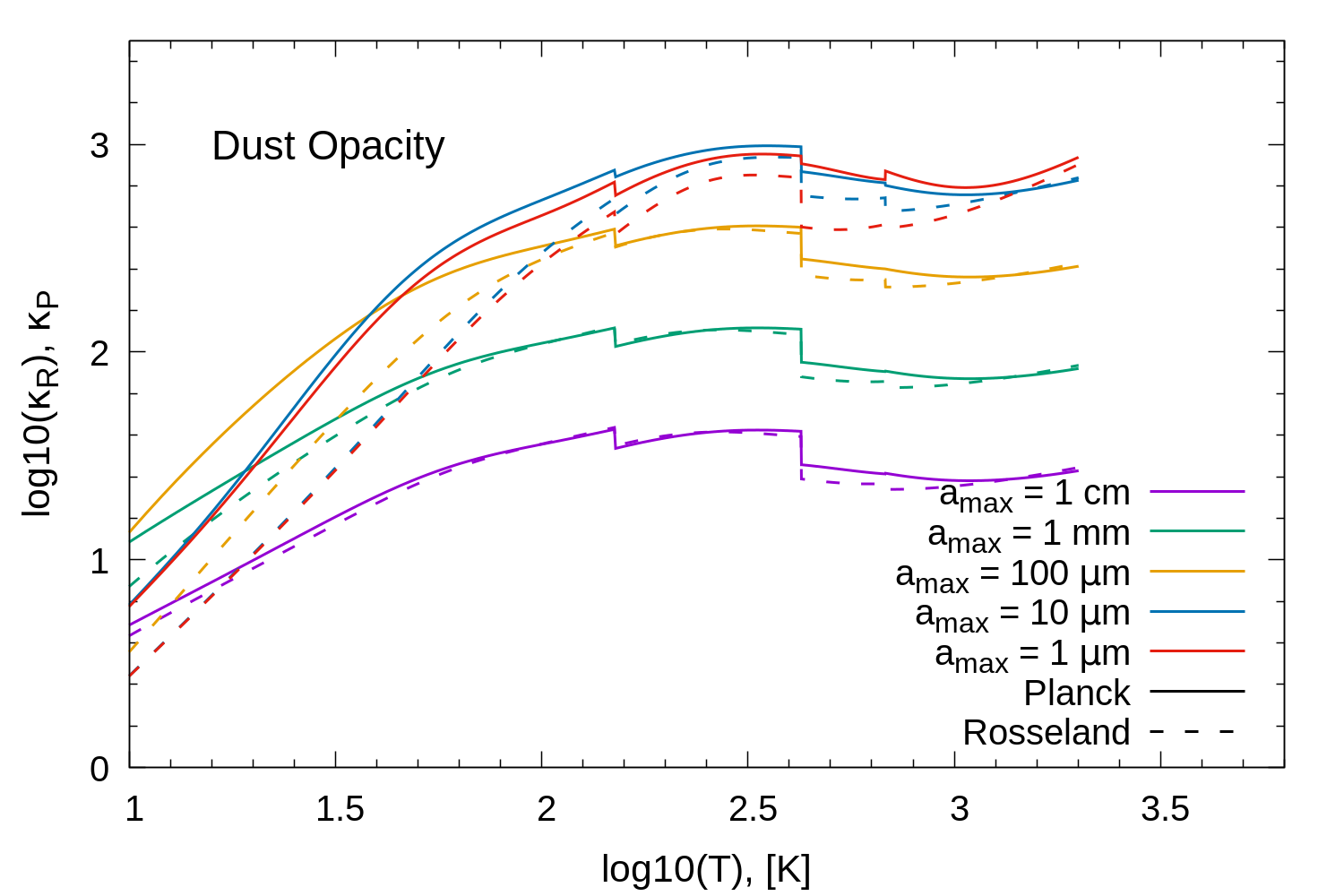}
    \includegraphics[width=1\columnwidth]{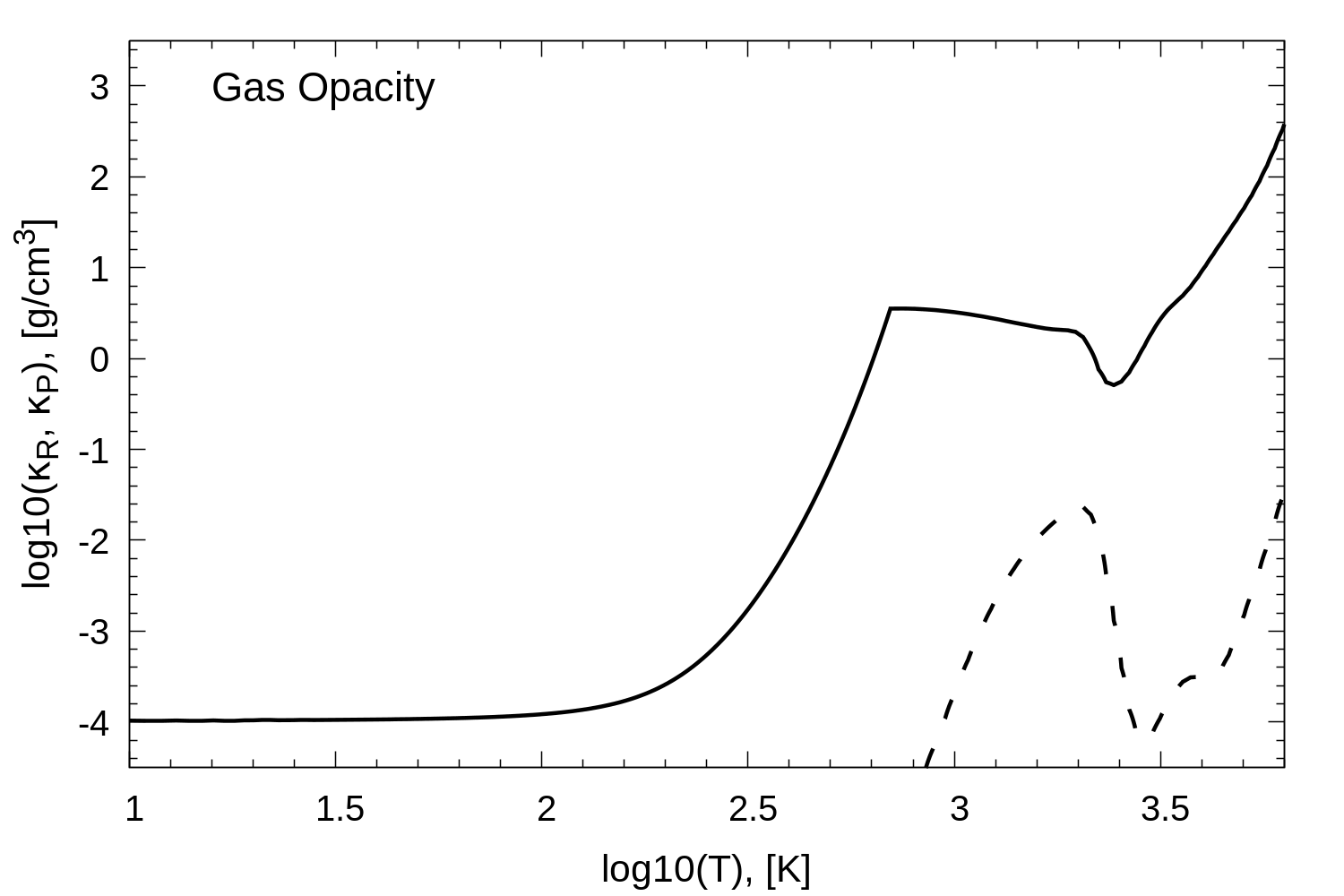}
    \caption{Mean dust and gas opacities adopted for this study. \textit{Top}: Comparison of the mean dust opacities as a function of temperature calculated using the OpTool code for various maximum grain sizes (1~cm, 1~mm, 1~$\mu$m, 10~$\mu$m, 100~$\mu$m). Sharp drops in the opacity are caused by sublimation of dust components, as described in Table~\ref{tab:abundances}. The water ice fraction by mass was set to 40\% in this particular case.
    \textit{Bottom}: Mean gas opacities as a function of temperature for the gas volume density of $10^{-10}$~g~cm$^{-3}$. In both panels, the dashed and solid lines represent the Rosseland mean and Planck mean opacities, respectively.}
    \label{fig:opacity}
\end{figure}

The optical depth to the disk midplane in Eq.~(\ref{eq:cool-heat}) is calculated as $\tau^{\rm d}_{\rm R,P} = 0.5 (\Sigma_{\rm d,gr} + \Sigma_{\rm d,sm}) \kappa^{\rm d}_{\rm R,P}$, where  $\kappa^{\rm d}_{\rm R}$ and $\kappa^{\rm d}_{\rm P}$ are the Rosseland and Planck mean opacities of dust  for the continuous dust size distribution from $a_{\rm min}$ to $a_{\rm max}$.
The values of $\kappa_{\rm P}$ and $\kappa_{\rm R}$  were precomputed using the OpTool  \citep{OpTool2021}, taking only dust absorption into account. The possible effects of scattering will be considered in follow-up studies.
The opacities are computed for a set of distinct dust size distributions. In most cases, the minimum size of dust is set equal to $a_{\rm min}=5\times 10^{-3}$~$\mu$m and the slope of the dust size distribution is always fixed at $p=3.5$. Since $a_{\rm max}$ may vary in wide limits in our hydrodynamic models, we considered a range of values for $a_{\rm max}$ from 1.0~$\mu$m to 10~cm with an increment by a factor of 10. The resulting tabulated dust opacities were used when calculating the optical depths in Eq.~(\ref{eq:cool-heat}). In practice, we choose the opacity table corresponding to the nearest tabulated value of $a_{\rm max}$ (no additional interpolation). 

The adopted composition of dust grains at different temperature bands is listed in Table~\ref{tab:abundances}.  For temperatures below 150 K, the composition of the refractory dust core in our model (40\% astrosil, 10\% FeS, 50\% CHON) is nearly identical to that of \citet{Semenov2003} after excluding water ice (40\% silicates, 9\% troilite, 49\% refractory organics). The main differences are that (i) our model does not include metallic iron as a separate component, and (ii) at temperatures above 425 K, we assume a larger fraction of silicates (80-100\%) compared to \citet{Semenov2003} ($\sim$40\%), who instead retain a significant fraction of metallic iron. These differences are not expected to qualitatively change the opacity behavior, as both models predict that silicates dominate the dust mass at high temperatures.
The sensitivity of dust opacities to the adopted composition is further analyzed in Appendix~\ref{App:opacities}.
Since we explicitly compute the evolution of volatiles in our models (see Eqs~\ref{eq:sig1}--\ref{eq:sig3}), we also consider the contribution of water ice to dust absorption opacities. An additional set of opacity tables is generated with relative contributions of H$_2$O ice to the total mass of dust grains ranging from 0 to 90\% with an increment of 10\%.

The optical depths due to gas opacity are $\tau^{\rm g}_{\rm R,P} = 0.5 \Sigma_{\rm g} \kappa^{\rm g}_{\rm R,P}$, where the Rosseland and Planck mean opacities of gas, $\kappa^{\rm g}_{\rm R}$ and $\kappa^{\rm g}_{\rm P}$, are taken from \citet{Malygin2014}. We note that both gas opacities were calculated down to 700~K, while the disk temperature may be as low as 10~K. Therefore, we artificially decrease the Planck opacity at $T<700$~K to  $10^{-4}$~cm$^2$~g$^{-1}$, a value that ensures small input of gas optical depth compared to that of dust, to avoid the possible uncertainty with its value at lower temperatures. The optical depths in Eq.~(\ref{eq:cool-heat}) are the sum of the corresponding gas and dust optical depths.

Figure \ref{fig:kappavs_size_Temp} shows the Rosseland and Planck mean opacities for dust as a function of the maximum dust size, while Fig.~\ref{fig:opacity} presents the corresponding opacities for dust and gas as a function of temperature. The dust opacity is shown per unit mass of dust while the gas opacity is per unit mass of gas. 
We note that FEOSAD does not consider explicitly dust evaporation in the dust dynamics Eqs.~(\ref{eq:contDsmall}) and (\ref{eq:contDlarge}). Therefore, we must truncate dust opacities above the dust evaporation temperature in order to avoid nonphysical values of $\tau^{\rm d}_{\rm R,P}$ in the hot regions of the disk. We find the dust evaporation temperature using a simple model from \citet{Das2025}, which is based on the ggCHEM astrochemistry code calculations \citep{Woitke2018}. The resulting dust opacities sharply decline at temperatures above the dust evaporation threshold; the exact value of the critical temperature depends on the gas density (see Appendix~D in \citealt{Das2025}).

Figure~\ref{fig:kappavs_size_Temp} demonstrates that at temperatures $T=10$~K and $20$~K the mean dust opacities are insensitive to the choice of  $a_{\rm min}$. Furthermore, they are insensitive to the values of $a_{\rm max}$ up to $a_{\rm max} \approx 10$~$\mu$m. At these low temperatures, the peak of thermal radiation falls at 150-300~$\mu$m, according to Wien's displacement law $\lambda_{\rm max} [\mu m] \approx 3\times 10^3 / (T [K])$. The dust sizes are therefore in the Rayleigh limit with $\lambda_{\rm max} \gg 2 \pi a_{\rm max}$, in which case $\kappa_\nu$ and hence the mean Planck and Rosseland opacities are independent of $a_{\rm max}$.

As $a_{\rm max}$ continues to increase from hundreds of micron to millimeters and beyond, dust grains enter the Mie regime with $\lambda_{\rm max} \ll 2 \pi a_{\rm max}$. After a local increase in the mean opacity around $\lambda_{\rm max} \approx 2 \pi a_{\rm max}$, they begin to decline with growing $a_{\rm max}$, but this change occurs sooner at higher temperatures than at lower ones. This trend is caused by global redistribution of dust grains from smaller to larger sizes as $a_{\rm max}$ increases. The decline in $\kappa_P$ and $\kappa_R$ occurs sooner for higher temperatures because short wavelength radiation, typical of high temperatures, is mostly absorbed by smallest grains, which deplete fastest.

The top panel in Fig.~\ref{fig:opacity}  reflects the trends described above. At moderate and  high temperatures of $T \ge 100$~K,  dust opacity begins to decrease already for $a_{\rm max} \ge 10$~$\mu$m, because dust grains quickly enter the Mie regime.   
At low temperatures ($\le 50$~K), the behavior of dust opacity with increasing $a_{\rm max}$ is more complicated. Dust grains first stay in the Rayleigh regime where they are independent of $a_{\rm max}$, then cross to the Mie regime, showing a temporal rise followed by ultimate decline.

\begin{table*}
\centering
\small
\caption{Model parameters. }
\label{tab:model}
\begin{tabular*}{\textwidth}{@{\extracolsep{\fill}}ccccccccc@{}}
\hline\noalign{\smallskip}
 $M_{\rm core}$ & $\beta$ & $T_{\rm init}$ & $\Omega_{\rm 0}$ & $\Sigma_{\rm 0}$ & $r_{\rm 0}$ & $R_{\rm out}$ & $M_{\rm \star}$  & $M_{\rm disk}$  \\
& & & & & & & (0.6 Myr) & (0.6 Myr) \\
 ($M_{\odot}$) & (\%) & (K) & (km s$^{-1}$ pc$^{-1}$) & (g cm$^{-2}$) & (au) & (au) & ($M_{\odot}$) & ($M_{\odot}$) \\
\hline\noalign{\smallskip}
1.00 & 0.1 & 15 & 0.93 & 0.12 & 1540 & 9282 & 0.70 & 0.30 \\
\noalign{\smallskip}\hline
\end{tabular*}
\tablefoot{$M_{\rm core}$ is the initial core mass; $\beta$ is the ratio of rotational energy to gravitational energy; $T_{\rm init}$ is the initial gas temperature equal to the background radiation temperature, $T_{\rm bg}$; $\Omega_{\rm 0}$ is the angular velocity of the core center; $\Sigma_{\rm 0}$ is the gas surface density at the core center; $r_{\rm 0}$ is the radius of the central plateau in the initial core; and $R_{\rm out}$ is the outer boundary of the computational domain. $M_{\rm \star}$ and $M_{\rm disk}$ are the masses of the central star and the disk at the end of the simulation (600\,kyr).}
\end{table*}

Regarding the minimum dust size $a_{\rm min}$, Fig.~\ref{fig:kappavs_size_Temp} demonstrates that the dust opacities are weakly sensitive to its value for as long as temperature stays below 100~K. The deviation becomes notable only for $T\ge 100$~K. Nevertheless, in this study we fixed $a_{\rm min}$ at $5\times 10^{-3}$~$\mu$m, and we will consider the effects of the varying $a_{\rm min}$ on the snow lines in a future study.

\subsection{Initial and boundary conditions}
Simulations start from the gravitational collapse of a flattened prestellar cloud core with a mass of 1.0~$M_\odot$, consisting of gas, dust, and volatiles. The core rotation is introduced by setting the ratio of rotational-to-gravitational energy $\beta = 10^{-3}$, which is within the limits inferred from prestellar cloud cores \citep{2002Caselli}.
All dust is initially in the form of small dust grains and all volatiles are on the surface of small dust grains. Initial abundances of volatile species were taken from table~1 in \citet{Eistrup2016}. The initial dust-to-gas ratio is 1:100. 
The gas surface density and angular velocity of the natal prestellar core are distributed as follows \citep{1997Basu}:
\begin{equation}
    \Sigma_{\rm g}(r)=\frac{r_0\Sigma_{\rm 0}}{\sqrt{r^2+r_0^2}},
\label{eq:coredens}
\end{equation}
\begin{equation}
    \Omega_{\rm g}(r)=2\Omega_{\rm 0}\bigg(\frac{r_0}{r}\bigg)^2\left[\sqrt{1+\left(\frac{r}{r_0}\right)^2}-1\right],
\label{eq:corevel}
\end{equation}
where $\Sigma_{\rm 0} = 0.12$~g~cm$^{-2}$ is the surface density and $\Omega_{\rm 0} = 0.93$~km~s$^{-1}$~pc$^{-1}$ is the angular velocity, both of which are defined at the center of the cloud core. The radius of the near-uniform region in the center of the core is $r_{0} = 1540$~au.  The parameters of the cloud core are summarized in Table~\ref{tab:model}.

As the core contracts gravitationally, it spins up and a centrifugal disk forms when the in-spiralling gas hits the centrifugal barrier near the inner computational boundary, which is set equal to the radius of the sink cell $r_{\rm sc}=0.2$~au.  Subsequently, the disk grows in size and mass owing to infall from the cloud, while the central star gains mass via accretion through the inner computational boundary. Free mass exchange (inflow and outflow) across the sink-disk interface is introduced as the inner boundary condition \citep[see][for details]{2018VorobyovAkimkin}. 
The outer boundary condition allows free mass outflow, but mass inflow from outside the computational domain is prohibited.

\begin{figure}    
\includegraphics[width=1\columnwidth]{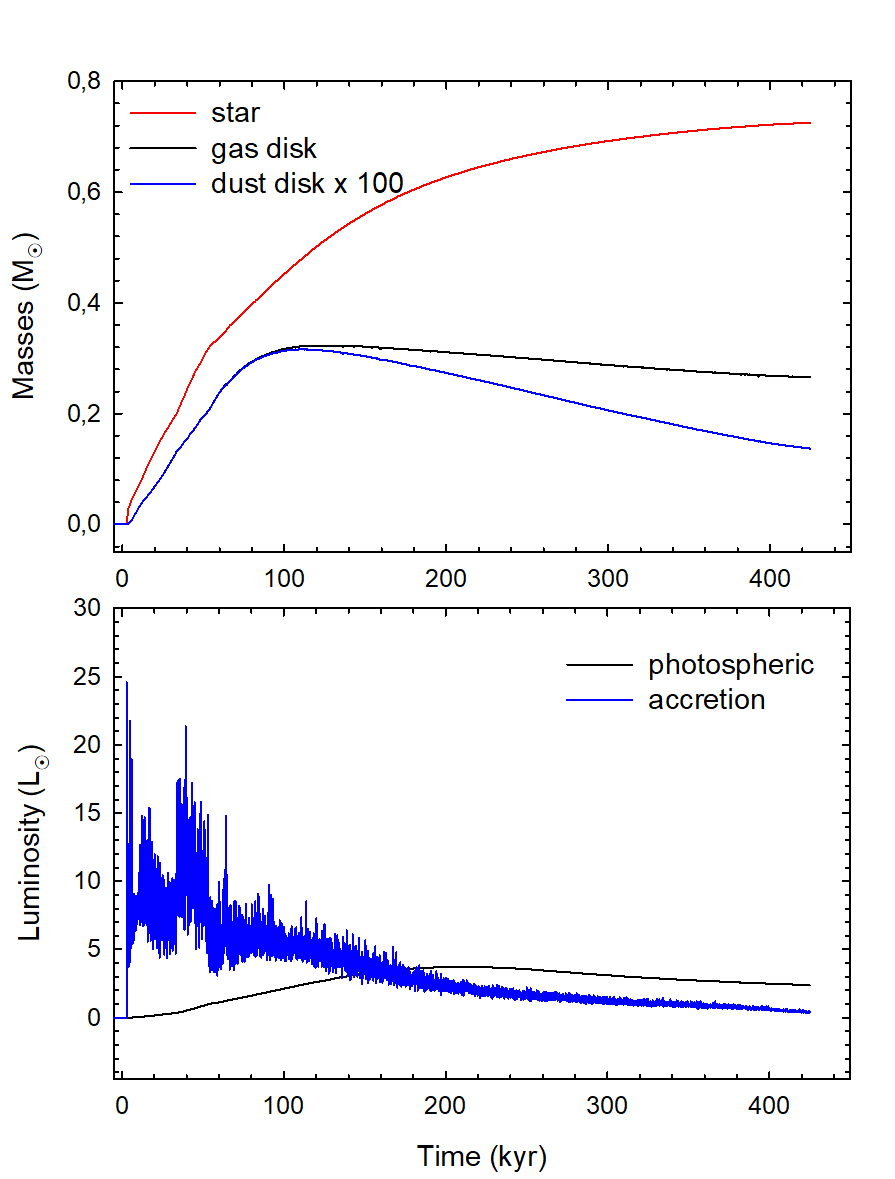}
\caption{Time evolution of the stellar and protoplanetary disk properties. \textit{Top}: Stellar mass, gas disk mass, and dust disk mass (multiplied by 100). \textit{Bottom}: Accretion and photospheric luminosities of the protostar. The time is counted from the onset of the protostar formation. }
\label{fig:lum-mass}
\end{figure}

\begin{figure*}    
\includegraphics[width=2\columnwidth]{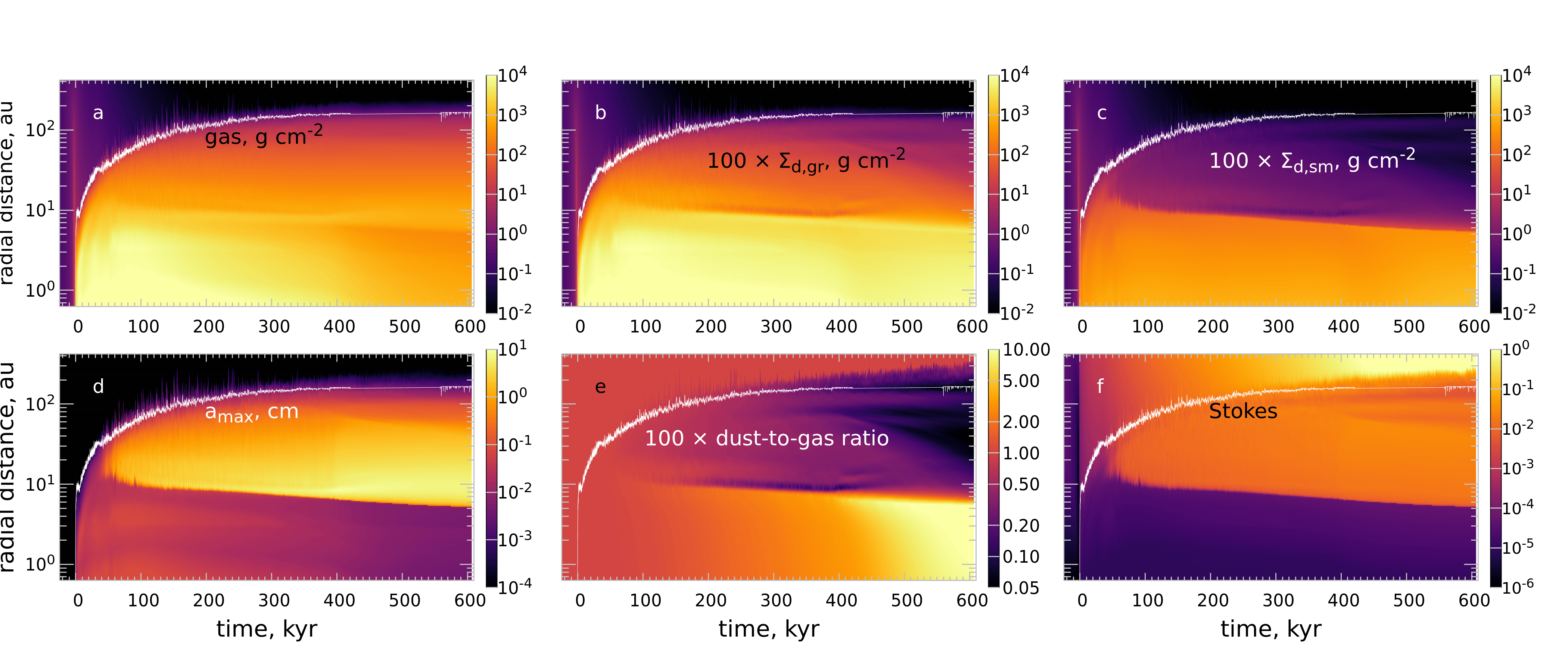}
\caption{Evolution of the azimuthally averaged radial distributions of gas and dust. The panels show the (a) surface density of gas, (b) surface density of grown dust, (c) surface density of small dust, (d) maximum dust size, (e) dust-to-gas mass ratio, and (f) Stokes number. The white contour line shows the outer radius of the gas disk defined as a distance where $\Sigma_{\rm g}$ drops below 0.1~g~cm$^{-2}$. }
\label{fig:diskevol}
\end{figure*}

\section{Dust-size-dependent opacity model}
\label{Sect:results}
In Sects.~\ref{Sect:star-disk}--\ref{Sect:shape-snowline}, we consider the evolution of the fiducial model, in which dust-size-dependent opacities are used to  calculate the cooling and heating rates $\Lambda$ and $\Gamma$ in Eq.~(\ref{eq:energ}). A comparison model that uses simplified dust opacities, which do not depend on the dust size distribution, is also presented in Sect.~\ref{Sect:fixed-amax}. Implications of dust-size-dependent opacities for pebbles and planetesimal formation are discussed in Sect.~\ref{Sect:Discuss}.

\subsection{Evolution of stellar and disk characteristics}
\label{Sect:star-disk}

We start with showing in Fig.~\ref{fig:lum-mass} the time evolution of the integrated masses of gas and dust in the disk, along with the mass and luminosity of the protostar.
The stellar mass increases gradually via accretion from the disk until it reaches $\approx 0.74~M_\odot$ when we stop the simulations. The initial increase in the mass of the gas and dust disks is followed by their gradual decline after the disk infall phase ends. In particular, the mass of dust in the disk declines faster than that of gas, signaling an onset of radial dust drift due to dust growth and decoupled dust dynamics. The stellar accretion luminosity initially dominates that of the photospheric luminosity, and also shows moderate variability caused by the perturbing influence of spiral density waves in the initially gravitationally unstable disk \citep{2016ARep...60..879E}. The photospheric luminosity begins to dominate after approx 160~kyr from the epoch of star formation. After $t\approx 80$~kyr, the net luminosity of the protostar gradually declines with time, providing also less heating to the disk with time.

\subsection{Evolution of azimuthally averaged disk properties}

Figure~\ref{fig:diskevol} presents the space-time diagrams showing the evolution of azimuth-averaged disk properties in the fiducial model. The time is counted from the instance of the central star formation. The disk forms about 1.0~kyr later as manifested by a sharp increase in the gas and dust surface densities in the inner 10~au.  By the end of simulations, the disk extends to $\approx 150$~au as indicted by the white curve. 

Notable changes in the values of dust surface density, dust maximum size, dust-to-gas mass ratio, and Stokes number, defined as $\mathrm{St}=\Omega_{\rm K} t_{\rm stop}$, where $\Omega_{\rm K}$ is the Keplerian angular velocity and $t_{\rm stop}$ is the stopping time (see Eq.~\ref{Eq:stop-time}), develop in the course of time in the vicinity of 10~au. 
This phenomenon is consistent with the ``many-seeds'' model of dust grains\citep{Schoonenberg2017,2017A&A...605L...2S}, in which large dust aggregates are held together by a water ice matrix. When these aggregates cross the snow line, the water ice sublimates, causing the aggregate to quickly break apart into its constituent silicate monomers. This process is implemented in our model by instantly reducing the maximum dust size $a_{\mathrm{max}}$ to match the lower fragmentation limit for bare grains when the water ice mantles are lost (see Sect.~\ref{sect:dustgrowth} for details). The Stokes number of bare grains drops below $10^{-3}$, greatly decelerating dust inward drift and causing  dust particles to pile up interior to the water snow line. The effect is also evident in the dust-to-gas plot, revealing 
dust depletion beyond the front and dust accumulation interior to it. The enhancement of dust interior to the water snow line builds up with time and the position of the front gradually shifts toward the star because of the overall disk cooling with time.

\begin{figure}   
    \centering
    \includegraphics[width=1\columnwidth]{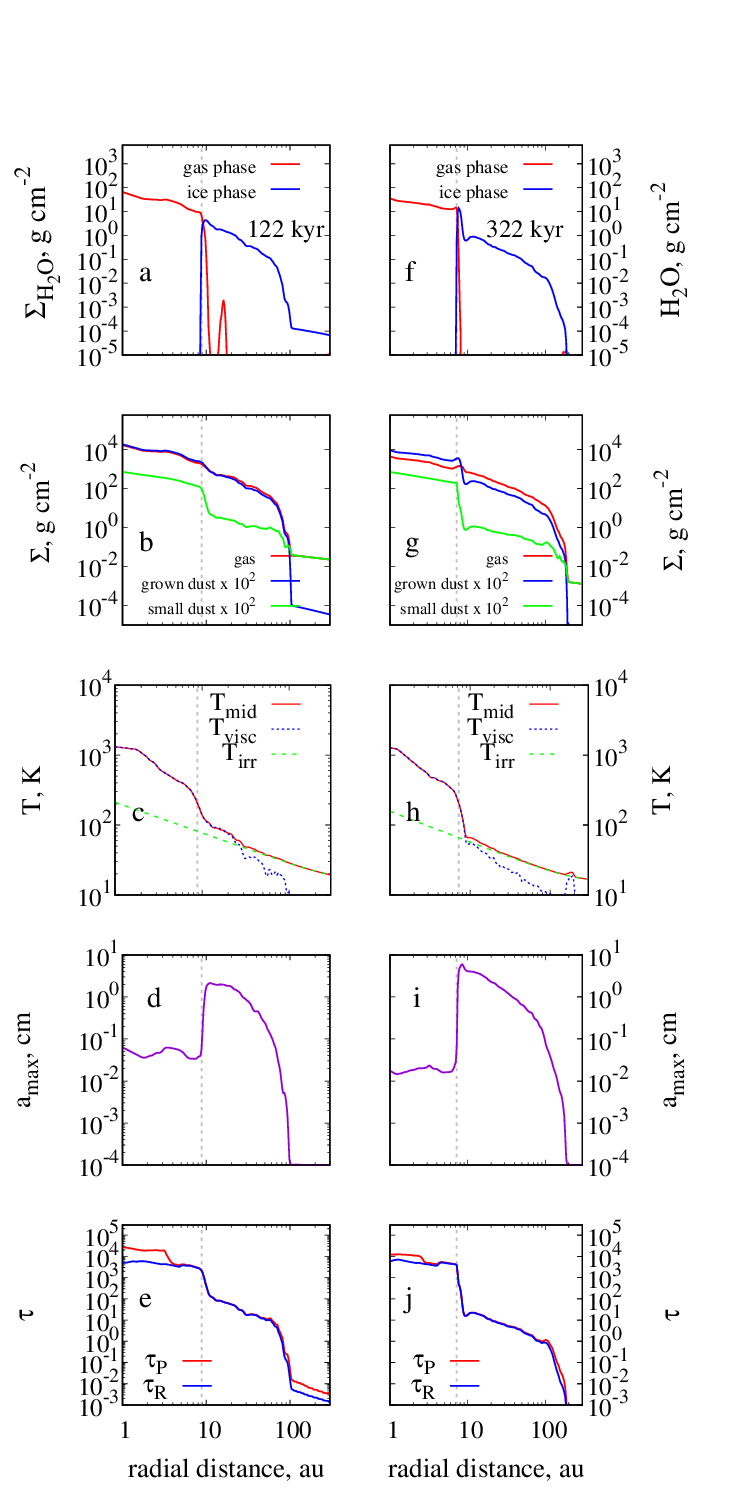}
    \caption{Radial profiles of gas, dust, and H$_2$O in the young (122~kyr, \textit{left column}) and evolved (322~kyr, right column) disks. Panels from top to bottom show the surface densities of water vapor and ice (a and f), the surface densities of gas, small and grown dust (b and g; note that dust is scaled up by a factor of 100), the gas temperature (c and h), the maximum size of dust grains (d and i), and the optical depth to the disk midplane (e and j). The vertical dotted lines designate the position of the water snow line. Note that the water snow line is located in the part of the disk where viscous heating dominates.} 
    \label{fig:radial}
\end{figure}

To quantify the trends described above, Fig.~\ref{fig:radial}  shows the azimuthally averaged radial profiles of various disk properties at two fixed time instances after the epoch of star formation: $t=122$~kyr and $t=322$~kyr. For our in-depth analysis, we chose a young disk because planet formation can occur within the first few hundred thousand years, as suggested by presence of gas and rings in young T~Tauri disks \citep{2015ALMABrogan}.
Later evolution times are briefly analyzed in Appendix~\ref{App:later-evolution} for consistency.
The vertical dotted lines mark the position of the water snow line defined here as the radial distance where the azimuthally averaged surface densities of water vapor ($\overline{\Sigma}_{\rm H2O}^{\rm gas}$) and water ice ($\overline{\Sigma}_{\rm H2O}^{\rm dust}$) become equal each other (see the top panels). 
In the younger disk,  grown dust is slightly depleted  beyond the water snow line (with respect to the initial 1:100 ratio) and enhanced interior to it, indicating that dust drift has only just begun to influence the radial distribution of dust, but little features are yet visible at the snow line. On the contrary, small dust has already developed a notable drop in surface density by about an order of magnitude across the water snow line. We note, however, that only a small fraction of total dust mass resides in the small dust population.
In the older disk at $t=322$~kyr,
both small and grown dust develop a strong drop in surface density across the snow line. Notable dust enhancement interior of the snow line and a strong depletion exterior to it develop at this evolution stage. 

\begin{figure}   
    \centering
    \includegraphics[width=1\columnwidth]{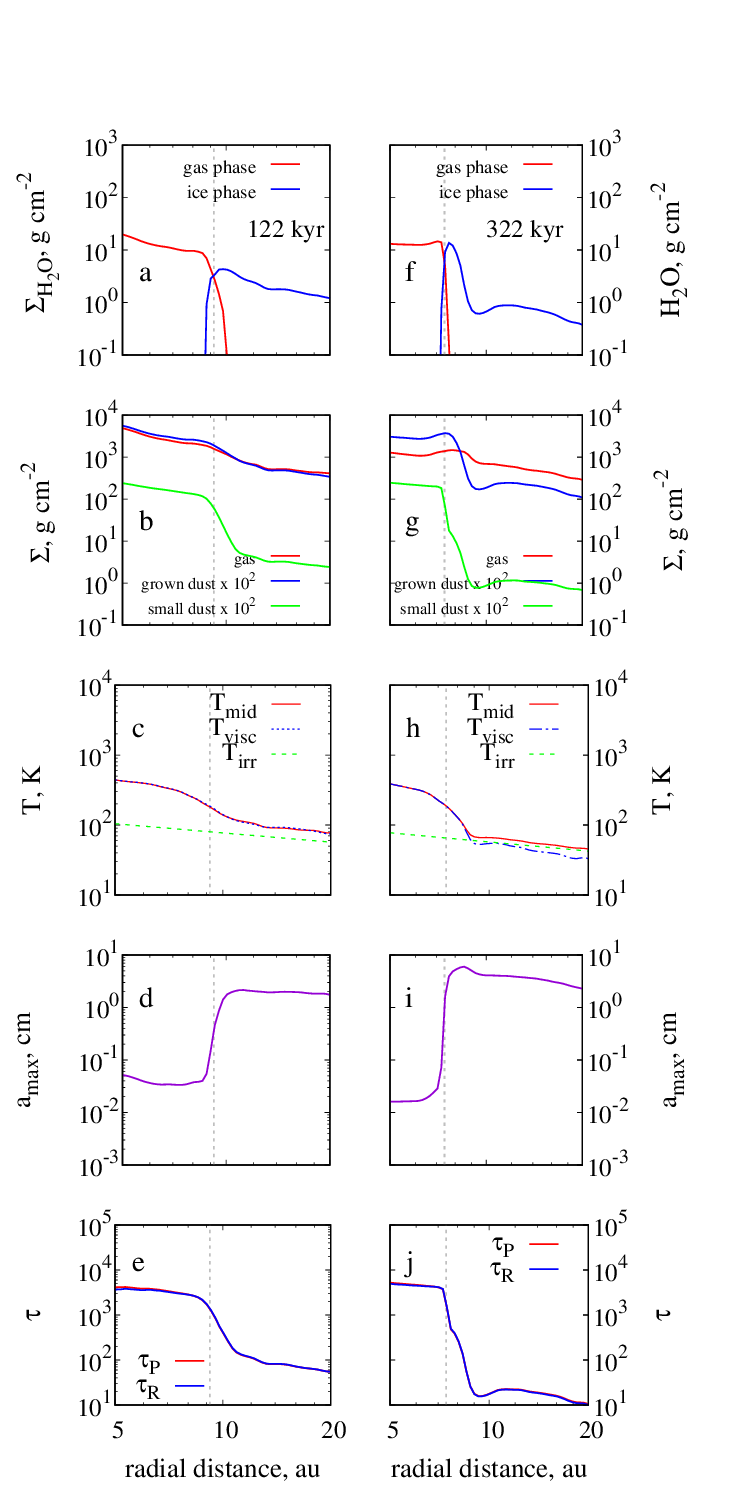}
    \caption{Similar to Fig.~\ref{fig:radial} but zoomed in to the vicinity of the water snow line. }
    \label{fig:radial-zoom}
\end{figure}

Other dust and gas disk properties also develop  specific features in the vicinity of the water snow line with time. The maximum size of dust grains sharply rises across the water snow line and the magnitude of the effect increases with time (2.5 orders of magnitude at $t=322$~kyr against 1.5 orders of magnitude at $t=122$~kyr). The gas temperature steeply drops exterior to the snow line and this effect is caused by the corresponding behavior in the disk optical depth.  The entire disk, apart from outermost parts near the disk outer edge, is optically thick in the Rosseland and Planck sense at these evolutionary stages (but see Appendix~\ref{App:later-evolution}), and interior to the snow line the optical depth is particularly high (see the bottom panels). Exterior to the snow line, however, the optical depth quickly decreases. This effect is partly caused by decreasing surface density of dust, but a drop in the dust opacity as the maximum dust size increases behind the snow line also contributes to this effect. Lower optical depth makes it easier for the heat generated in the disk midplane by viscosity and PdV work to escape the disk, thus lowering the disk temperature behind the water snow line.

\begin{figure}   
\includegraphics[width=1\columnwidth]{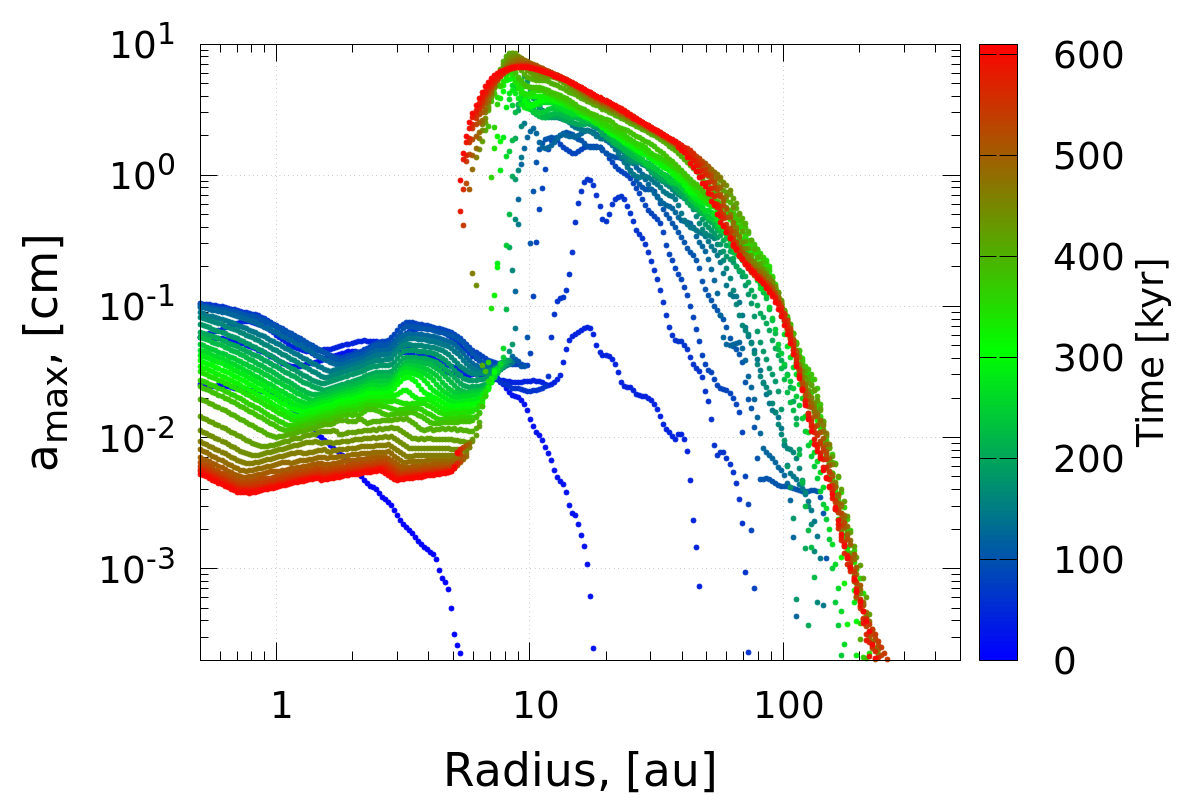}
\caption{Radial profiles of the maximum dust size at different time instances during the disk evolution. A sharp rise in $a_{\rm max}$ develops across the water snow line.} 
\label{fig:dust-size-evol}
\end{figure}

\begin{figure*}   
\includegraphics[width=2\columnwidth]{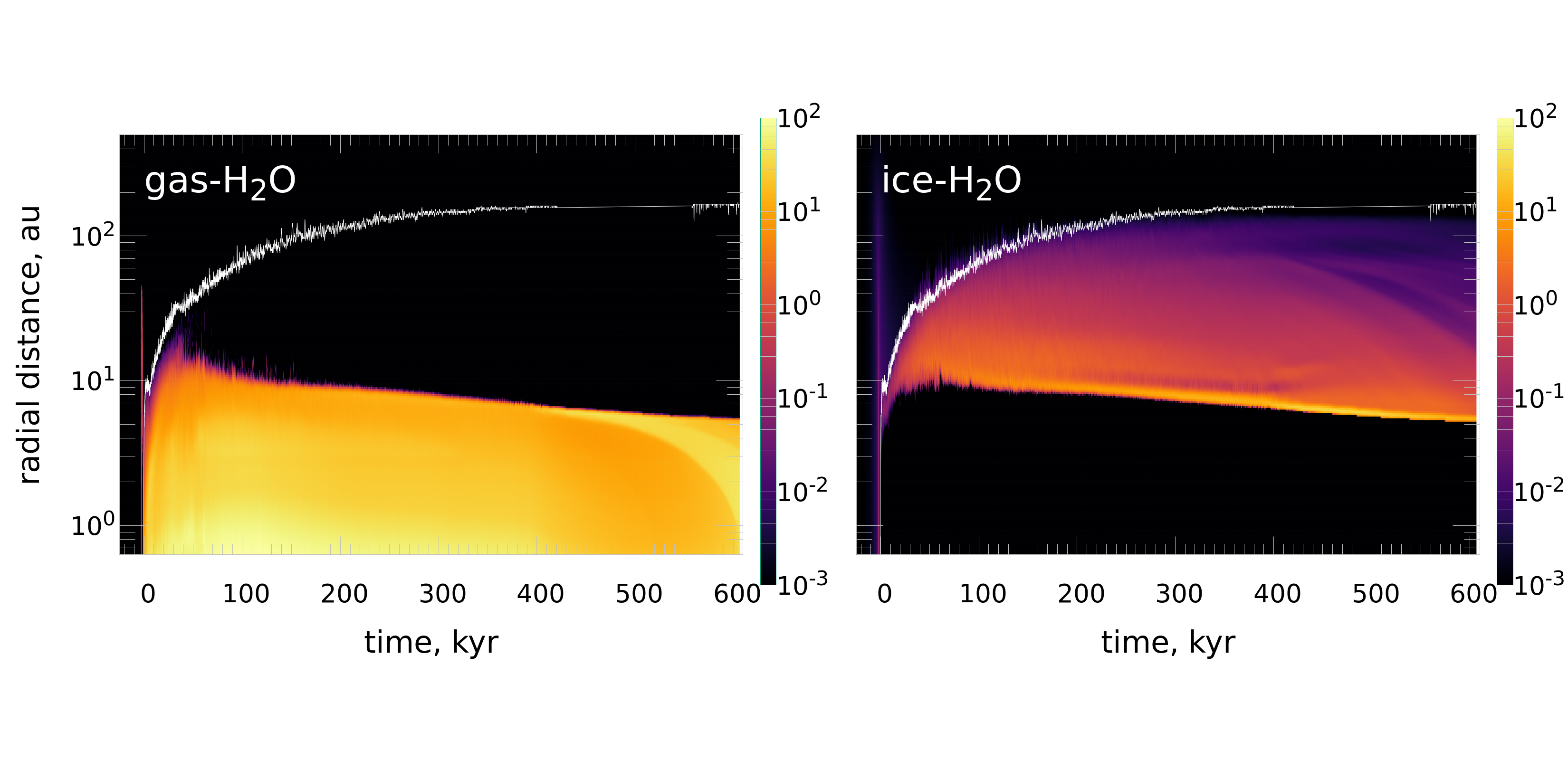}
\caption{Time evolution of the azimuthally averaged surface densities of water in the gas (\textit{left}) and ice (\textit{right}) phases. The white curve outlines the radial extent of the gas disk. The scale bars are in gram per centimeter squared. }
\label{fig:chem}
\end{figure*}

Figure~\ref{fig:radial-zoom} shows a zooms-in on the radial profiles in the vicinity of the water snow line to illustrate additional features. In particular, local maxima in the surface density of water ice, grown dust, and dust size develop just exterior to the snow line.  These peaks are caused by small dust and gas-phase water diffusing  across the snow line in the direction opposite to the usual inward dust drift \citep[e.g.,][]{Schoonenberg2017}. The physical mechanisms behind diffusion can be local turbulence, set in our model with $\alpha_{\rm visc}=10^{-3}$, or  azimuthal variations in the radial velocities of dust and volatiles caused by spiral density waves \citep{Molyarova2021}. Once water vapor finds itself behind the snow line, it freezes out onto dust grains and promotes further growth of dust particles. Small dust particles diffusing behind the snow line also provide building blocks for further dust growth.
The accumulation of water ice behind the snow line is particularly prominent (see the top panels), reaching almost an order of magnitude in the 322-yr old disk.

Panels (c) and (h) in Figs.~\ref{fig:radial} and~\ref{fig:radial-zoom} show the comparison of the gas midplane temperature obtained in our hydrodynamic model (red solid line) with the temperature expected from analytical estimates based on viscous heating alone, $T_{\rm vis}$ (blue dotted line). The latter is derived by balancing the viscous heating rate per disk unit area, $Q_{\rm vis}^+ = (9/4) \nu \Sigma_{\rm g} \Omega_{\rm K}^2$, with the radiative cooling rate from Eq.~(\ref{eq:cool-heat}), assuming no other heating/cooling sources. The green dashed line shows the irradiation temperature at the disk surface, $T_{\rm irr}$, which is calculated from Eqs.~(\ref{fluxCS}) and~(\ref{fluxF}) following \citet{2018VorobyovAkimkin}.

At both evolutionary times (122 kyr and 322 kyr), the inner disk region ($r \lesssim 10-30$~au) is clearly dominated by viscous heating, as $T_{\rm gas}$ closely follows $T_{\rm vis}$ and significantly exceeds $T_{\rm irr}$. In contrast, the outer disk ($r \gtrsim 10-30$~au) is irradiation-dominated, with $T_{\rm gas}$ approaching $T_{\rm irr}$. The transition region between these two regimes is marked by a flattening or change in the slope of the temperature radial profile. At an earlier time (122 kyr), the region of viscous heating dominance extends to 25-30 au, and at a later time (322 kyr), only to 8-9 au. This is due to the fact that the disk mass decreases over time after $t\approx 100$~kyr (see Fig.~\ref{fig:lum-mass}), and therefore the gas surface density $\Sigma_{\rm g}$ also decreases. As a consequence, the contribution of viscous heating ($Q_{\rm vis}^+=(9/4) \alpha_{\rm visc} c_{\rm s}^2 \Sigma_{\rm g} \Omega_{\rm K}$) to the overall energy balance in the disk decreases. This causes a negative feedback loop, further reducing the viscous heating due to decreasing disk temperature ($T \propto c_{\rm s}^2$). An increase in $\Omega_{\rm K}~\propto \sqrt{M_\ast}$ over time cannot reverse the overall trend. 

The vertical dotted lines in Figs.~\ref{fig:radial} and~\ref{fig:radial-zoom} indicate the position of the water snow line (where the surface density of water ice and vapor are equal). Notably, the snow line is located firmly within the viscously heated, inner region of the disk at both times. This trend continues to the further evolution of the disk; see Appendix~\ref{App:later-evolution}. 
The water snow line's position is therefore not directly set by stellar irradiation, at least at the considered stages of disk evolution, but by the interplay of viscous heating, which is determined by disk dynamical and thermal properties, and radiative cooling,  which is sensitive to the dust size distribution through its effect on dust opacity (as shown in Fig.~\ref{fig:opacity}). This emphasizes the complex feedback loop between dust growth, dust opacity, disk dynamics, thermal structure, and the location of the water snow line in the FEOSAD model.   

Figure~\ref{fig:dust-size-evol} illustrates the evolution of the maximum dust size $a_{\rm max}$ over time. We recall that initially $a_{\rm max} = 1.0~\mu$m, but dust quickly grows as disk forms and evolves, an effect also seen in three-dimensional disk formation models \citep{Lebreuilly2020,Bate2022,Vorobyov2025}.  A steep rise in $a_{\rm max}$ quickly develops at the position of the water snow line around $r=6-15$~au and its magnitude increases with time. By the end of simulations, the contrast in $a_{\rm max}$ across the water snow line amounts to three orders of magnitude, while the maximum dust size just behind the snow line reaches 10~cm and drops to 0.1~mm just interior to it. Such a sharp transition in $a_{\rm max}$ across the water snow line was also reported in simplified one-dimensional disk evolution models \citep[e.g.,][]{Pinilla2017}.

Figure~\ref{fig:chem} displays the space-time diagram of azimuthally averaged surface density of H$_2$O in the gas and ice phases.  
In the initial 10-20~kyr of disk evolution, the separation front between the water vapor and water ice closely follows the disk's outer edge, 
which is outlined by the white curve. This
indicates that the young disk is compact and hot, so that the water snow line shifts to the outer disk regions. In the subsequent evolution, however, the radial positions of the disk's outer edge and water snow line diverge, with the former continuing to increase due to mass (and angular momentum) infall from the envelope and viscous spreading, while the latter is gradually moving inward from about 15~au to 6~au due to disk cooling.  Water is brought to the disk outer regions with accreted envelope material; it freezes out onto dust grains and drifts inward with grown dust.  As icy dust grains cross the water snow line, the water ice turns into vapor, diffuses back, and freezes onto dust grains again, forming a notable enhancement in water ice just exterior to the water snow line. This enhancement builds up over time and becomes clearly visible in the right panel at $t>100$~kyr.

\begin{figure*}    
\includegraphics[width=2\columnwidth]{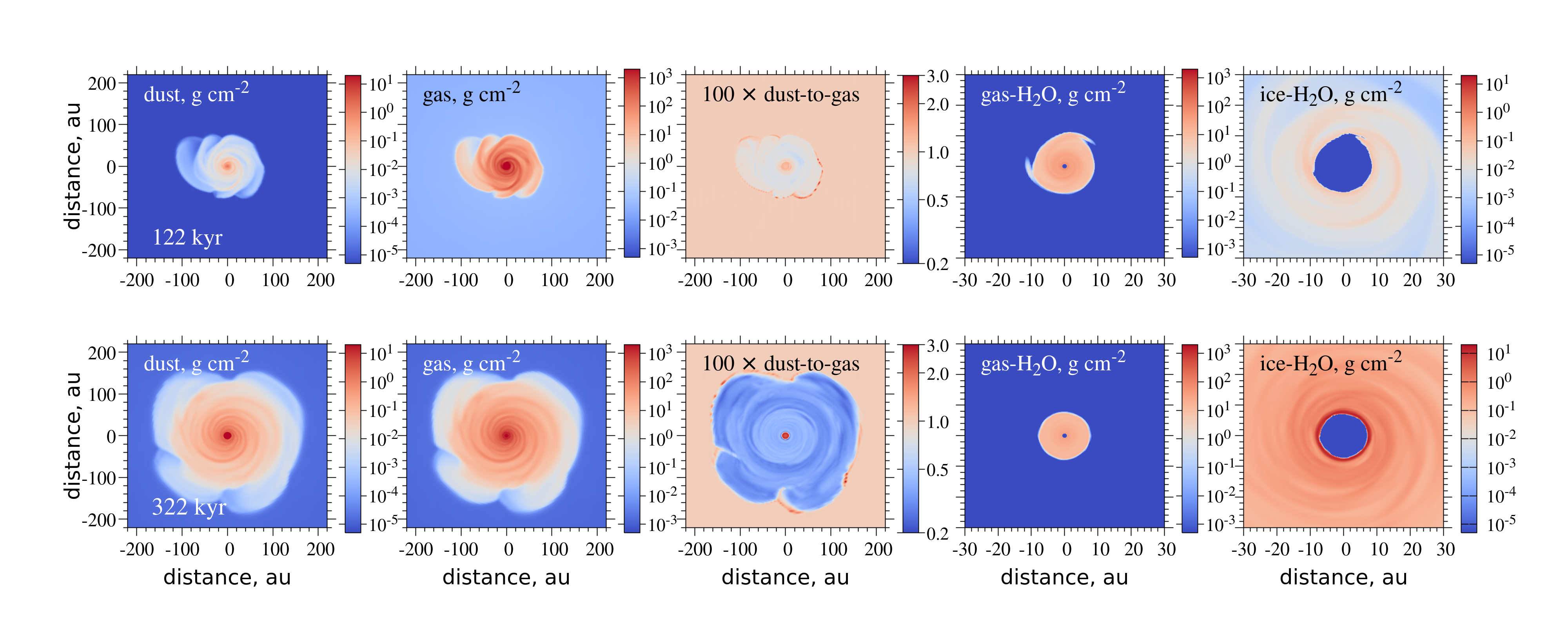}
    \caption{Spatial distributions of gas, dust, and water ice at t=122~kyr (\textit{top row}) and 322~kyr (\textit{bottom row}). Panels from left to right show the gas surface density, dust surface density, dust-to-gas mass ratio (multiplied by 100), surface density of water in the gas phase, and the surface density of water ice on dust grains. Note the difference in the spatial scales between the panels. The star is located in the coordinate center. The scale bars for surface densities are in gram per centimeter squared. } 
    \label{fig:2D}
\end{figure*}

\subsection{Two-dimensional shape of the water snow line}
\label{Sect:shape-snowline}
In the previous section we presented the azimuthally averaged disk properties as a function of time and radial distance from the star. However, the disk in our models is generally not axisymmetric owing to the development of gravitational instability at the early stages of evolution. The variations in disk properties in the azimuthal direction that naturally occur in gravitationally unstable disks can lead to interesting effects on the geometrical shape of the water snow line in the disk midplane\footnote{We use this term as a general definition of the disk region where water ice turns into water vapor.}, which we discuss below.

The two-dimensional spatial distribution of gas, dust, and water is shown in Fig.~\ref{fig:2D}  at the same time instances $t=122$ and 322~kyr as in Figs.~\ref{fig:radial} and \ref{fig:radial-zoom}. The disk is gravitationally unstable at both evolutionary times, but the younger disk has a sharper spiral pattern than the older one, with a larger contrast in gas and dust density between the arms and the inter-arm regions. This indicates that the gravitational instability, sustained by continuing mass loading from the infalling envelope \citep{VorobyovBasu2005}, begins to weaken by $t=322$~kyr when the envelope has mostly dissipated. The younger disk shows only first signatures of spatial differentiation in the dust-to-gas mass ratio across the water snow line, while the older disk is already strongly depleted in dust behind the snow line and is enhanced in dust interior to it. 

Interestingly, the water sublimation front, as indicated by the sharp drop in the surface density of icy H$_2$O on dust grains in the right-hand-side panel of Fig.~\ref{fig:2D}, has a profoundly noncircular shape in the younger disk. This effect, also emphasized in \citet{Molyarova2021}, is caused by the non-axisymmetric temperature distribution in the gravitationally unstable disk heated by spiral density waves. Enhancement in water ice  near the water sublimation front becomes apparent only when the front attains a circular shape in the older disk; see the bottom-right panel. The lack of such a strong enhancement in the younger disk; see the top-right panel, is partly because dust drift has not yet brought enough water ice to the snow line and partly due to the agile nature of the water sublimation front itself, constantly changing its shape to adjust to the spiral pattern varying on orbital timescales.

\begin{figure}  
    \centering
    \includegraphics[width=\columnwidth]{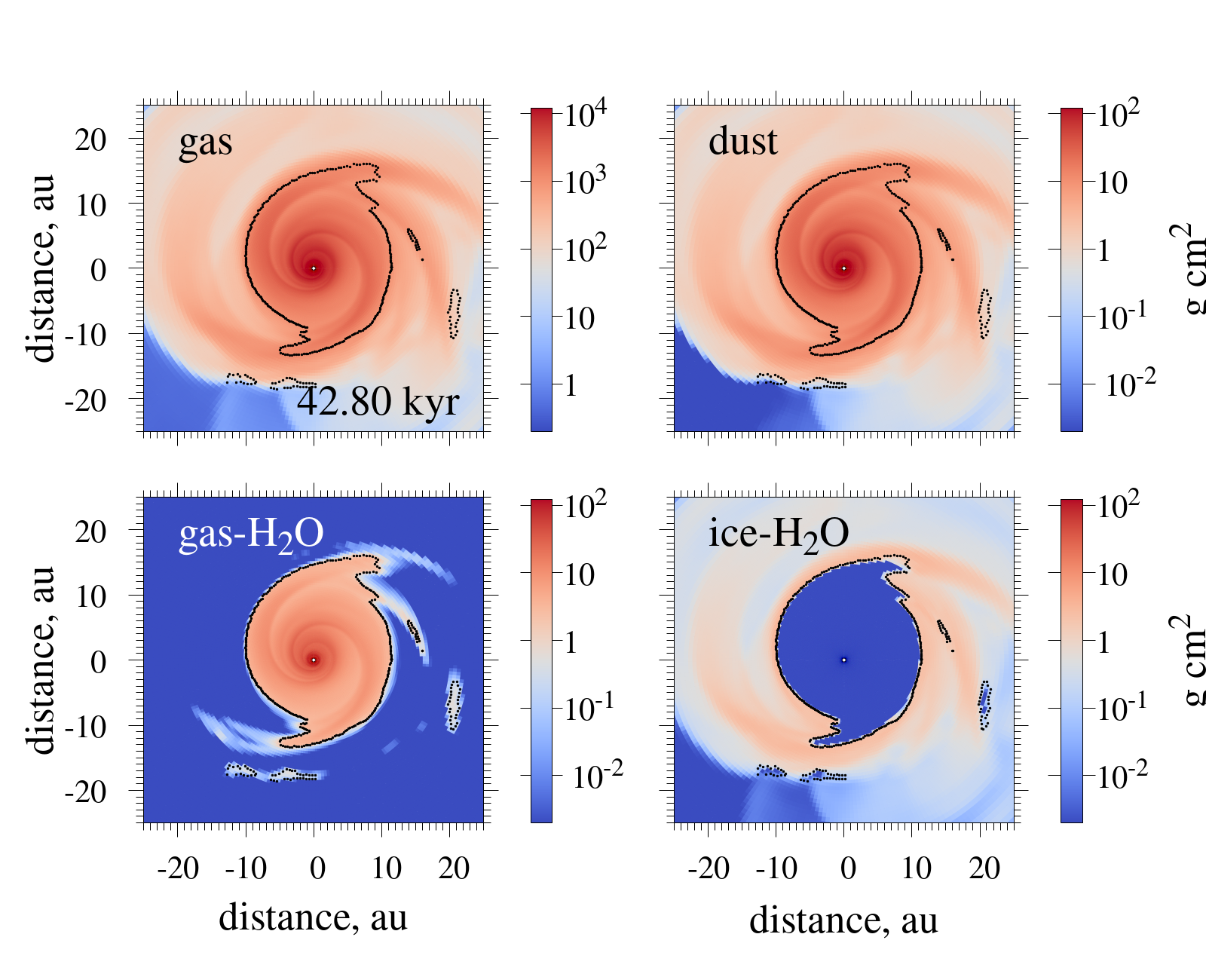}
    \caption{Spatial distribution of the surface densities of gas (\textit{top left}), grown dust (\textit{top right}), water vapor (\textit{bottom left}), and water ice (\textit{bottom right}) in a 42.8-kyr old disk. The black contours outline the position of the water snow lines. The scale bars are in gram per centimeter squared.}
    \label{fig:2Dzoom}
\end{figure}

\begin{figure*}  
    \centering
    \includegraphics[width=2\columnwidth]{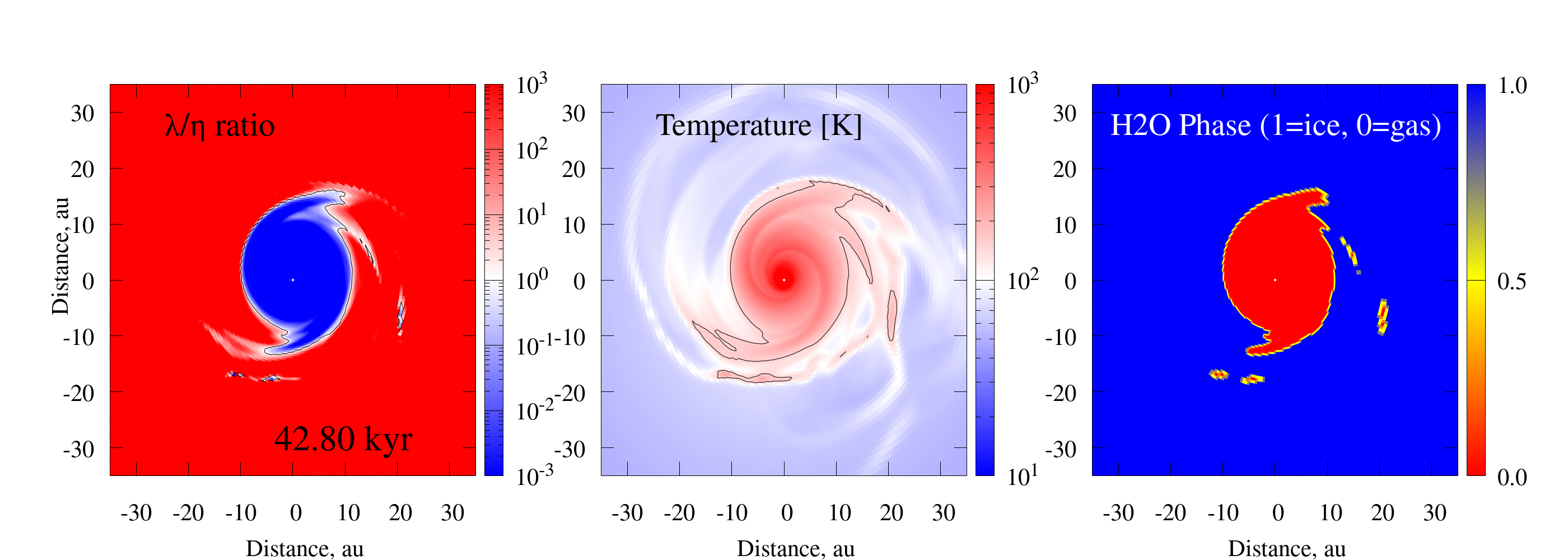}
    \caption{Origins of the secondary water vapor patches in the young ($t=42.80$~kyr) gravitationally unstable disk. \textit{Left}: Ratio of the adsorption rate to the thermal desorption rate for grown dust, $\lambda/\eta$. Red ($\lambda/\eta>1$) indicates regions where adsorption dominates, favoring water ice. Blue ($\lambda/\eta<1$) indicates regions where thermal desorption dominates, favoring water vapor. \textit{Middle}: Gas temperature distribution.  \textit{Right}: Fraction of water in the ice, $\Sigma_{\rm H_{2}O,ice}/(\Sigma_{\rm H_{2}O,gas} + \Sigma_{\rm H_{2}O,ice})$. The black contours indicate the position of the water snow line where $\Sigma_{\rm H_{2}O,ice} = \Sigma_{\rm H_{2}O,gas}$. }
    \label{fig:2Dads_des}
\end{figure*}

In figures that illustrate this feature (e.g., Figs~\ref{fig:2Dzoom} and \ref{fig:2Dads_des}), we calculate the two-dimensional position of the snow line at each azimuthal angle of the polar coordinate system as a radial distance where the local surface density of water ice on small and grown grains equals that of water vapor.

Figure~\ref{fig:2Dzoom} shows a zooms-in in on the gas, dust, and water spatial distribution in a young 43-kyr-old disk where the azimuthal variations in the water sublimation front are particularly profound. In particular, we calculate the two-dimensional position of the snow line at each azimuthal angle of the polar coordinate system as a radial distance where the local surface density of water ice on small and grown grains equals that of water vapor. The black curve outlines the resulting snow line  and its shape is profoundly non-axisymmetric.
We note that the star is located in the coordinate center, which highlights the noticeable deviation of the snow line from the conventional circular geometry.
In addition, there are several secondary water vapor spots located at disk regions exterior to the main snow line and surrounded by their own secondary snow lines.

The origin of these secondary water vapor spots can be understood by analyzing the rates of adsorption ($\lambda$) and thermal desorption ($\eta$), which govern the phase transition of water \citep[see, e.g., Eqs.~20 and~27 in][]{Topchieva2024}. Figure~\ref{fig:2Dads_des} shows a multi-panel view of the same time instance $t=42.8$~kyr, including the ratio $\lambda/\eta$ for grown dust and the gas temperature distribution. The comparison reveals a perfect correlation: every secondary water vapor patch (red areas beyond the main snow line in the right panel) corresponds exactly to a region where the $\lambda/\eta$ ratio drops below unity (blue spots in the left panel). This indicates that in these locations thermal desorption of water ice locally dominates over adsorption of water vapor. By comparing with the middle panel, we see that these desorption-dominated regions are precisely the locations where the gas temperature is locally elevated due to compressional heating in spiral arms. Thus, the non-axisymmetric temperature structure of the gravitationally unstable disk creates localized ``hot spots'' where water ice can be re-evaporated, forming isolated vapor islands well beyond the main snow line.

\begin{figure}  
    \centering
    \includegraphics[width=1\columnwidth]{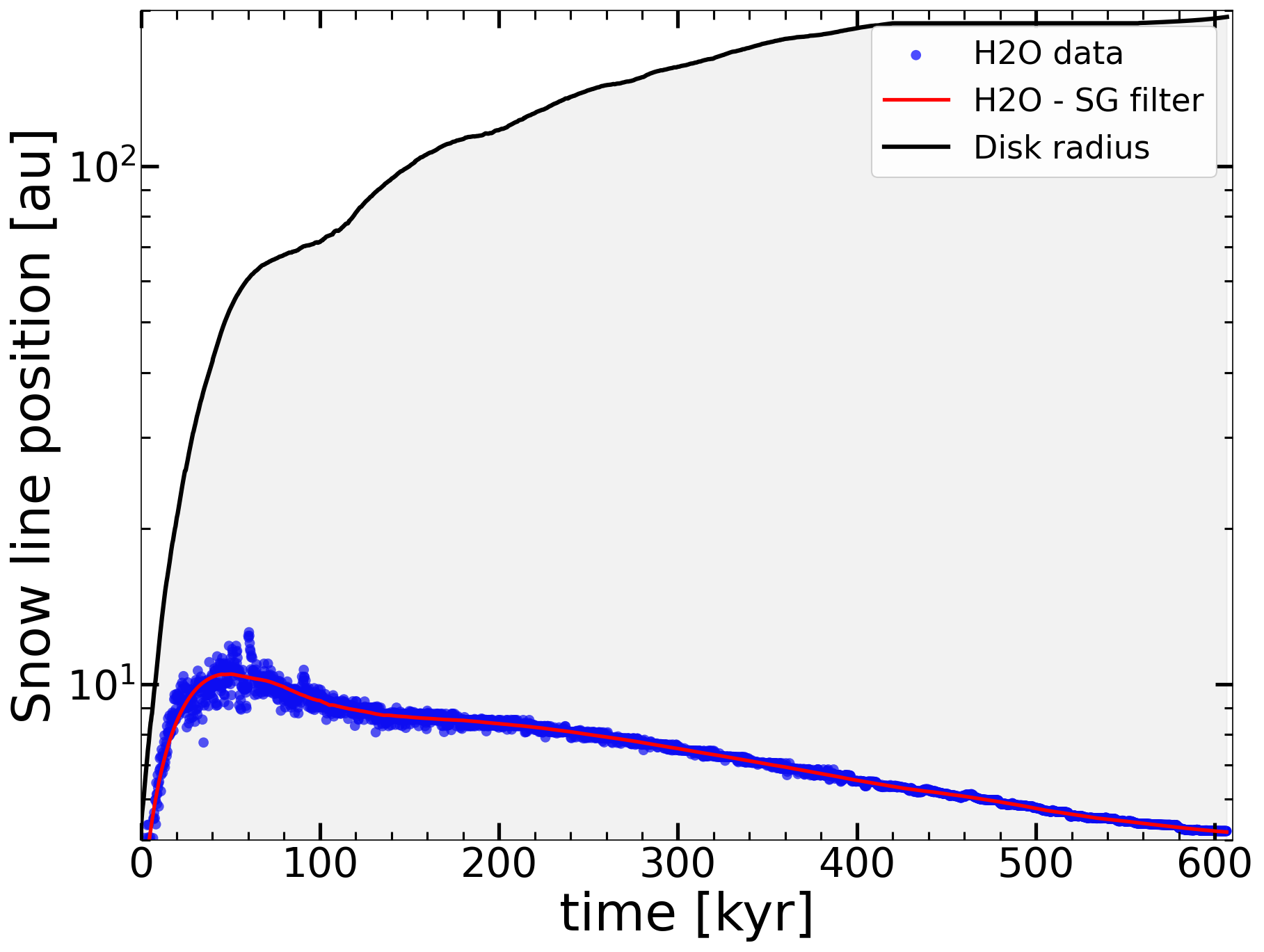} 
    \caption{Azimuthally averaged distance to the main water snow line as a function time passed since the star formation epoch in the dust-size-dependent opacity model. }  
    \label{fig:snow lines-all}
\end{figure}

Figure~\ref{fig:snow lines-all} shows the azimuth-averaged position of the main water snow line during the disk evolution, excluding the contribution of the secondary water spots. The average snow line position is calculated by finding the radial distances at each azimuthal angle of our polar grid where $\Sigma_{\rm H2O}^{\rm dust}=\Sigma_{\rm H2O}^{\rm gas}$ and then azimuthally averaging the corresponding values along the polar circumference.
Clearly, the average position of the water snow line is not stationary but notably evolves with time. The snow line forms concurrently with the disk and first moves outward to $\approx 10$~au during the initial 50~kyr of disk evolution. Then the water snow line gradually retreats back to $\approx 6.0$~au in the subsequent 400~kyr of evolution, as the disk mass declines and the disk cools. As was already noted in Fig.~\ref{fig:chem}, initially both the disk radius and the average position of the snow line increase with time, but already at a few thousand years this trend breaks, and the snow line begins to retreat while the disk radius continues to increase.

\subsection{Comparison with the $a_{\rm max}=1.0$~$\mu$m opacity model}
\label{Sect:fixed-amax}

In this section we evaluate the impact of opacity changes due to dust growth on the properties of the protoplanetary disk and, in particular, on the position of the water snow line. For this purpose,  we consider another disk evolution model in which we artificially set the Rosseland and Planck mean dust opacities equal to the case with $a_{\rm max}=1.0$~$\mu$m; see the red lines in  Fig.~\ref{fig:opacity}. No matter how big dust grains can grow in the course of disk evolution in this comparison model, the dust opacity is always determined by the red lines corresponding to a dust size distribution with $a_{\rm max}=1.0~\mu$m. 
This case is similar to using, for example, the \citet{Semenov2003} Rosseland and Planck mean opacities, which also do not depend on dust growth in the disk and are typical of the dust size distribution in star-forming regions with a maximum dust size of a few microns. We emphasize that Semenov's opacities (or similar) are still widely used in numerical simulations of protoplanetary disks. 

\begin{figure}  
    \centering
    \includegraphics[width=1\columnwidth]{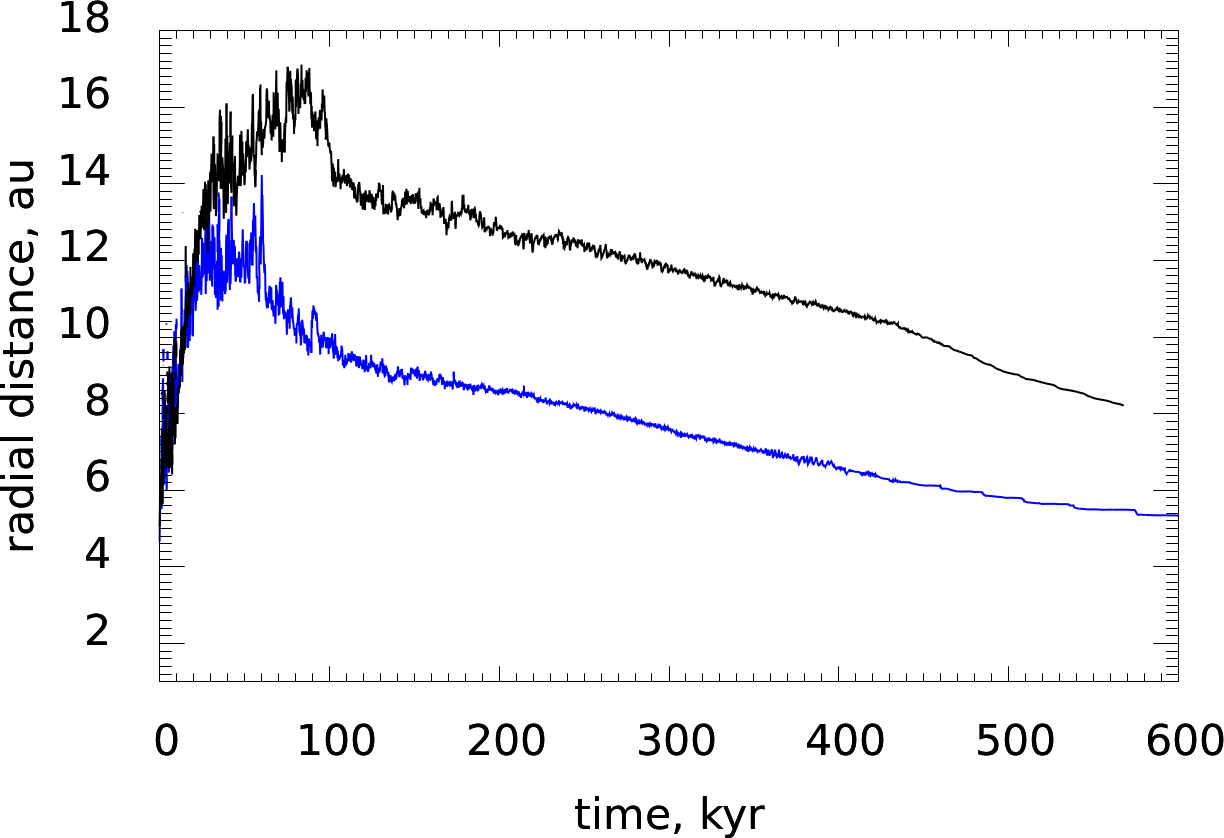}
    \caption{Position of the water snow line in the model with dust-size-dependent opacity (blue line) and the model with dust opacity corresponding to the fixed value of $a_{\rm max}=1.0$~$\mu$m (black line).}
    \label{fig:snow lines-compare}
\end{figure}

Figure~\ref{fig:snow lines-compare} compares the average position of the water snow line in models with and without opacity changes due to dust growth. Clearly,
the location of the snow line is notably affected by opacity changes due to dust growth, particularly after the initial disk formation stage. In the earliest several tens of kyr, both snow lines move outward due to the general disk warming during its build-up stage and their radial locations almost coincide.
In the subsequent evolution, however, the positions of snow lines diverge. 
The snow line in the $a_{\rm max}=1.0$~$\mu$m opacity model continues to move away from the star to $r \approx 16$~au during another 50~kyr, before reversing its direction of motion after $t\approx 90$~kyr.
The snow line in the model with dust-size-dependent opacity lingers at $r \approx 12$~au for several tens of kyr and begins to irreversibly retreat toward the star after $t\approx 60$~kyr.  The retreat reflects the gradual cooling of the disk as it evolves and loses mass via accretion on the star. 
While the snow line in the $a_{\rm max}=1.0$~$\mu$m opacity model retreats to $r \approx 10$~au at the end of simulations, the snow line in the model with dust-size-dependent opacities finds itself at $r\approx 6$~au at the same evolutionary instance. 
Therefore, using more realistic dust-size-dependent opacities shifts the water snow line by almost a factor of two closer to the star.

\begin{figure}   
    \centering
    \includegraphics[width=9 cm]{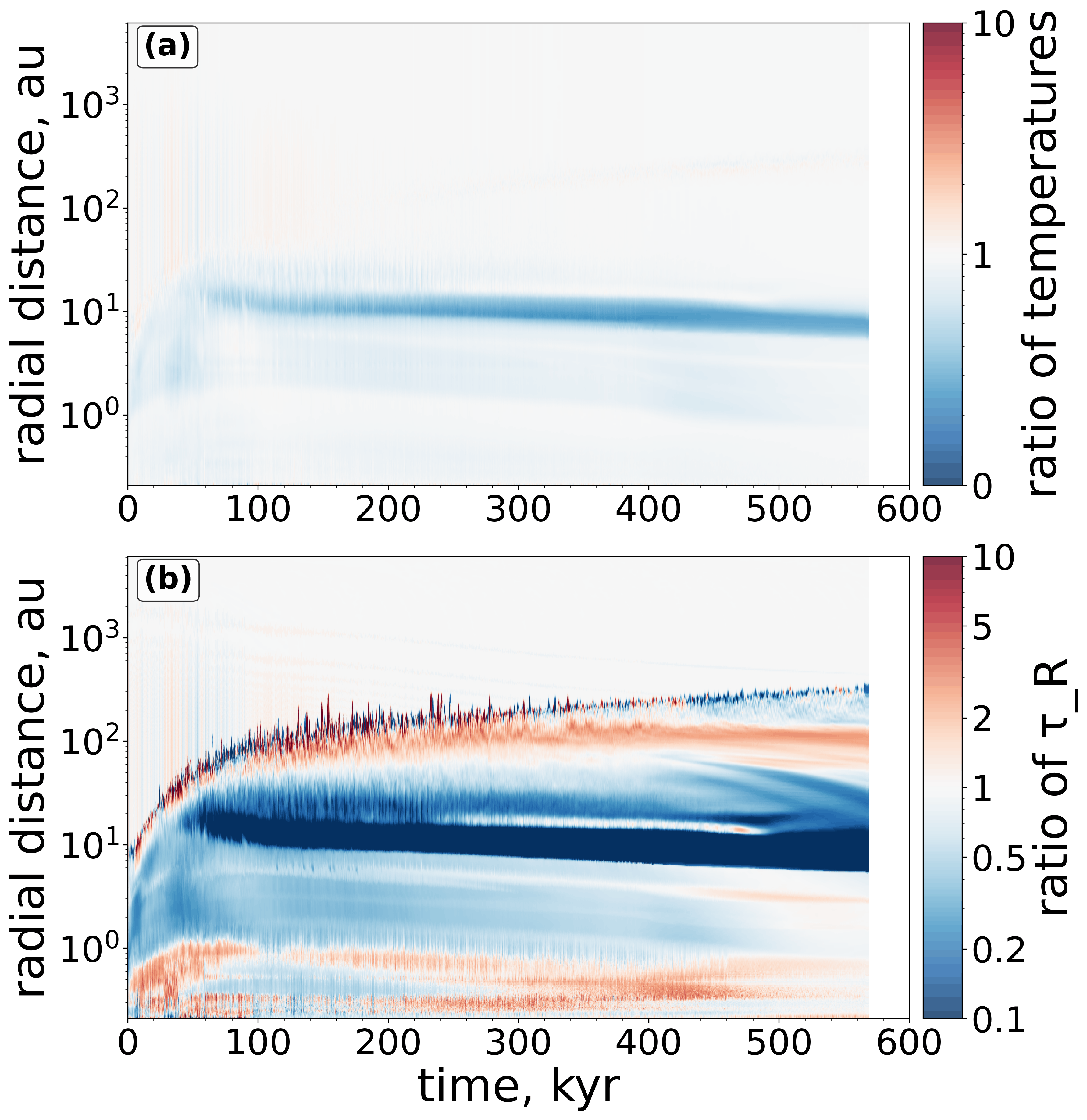}
    \caption{Ratio of the midplane temperature (\textit{panel a}) and the Rosseland mean optical depth (\textit{panel b}) in the dust-size-dependent model to the corresponding values in the fixed $a_{\rm max}=1.0~\mu$m model. Time is counted from stellar birth. }
    \label{fig:radialss}
\end{figure}

The reason why the dust-size-dependent opacities favor more compact  water snow lines is clarified in Fig.~\ref{fig:radialss} showing the ratio of gas temperature and Rosseland mean dust opacity in the dust-size-dependent model to the corresponding values in the fixed $a_{\rm max}=1.0$~$\mu$m opacity model. The top panels reveals that the inner regions of the disk in the dust-size-dependent model are systematically colder than the corresponding regions
in the $a_{\rm max}=1.0~\mu$m opacity model. This trend is particularly evident after $t\approx 50$~kyr and it only intensifies with time. The bottom panel shows that the disk optical depth  in the model with dust-size-dependent opacity is systematically lower than in its counterpart model in the regions up to 30-40~au. This facilitates the loss of heat generated by viscous heating and PdV work in the disk midplane, making the disk cooler precisely in those areas where the water snow line should be.  The outer disk regions, on the contrary, are characterized by higher optical depths in the model with dust-size-dependent opacities. This trend can also be explained by a local increase in dust opacity around the so-called "opacity cliff" at the maximum dust sizes on the order of 0.1-1.0~mm at low temperatures (10-20~K); see Fig.~\ref{fig:kappavs_size_Temp}.  However, the expected increase in gas temperature is negligible because the thermodynamics of the outer disk regions is determined by stellar irradiation rather than by viscous heating; see Fig.~\ref{fig:radial}.

\begin{figure}[!ht]
    \centering
    \includegraphics[width=7 cm]{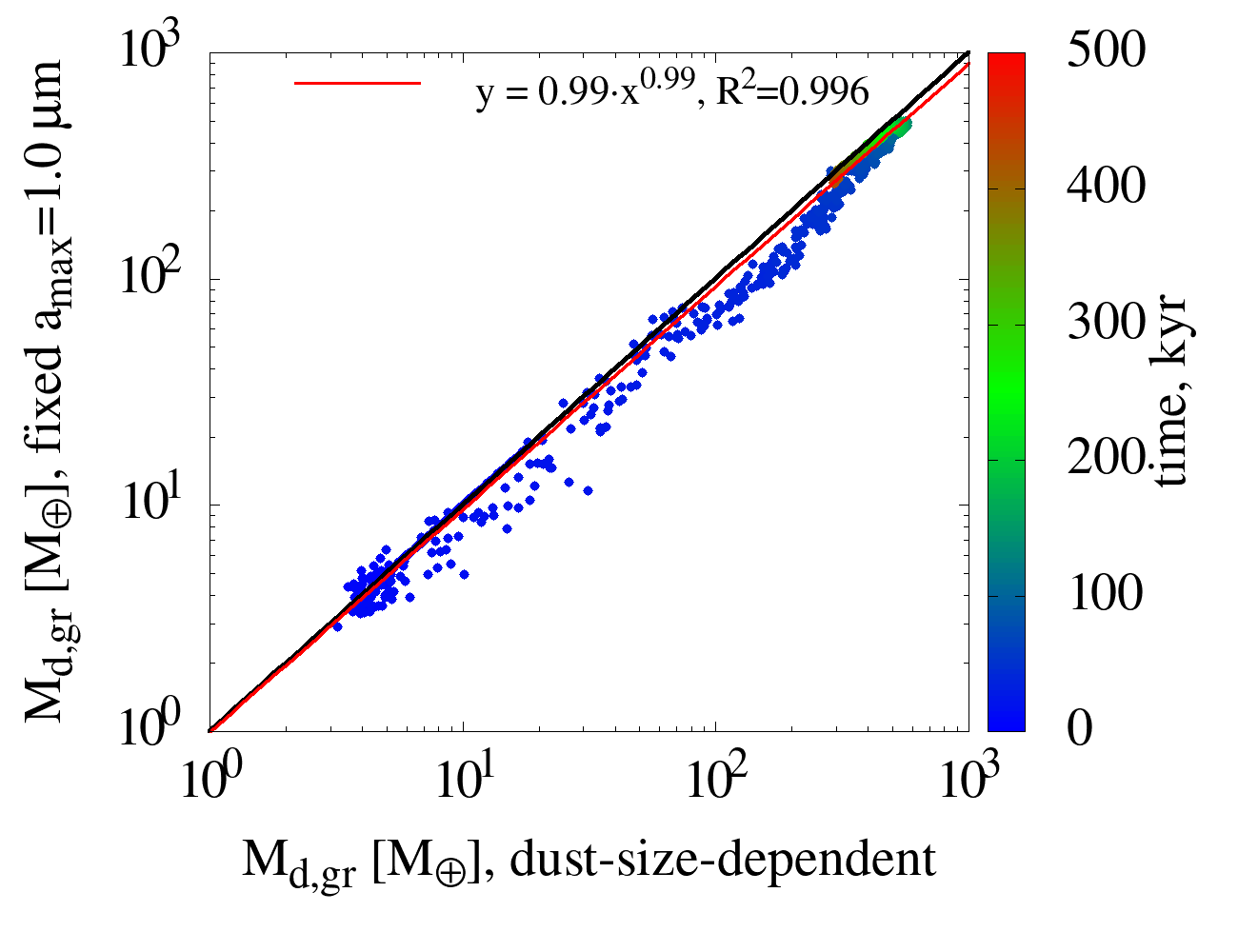}
    \includegraphics[width=7 cm]{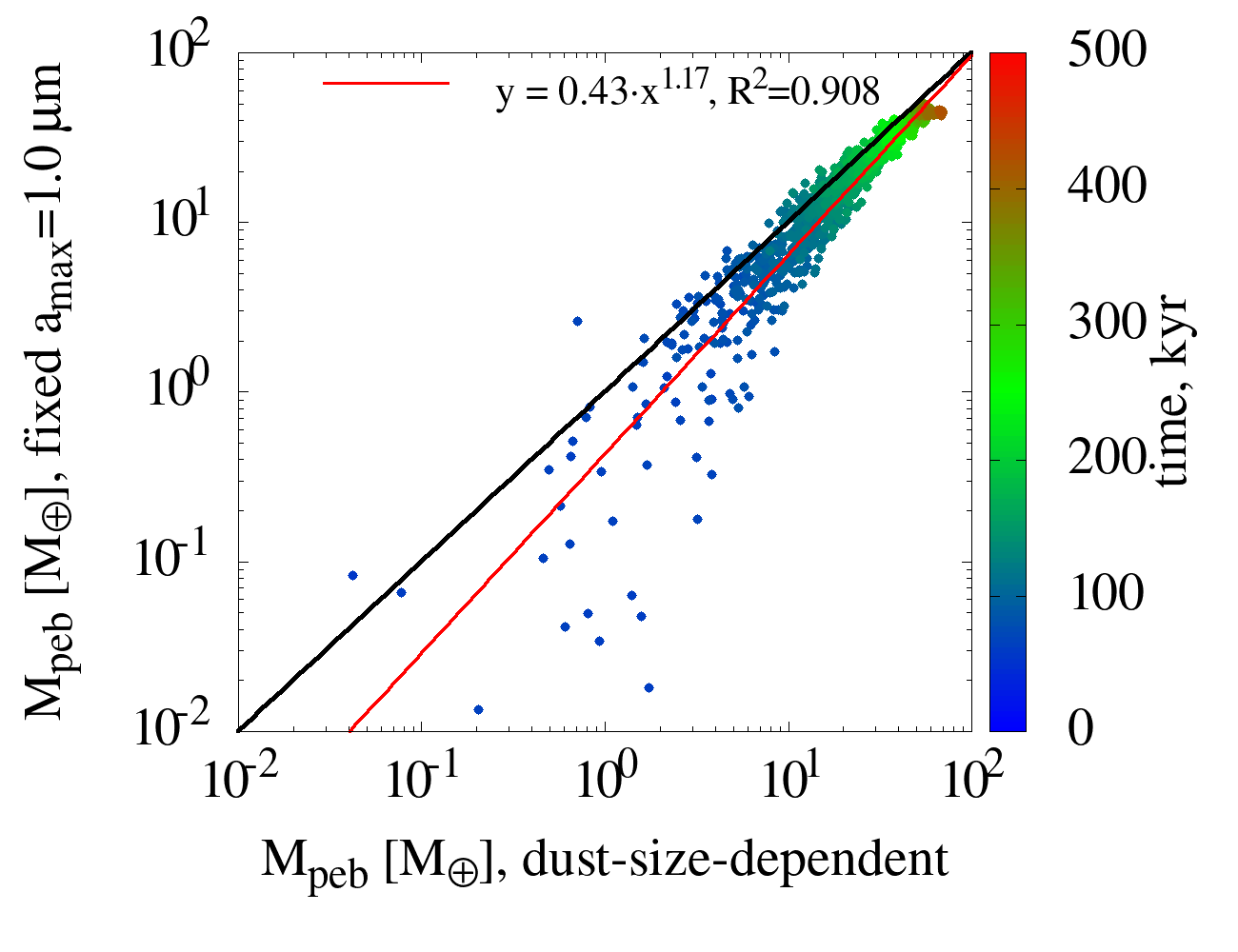}
    \includegraphics[width=7 cm]{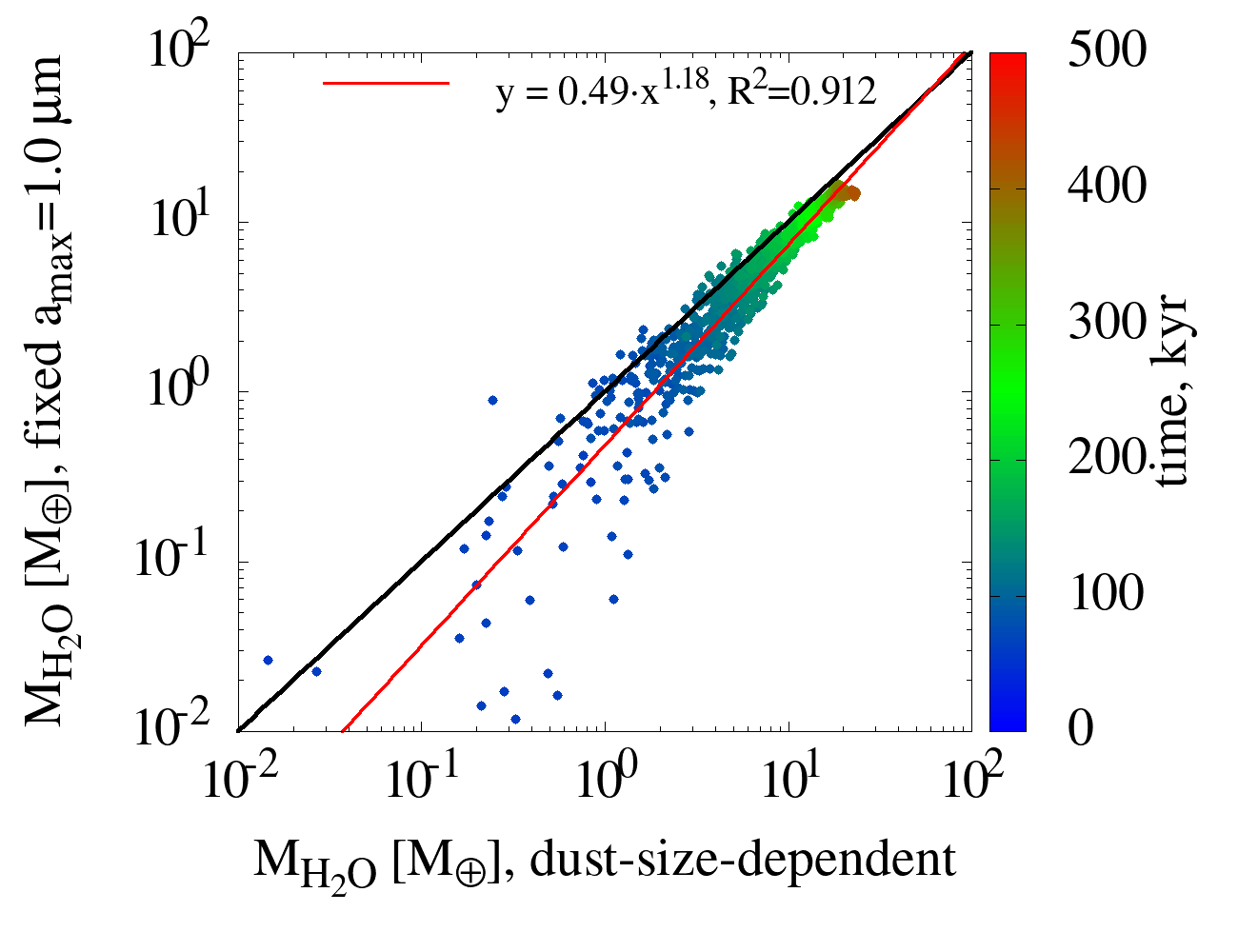}
    \caption{Comparison between total masses of grown dust (\textit{top}),  pebbles (\textit{middle}), and water ice (\textit{bottom}) in the models considered. Time is counted from the instance of star formation. The red lines present the fits of the model data to a power law function.}
    \label{fig:ratio}
\end{figure}

\section{Discussion}
\label{Sect:Discuss}

In this section we discuss the effects that the dust opacity changes due to dust growth may have on
the resulting amounts of grown dust, pebbles and water ice in the disk.  We also discuss the limitations of considering only one set of models and the difficulties with observing the position of the water snow line in protoplanetary disks.

\paragraph{Implications for pebbles and planetesimal formation. }  
Grown dust and, in particular, pebbles are important building blocks of planetesimals and protoplanetary cores in the streaming instability and pebble accretion models \citep{Yang2017,Umurhan2020,2017JohansenLambrechts}, while water ice is the key ingredient in water delivery to terrestrial planets. Figure~\ref{fig:ratio} compares the total masses 
of grown dust, pebbles, and water ice at similar evolution times  during 
the entire simulation. The mass of grown dust and water ice are directly derived from the corresponding surface densities of grown dust and water ice on dust grains, while the mass of pebbles is derived assuming that pebbles are grown dust with the Stokes number and size exceeding a particular threshold; see \citet{Topchieva2024} for more details. 

Water ice occurs concurrently with the disk formation epoch and its mass  increase over time, which can be explained by the water snow line moving closer to the star so that the area of the disk covered with water ice becomes larger. The mass of pebbles also increases as conditions behind the water snow line become more favorable to their formation due to continuing dust growth with time; see Fig.~\ref{fig:dust-size-evol}. 
We note, however, that pebbles do not occur concurrently with the disk formation epoch but rather with some time delay. The mass of grown dust increases with time until pebbles begin to form, so that part of grown dust converts into pebbles. This explains a decrease of the grown dust mass closer to the end of simulations.
Clearly, pebbles and water ice are more abundant in the model that considers opacity changes due to dust growth. This trend persists during the entire considered evolution period.

\paragraph{Model limitations.}
We demonstrated in Sect.~\ref{Sect:fixed-amax} that accounting for dust-size-dependent opacities shifts the water snow line inward by nearly a factor of two compared to the model that adopts fixed micron-sized dust opacities. The magnitude of this effect may, however, depend on the initial and free parameters of our model. As was shown in Figs.~\ref{fig:radial} and \ref{fig:radial-zoom} (see also Fig.~\ref{fig:radial-add}), the water snow line lies in the optically thick disk regions dominated by viscous heating. The midplane temperature of these regions is determined by the balance between viscous heat generation in the disk midplane (${\bl \nabla v} : {\bl \Pi}$) and dust cooling from disk's surface ($\Lambda$), the latter being sensitive to the dust optical depth (see Eq.~\ref{eq:cool-heat}). When dust begins to grow, its opacity at the temperatures corresponding to these regions drops, making dust cooling more efficient, lowering the resulting temperature, and shifting the water snow line closer to the star.  

The effect may diminish if the water snow line happens to be in the disk regions where stellar irradiation dominates over viscous heating and the midplane temperature is determined by the balance of stellar irradiation heating ($\Gamma$) and dust cooling ($\Lambda$). In this case, the dust optical depth is expected to have little effect on the midplane temperature (optical depth cancels out in Eq.~\ref{eq:cool-heat}), and dust growth is expected to have a weaker (if any) effect on the position of the snow line. This may occur in protoplanetary disks with $\alpha_{\rm visc} \ll 10^{-3}$ and/or low gas surface density, because both reduce the dynamical viscosity $\mu$ that enters the viscous stress tensor ${\bl \Pi}$.

Alternatively, protoplanetary disks around higher mass stars are characterized by a stronger irradiation flux (see Eq.~\ref{fluxF}). In such disks, the irradiation-dominated area may extend closer to the star, covering the snow line region. In this case, the effects of dust growth on the snow line position are expected to diminish.

\paragraph{Detectability of the water snow line.}
As shown in Sect.~\ref{Sect:shape-snowline}, the water snow line in the midplane of a young gravitationally unstable disk may have a complex noncircular shape.  However, detecting such a non-axisymmetric morphology may not be easy. Dust thermal emission at relevant wavelengths ($\lambda < 3.0$~mm) may be optically thick at the position of the snow line ($r\le 10$~au), which hinders a direct detection of the water line emission from the disk's midplane \citep{Bergin2019,Facchini2024}.  Luminosity bursts in FU Orionis-type young protostars can shift the snow line to a larger radial distance where direct detection of water emission becomes feasible \citep{Tobin2023,Nakasone2026}. However, disk warming by the burst and associated disk expansion can erase the signatures of gravitational instability \citep{Vorobyov2020}, thereby likely imposing a circular shape to the spatial morphology of the snow line.

Other indirect tracers of the water snow line, such as the dust spectral index and proxy molecules like HCO$^+$ and methanol, can also be used to infer the position of the water snow line \citep{Cieza2016,vanHoff2018,Leemker2021}. However, they also suffer from high optical depths  and work best in the disks of FU~Orionis-type objects.   
Calculations of spatial maps of synthetic water line emission and dust spectral index \citep[e.g.,][]{Pavlyuchenkov2019,Sai2026} are required to set the limits on sensitivity and spatial resolution required to probe the non-axisymmetric morphology of the water snow line in gravitationally unstable disks.

\section{Conclusions}
\label{Sect:conclusions}
In this paper, we have studied the evolution of a protoplanetary disk, starting from its formation to an age of approximately 0.6~Myr. We used the FEOSAD numerical hydrodynamics code, which accounts for the coevolution of gas, dust, and volatiles in the two-dimensional thin-disk limit $(r,\phi)$, including dust growth, phase transitions of volatiles, and transport of volatiles with gas or on the dust surface. We focused on the position, shape, and time evolution of the water snow line in the disk midplane. Unlike many previous numerical studies, we were particularly interested in the effects that the dust opacity changes due to dust growth may have on the position of the snow line in the course of disk evolution. To this end, we considered two disk models: one that takes into account the dependence of dust opacity on the maximum dust grain size $a_{\rm max}$ when calculating the thermal balance in the disk and another that always employs dust opacities for a maximum dust size of $a_{\rm max}=1.0$~$\mu$m, no matter how big the dust grains can actually grow during the disk evolution. The dust-size-dependent opacities were calculated using the OpTool \citep{OpTool2021} for the standard dust composition listed in Table~\ref{tab:abundances}. The input from water ice (time and space variable) was also considered when calculating the dust opacities. Our findings can be summarized as follows.
\begin{itemize}
    \item We observed quick dust growth as the disk forms and evolves from the initial value of $a_{\rm max}=1.0~\mu$m in the prestellar cloud to almost 10~cm just behind the water snow line. Further out in the disk, $a_{\rm max}$ gradually declines to the initial value at the disk-envelope interface. The maximum dust size interior to the water snow line  varies in the  $a_{\rm max}=0.1-1.0$~mm limits.
    
    \item As dust grows and $a_{\rm max}$ increases, the Rosseland and Planck mean dust opacities tend to decrease for temperatures $>50-100$~K. This has a substantial feedback effect on the thermal structure in the inner warm regions of the disk, lowering the disk temperature and shifting the water snow line closer to the star by almost a factor of two compared to the model where opacity changed due to dust growth are not considered.
    
    \item The geometry of the water snow line in the disk midplane may strongly deviate from a circular shape in the initial gravitationally unstable stages of disk evolution. 
    Additional water vapor patches episodically occur behind the main water snow line and are caused by compressional heating from spiral density waves.

\item  A noticeable increase in the amounts of dust and water ice and in the maximum size of dust grains just beyond the snow line occurs only after disk gravitational instability weakens and the snow line becomes nearly circular.
The absence of this effect in the early gravitationally unstable stage can be partly attributed to an increased circumference of the water snow line and partly to perturbations caused by spiral density waves traversing the snow line. 

\item
The radial position of the water snow line, averaged over the circumference, is highly nonsteady in the course of early disk evolution. During the disk build-up epoch, it increases over time, but this trend soon reverses, and the water snow line begins to retreat toward the star as the disk cools. The snow line position varies from just 1.0~au at the disk formation epoch to a maximum of 12~au around $t=50$~kyr to 6.0~au by the end of the simulations.

\end{itemize}

In the entire considered evolution period, the radial position of the water snow line is determined by viscous heating rather than by heating due to stellar irradiation.
Further studies beyond the fiducial model exploring a wider parameter space of turbulent viscosity and stellar and disk masses together with synthetic maps of water line emission
are required to assess the robustness of our model predictions.

\begin{acknowledgements} The authors are thankful to the anonymous referee for insightful comments that helped improve the manuscript.
Simulations were performed on the Austrian Scientific Cluster (\href{https://asc.ac.at/}{https://asc.ac.at/}). 
\end{acknowledgements}

\bibliographystyle{aa}
\bibliography{refs}

@ARTICLE{2018VorobyovAkimkin,
   author = {{Vorobyov}, E.~I. and {Akimkin}, V. and {Stoyanovskaya}, O. and 
	{Pavlyuchenkov}, Y. and {Liu}, H.~B.},
    title = "{Early evolution of viscous and self-gravitating circumstellar disks with a dust component}",
  journal = {\aap},
archivePrefix = "arXiv",
   eprint = {1801.06898},
 primaryClass = "astro-ph.EP",
     year = 2018,
    month = jun,
   volume = 614,
      eid = {A98},
    pages = {A98},
      doi = {10.1051/0004-6361/201731690},
   adsurl = {http://adsabs.harvard.edu/abs/2018A%26A...614A..98V}
}

@Article{VorobyovBasu2009,
  author        = {{Vorobyov}, E.~I. and {Basu}, S.},
  title         = {{Secular evolution of viscous and self-gravitating circumstellar discs}},
  journal       = {\mnras},
  year          = {2009},
  volume        = {393},
  pages         = {822-837},
  month         = mar,
  adsurl        = {http://adsabs.harvard.edu/abs/2009MNRAS.393..822V},
  archiveprefix = {arXiv},
  doi           = {10.1111/j.1365-2966.2008.14376.x},
  eprint        = {0812.1306},
}

@Article{2010VorobyovBasu,
  author        = {{Vorobyov}, E.~I. and {Basu}, S.},
  title         = {{The Burst Mode of Accretion and Disk Fragmentation in the Early Embedded Stages of Star Formation}},
  journal       = {ApJ},
  year          = {2010},
  volume        = {719},
  pages         = {1896-1911},
  month         = aug,
  adsurl        = {http://adsabs.harvard.edu/abs/2010ApJ...719.1896V},
  archiveprefix = {arXiv},
  doi           = {10.1088/0004-637X/719/2/1896},
  eprint        = {1007.2993},
  primaryclass  = {astro-ph.SR},
}

@Article{1973ShakuraSunyaev,
  author  = {{Shakura}, N.~I. and {Sunyaev}, R.~A.},
  title   = {{Black holes in binary systems. Observational appearance.}},
  journal = {\aap},
  year    = {1973},
  volume  = {24},
  pages   = {337-355},
  adsurl  = {http://adsabs.harvard.edu/abs/1973A%26A....24..337S},
}

@Article{1997Basu,
  author   = {{Basu}, S.},
  title    = {{A Semianalytic Model for Supercritical Core Collapse: Self-Similar Evolution and the Approach to Protostar Formation}},
  journal  = {ApJ},
  year     = {1997},
  volume   = {485},
  pages    = {240-253},
  month    = aug,
  adsurl   = {http://adsabs.harvard.edu/abs/1997ApJ...485..240B},
}

@ARTICLE{2017Drazkowska,
   author = {{Dr{\c a}{\.z}kowska}, J. and {Alibert}, Y.},
    title = "{Planetesimal formation starts at the snow line}",
  journal = {\aap},
archivePrefix = "arXiv",
   eprint = {1710.00009},
 primaryClass = "astro-ph.EP",
     year = 2017,
    month = dec,
   volume = 608,
      eid = {A92},
    pages = {A92},
      doi = {10.1051/0004-6361/201731491},
   adsurl = {http://adsabs.harvard.edu/abs/2017A%26A...608A..92D}
}

@ARTICLE{2016Birnstiel,
   author = {{Birnstiel}, T. and {Fang}, M. and {Johansen}, A.},
    title = "{Dust Evolution and the Formation of Planetesimals}",
  journal = {\ssr},
archivePrefix = "arXiv",
   eprint = {1604.02952},
 primaryClass = "astro-ph.SR",
     year = 2016,
    month = dec,
   volume = 205,
    pages = {41-75},
      doi = {10.1007/s11214-016-0256-1},
   adsurl = {http://adsabs.harvard.edu/abs/2016SSRv..205...41B}
}

@ARTICLE{2012Birnstiel,
   author = {{Birnstiel}, T. and {Klahr}, H. and {Ercolano}, B.},
    title = "{A simple model for the evolution of the dust population in protoplanetary disks}",
  journal = {\aap},
     year = "2012",
    month = "Mar",
   volume = {539},
      eid = {A148},
    pages = {A148},
      doi = {10.1051/0004-6361/201118136},
archivePrefix = {arXiv},
   eprint = {1201.5781},
primaryClass = {astro-ph.EP},
   adsurl = {https://ui.adsabs.harvard.edu/abs/2012A&A...539A.148B}
}

@ARTICLE{Okuzumi2016,
       author = {{Okuzumi}, Satoshi and {Momose}, Munetake and {Sirono}, Sin-iti and
         {Kobayashi}, Hiroshi and {Tanaka}, Hidekazu},
        title = "{Sintering-induced Dust Ring Formation in Protoplanetary Disks: Application to the HL Tau Disk}",
      journal = {\apj},
         year = 2016,
        month = apr,
       volume = {821},
       number = {2},
          eid = {82},
        pages = {82},
          doi = {10.3847/0004-637X/821/2/82},
archivePrefix = {arXiv},
       eprint = {1510.03556},
 primaryClass = {astro-ph.SR},
       adsurl = {https://ui.adsabs.harvard.edu/abs/2016ApJ...821...82O}
}

@ARTICLE{Pinilla2017,
       author = {{Pinilla}, P. and {Pohl}, A. and {Stammler}, S.~M. and {Birnstiel}, T.},
        title = "{Dust Density Distribution and Imaging Analysis of Different Ice Lines in Protoplanetary Disks}",
      journal = {\apj},
         year = 2017,
        month = aug,
       volume = {845},
       number = {1},
          eid = {68},
        pages = {68},
          doi = {10.3847/1538-4357/aa7edb},
archivePrefix = {arXiv},
       eprint = {1707.02321},
 primaryClass = {astro-ph.EP},
       adsurl = {https://ui.adsabs.harvard.edu/abs/2017ApJ...845...68P}
}

@ARTICLE{Yang2017,
       author = {{Yang}, C. -C. and {Johansen}, A. and {Carrera}, D.},
        title = "{Concentrating small particles in protoplanetary disks through the streaming instability}",
      journal = {\aap},
         year = 2017,
        month = oct,
       volume = {606},
          eid = {A80},
        pages = {A80},
          doi = {10.1051/0004-6361/201630106},
archivePrefix = {arXiv},
       eprint = {1611.07014},
 primaryClass = {astro-ph.EP},
       adsurl = {https://ui.adsabs.harvard.edu/abs/2017A&A...606A..80Y}
}

@ARTICLE{Youdin2005,
       author = {{Youdin}, Andrew N. and {Goodman}, Jeremy},
        title = "{Streaming Instabilities in Protoplanetary Disks}",
      journal = {\apj},
         year = 2005,
        month = feb,
       volume = {620},
       number = {1},
        pages = {459-469},
          doi = {10.1086/426895},
archivePrefix = {arXiv},
       eprint = {astro-ph/0409263},
 primaryClass = {astro-ph},
       adsurl = {https://ui.adsabs.harvard.edu/abs/2005ApJ...620..459Y}
}

@ARTICLE{Molyarova2021,
       author = {{Molyarova}, Tamara and {Vorobyov}, Eduard I. and {Akimkin}, Vitaly and {Skliarevskii}, Aleksandr and {Wiebe}, Dmitri and {G{\"u}del}, Manuel},
        title = "{Gravitoviscous Protoplanetary Disks with a Dust Component. V. The Dynamic Model for Freeze-out and Sublimation of Volatiles}",
      journal = {\apj},
         year = 2021,
        month = apr,
       volume = {910},
       number = {2},
          eid = {153},
        pages = {153},
          doi = {10.3847/1538-4357/abe2b0},
archivePrefix = {arXiv},
       eprint = {2103.06045},
 primaryClass = {astro-ph.EP},
       adsurl = {https://ui.adsabs.harvard.edu/abs/2021ApJ...910..153M}
}

@ARTICLE{Weidenschilling1977,
       author = {{Weidenschilling}, S.~J.},
        title = "{Aerodynamics of solid bodies in the solar nebula.}",
      journal = {\mnras},
         year = 1977,
        month = jul,
       volume = {180},
        pages = {57-70},
          doi = {10.1093/mnras/180.2.57},
       adsurl = {https://ui.adsabs.harvard.edu/abs/1977MNRAS.180...57W}
}

@ARTICLE{Stoyanovskaya2020,
       author = {{Stoyanovskaya}, O.~P. and {Okladnikov}, F.~A. and {Vorobyov}, E.~I. and {Pavlyuchenkov}, Ya. N. and {Akimkin}, V.~V.},
        title = "{Simulations of Dynamical Gas-Dust Circumstellar Disks: Going Beyond the Epstein Regime}",
      journal = {Astronomy Reports},
         year = 2020,
        month = mar,
       volume = {64},
       number = {2},
        pages = {107-125},
          doi = {10.1134/S1063772920010072},
archivePrefix = {arXiv},
       eprint = {2102.09155},
 primaryClass = {astro-ph.EP},
       adsurl = {https://ui.adsabs.harvard.edu/abs/2020ARep...64..107S}
}

@ARTICLE{Vorobyov2022,
       author = {{Vorobyov}, Eduard I. and {Skliarevskii}, Aleksandr M. and {Molyarova}, Tamara and {Akimkin}, Vitaly and {Pavlyuchenkov}, Yaroslav and {K{\'o}sp{\'a}l}, {\'A}gnes and {Liu}, Hauyu Baobab and {Takami}, Michihiro and {Topchieva}, Anastasiia},
        title = "{Evolution of dust in protoplanetary disks of eruptive stars}",
      journal = {\aap},
         year = 2022,
        month = feb,
       volume = {658},
          eid = {A191},
        pages = {A191},
          doi = {10.1051/0004-6361/202141932},
archivePrefix = {arXiv},
       eprint = {2112.06004},
 primaryClass = {astro-ph.EP},
       adsurl = {https://ui.adsabs.harvard.edu/abs/2022A&A...658A.191V}
}

@ARTICLE{Vorobyov2020,
       author = {{Vorobyov}, Eduard I. and {Elbakyan}, Vardan G. and {Takami}, Michihiro and {Liu}, Hauyu B.},
        title = "{Effect of luminosity outbursts on protoplanetary disk dynamics}",
      journal = {\aap},
         year = 2020,
        month = nov,
       volume = {643},
          eid = {A13},
        pages = {A13},
          doi = {10.1051/0004-6361/202038122},
archivePrefix = {arXiv},
       eprint = {2009.01888},
 primaryClass = {astro-ph.SR},
       adsurl = {https://ui.adsabs.harvard.edu/abs/2020A&A...643A..13V}
}

@ARTICLE{Henderson1976,
       author = {{Henderson}, C.~B.},
        title = "{Drag coefficient of spheres in continuum and rarefied flows}",
      journal = {AIAA Journal},
         year = 1976,
        month = jan,
       volume = {14},
        pages = {707-707},
          doi = {10.2514/3.61409},
       adsurl = {https://ui.adsabs.harvard.edu/abs/1976AIAAJ..14..707H}
}

@ARTICLE{Umurhan2020,
       author = {{Umurhan}, Orkan M. and {Estrada}, Paul R. and {Cuzzi}, Jeffrey N.},
        title = "{Streaming Instability in Turbulent Protoplanetary Disks}",
      journal = {\apj},
         year = 2020,
        month = may,
       volume = {895},
       number = {1},
          eid = {4},
        pages = {4},
          doi = {10.3847/1538-4357/ab899d},
archivePrefix = {arXiv},
       eprint = {1906.05371},
 primaryClass = {astro-ph.EP},
       adsurl = {https://ui.adsabs.harvard.edu/abs/2020ApJ...895....4U}
}

@ARTICLE{Vorobyov2023a,
       author = {{Vorobyov}, Eduard I. and {Elbakyan}, Vardan G. and {Johansen}, Anders and {Lambrechts}, Michiel and {Skliarevskii}, Aleksandr M. and {Stoyanovskaya}, O.~P.},
        title = "{Formation of pebbles in (gravito-)viscous protoplanetary disks with various turbulent strengths}",
      journal = {\aap},
         year = 2023,
        month = feb,
       volume = {670},
          eid = {A81},
        pages = {A81},
          doi = {10.1051/0004-6361/202244500},
archivePrefix = {arXiv},
       eprint = {2212.01023},
 primaryClass = {astro-ph.EP},
       adsurl = {https://ui.adsabs.harvard.edu/abs/2023A&A...670A..81V}
}

@ARTICLE{SN1992,
       author = {{Stone}, James M. and {Norman}, Michael L.},
        title = "{ZEUS-2D: A Radiation Magnetohydrodynamics Code for Astrophysical Flows in Two Space Dimensions. I. The Hydrodynamic Algorithms and Tests}",
      journal = {\apjs},
         year = 1992,
        month = jun,
       volume = {80},
        pages = {753},
          doi = {10.1086/191680},
       adsurl = {https://ui.adsabs.harvard.edu/abs/1992ApJS...80..753S}
}

@ARTICLE{2015ALMABrogan,
       author = {{ALMA Partnership} and {Brogan}, C.~L. and {P{\'e}rez}, L.~M. and {Hunter}, T.~R. and {Dent}, W.~R.~F. and {Hales}, A.~S. and {Hills}, R.~E. and {Corder}, S. and {Fomalont}, E.~B. and {Vlahakis}, C. and {Asaki}, Y. and {Barkats}, D. and {Hirota}, A. and {Hodge}, J.~A. and {Impellizzeri}, C.~M.~V. and {Kneissl}, R. and {Liuzzo}, E. and {Lucas}, R. and {Marcelino}, N. and {Matsushita}, S. and {Nakanishi}, K. and {Phillips}, N. and {Richards}, A.~M.~S. and {Toledo}, I. and {Aladro}, R. and {Broguiere}, D. and {Cortes}, J.~R. and {Cortes}, P.~C. and {Espada}, D. and {Galarza}, F. and {Garcia-Appadoo}, D. and {Guzman-Ramirez}, L. and {Humphreys}, E.~M. and {Jung}, T. and {Kameno}, S. and {Laing}, R.~A. and {Leon}, S. and {Marconi}, G. and {Mignano}, A. and {Nikolic}, B. and {Nyman}, L. -A. and {Radiszcz}, M. and {Remijan}, A. and {Rod{\'o}n}, J.~A. and {Sawada}, T. and {Takahashi}, S. and {Tilanus}, R.~P.~J. and {Vila Vilaro}, B. and {Watson}, L.~C. and {Wiklind}, T. and {Akiyama}, E. and {Chapillon}, E. and {de Gregorio-Monsalvo}, I. and {Di Francesco}, J. and {Gueth}, F. and {Kawamura}, A. and {Lee}, C. -F. and {Nguyen Luong}, Q. and {Mangum}, J. and {Pietu}, V. and {Sanhueza}, P. and {Saigo}, K. and {Takakuwa}, S. and {Ubach}, C. and {van Kempen}, T. and {Wootten}, A. and {Castro-Carrizo}, A. and {Francke}, H. and {Gallardo}, J. and {Garcia}, J. and {Gonzalez}, S. and {Hill}, T. and {Kaminski}, T. and {Kurono}, Y. and {Liu}, H. -Y. and {Lopez}, C. and {Morales}, F. and {Plarre}, K. and {Schieven}, G. and {Testi}, L. and {Videla}, L. and {Villard}, E. and {Andreani}, P. and {Hibbard}, J.~E. and {Tatematsu}, K.},
        title = "{The 2014 ALMA Long Baseline Campaign: First Results from High Angular Resolution Observations toward the HL Tau Region}",
      journal = {\apjl},
         year = 2015,
        month = jul,
       volume = {808},
       number = {1},
          eid = {L3},
        pages = {L3},
          doi = {10.1088/2041-8205/808/1/L3},
archivePrefix = {arXiv},
       eprint = {1503.02649},
 primaryClass = {astro-ph.SR},
       adsurl = {https://ui.adsabs.harvard.edu/abs/2015ApJ...808L...3A}
}

@ARTICLE{2012Lambrechts,
       author = {{Lambrechts}, M. and {Johansen}, A.},
        title = "{Rapid growth of gas-giant cores by pebble accretion}",
      journal = {\aap},
         year = "2012",
        month = "Aug",
       volume = {544},
          eid = {A32},
        pages = {A32},
          doi = {10.1051/0004-6361/201219127},
archivePrefix = {arXiv},
       eprint = {1205.3030},
 primaryClass = {astro-ph.EP},
       adsurl = {https://ui.adsabs.harvard.edu/abs/2012A&A...544A..32L}
}

@ARTICLE{2017JohansenLambrechts,
   author = {{Johansen}, Anders and {Lambrechts}, Michiel},
    title = "{Forming Planets via Pebble Accretion}",
  journal = {Annual Review of Earth and Planetary Sciences},
     year = "2017",
    month = "Aug",
   volume = {45},
   number = {1},
    pages = {359-387},
      doi = {10.1146/annurev-earth-063016-020226},
   adsurl = {https://ui.adsabs.harvard.edu/abs/2017AREPS..45..359J}
}

@ARTICLE{2022LauDrazkowska,
       author = {{Lau}, Tommy Chi Ho and {Dr{\k{a}}{\.z}kowska}, Joanna and {Stammler}, Sebastian M. and {Birnstiel}, Tilman and {Dullemond}, Cornelis P.},
        title = "{Rapid formation of massive planetary cores in a pressure bump}",
      journal = {\aap},
         year = 2022,
        month = dec,
       volume = {668},
          eid = {A170},
        pages = {A170},
          doi = {10.1051/0004-6361/202244864},
archivePrefix = {arXiv},
       eprint = {2211.04497},
 primaryClass = {astro-ph.EP},
       adsurl = {https://ui.adsabs.harvard.edu/abs/2022A&A...668A.170L}
}

@ARTICLE{2022Baehr,
       author = {{Baehr}, Hans and {Zhu}, Zhaohuan and {Yang}, Chao-Chin},
        title = "{Direct Formation of Planetary Embryos in Self-gravitating Disks}",
      journal = {\apj},
         year = 2022,
        month = jul,
       volume = {933},
       number = {1},
          eid = {100},
        pages = {100},
          doi = {10.3847/1538-4357/ac7228},
archivePrefix = {arXiv},
       eprint = {2204.13310},
 primaryClass = {astro-ph.EP},
       adsurl = {https://ui.adsabs.harvard.edu/abs/2022ApJ...933..100B}
}

@ARTICLE{2002Caselli,
       author = {{Caselli}, Paola and {Benson}, Priscilla J. and {Myers}, Philip C. and
         {Tafalla}, Mario},
        title = "{Dense Cores in Dark Clouds. XIV. N$_{2}$H$^{+}$ (1-0) Maps of Dense Cloud Cores}",
      journal = {\apj},
         year = "2002",
        month = "Jun",
       volume = {572},
       number = {1},
        pages = {238-263},
          doi = {10.1086/340195},
archivePrefix = {arXiv},
       eprint = {astro-ph/0202173},
 primaryClass = {astro-ph},
       adsurl = {https://ui.adsabs.harvard.edu/abs/2002ApJ...572..238C}
}

@ARTICLE{2023Rosotti,
       author = {{Rosotti}, Giovanni P.},
        title = "{Empirical constraints on turbulence in proto-planetary discs}",
      journal = {\nar},
         year = 2023,
        month = jun,
       volume = {96},
          eid = {101674},
        pages = {101674},
          doi = {10.1016/j.newar.2023.101674},
archivePrefix = {arXiv},
       eprint = {2302.01433},
 primaryClass = {astro-ph.EP},
       adsurl = {https://ui.adsabs.harvard.edu/abs/2023NewAR..9601674R}
}

@ARTICLE{VorobyovBasu2005,
       author = {{Vorobyov}, E.~I. and {Basu}, Shantanu},
        title = "{The Origin of Episodic Accretion Bursts in the Early Stages of Star Formation}",
      journal = {\apjl},
         year = 2005,
        month = nov,
       volume = {633},
       number = {2},
        pages = {L137-L140},
          doi = {10.1086/498303},
archivePrefix = {arXiv},
       eprint = {astro-ph/0510014},
 primaryClass = {astro-ph},
       adsurl = {https://ui.adsabs.harvard.edu/abs/2005ApJ...633L.137V}
}

@ARTICLE{Bate2022,
       author = {{Bate}, Matthew R.},
        title = "{Dust coagulation during the early stages of star formation: molecular cloud collapse and first hydrostatic core evolution}",
      journal = {\mnras},
         year = 2022,
        month = aug,
       volume = {514},
       number = {2},
        pages = {2145-2161},
          doi = {10.1093/mnras/stac1391},
archivePrefix = {arXiv},
       eprint = {2205.07681},
 primaryClass = {astro-ph.GA},
       adsurl = {https://ui.adsabs.harvard.edu/abs/2022MNRAS.514.2145B}
}

@ARTICLE{VorobyovKulikov2024,
       author = {{Vorobyov}, Eduard I. and {Kulikov}, Igor and {Elbakyan}, Vardan G. and {McKevitt}, James and {G{\"u}del}, Manuel},
        title = "{Dust growth and pebble formation in the initial stages of protoplanetary disk evolution}",
      journal = {\aap},
         year = 2024,
        month = mar,
       volume = {683},
          eid = {A202},
        pages = {A202},
          doi = {10.1051/0004-6361/202348023},
archivePrefix = {arXiv},
       eprint = {2401.02205},
 primaryClass = {astro-ph.EP},
       adsurl = {https://ui.adsabs.harvard.edu/abs/2024A&A...683A.202V}
}

@ARTICLE{Birnstiel2024,
       author = {{Birnstiel}, Tilman},
        title = "{Dust Growth and Evolution in Protoplanetary Disks}",
      journal = {\araa},
         year = 2024,
        month = sep,
       volume = {62},
       number = {1},
        pages = {157-202},
          doi = {10.1146/annurev-astro-071221-052705},
archivePrefix = {arXiv},
       eprint = {2312.13287},
 primaryClass = {astro-ph.EP},
       adsurl = {https://ui.adsabs.harvard.edu/abs/2024ARA&A..62..157B}
}

@ARTICLE{Cridland2022,
       author = {{Cridland}, Alex J. and {Rosotti}, Giovanni P. and {Tabone}, Beno{\^\i}t and {Tychoniec}, {\L}ukasz and {McClure}, Melissa and {Nazari}, Pooneh and {van Dishoeck}, Ewine F.},
        title = "{Early planet formation in embedded protostellar disks. Setting the stage for the first generation of planetesimals}",
      journal = {\aap},
         year = 2022,
        month = jun,
       volume = {662},
          eid = {A90},
        pages = {A90},
          doi = {10.1051/0004-6361/202142207},
archivePrefix = {arXiv},
       eprint = {2112.06734},
 primaryClass = {astro-ph.EP},
       adsurl = {https://ui.adsabs.harvard.edu/abs/2022A&A...662A..90C}
}

@ARTICLE{Schoonenberg2017,
       author = {{Schoonenberg}, Djoeke and {Ormel}, Chris W.},
        title = "{Planetesimal formation near the snowline: in or out?}",
      journal = {\aap},
         year = 2017,
        month = jun,
       volume = {602},
          eid = {A21},
        pages = {A21},
          doi = {10.1051/0004-6361/201630013},
archivePrefix = {arXiv},
       eprint = {1702.02151},
 primaryClass = {astro-ph.EP},
       adsurl = {https://ui.adsabs.harvard.edu/abs/2017A&A...602A..21S}
}

@BOOK{BT1987,
       author = {{Binney}, James and {Tremaine}, Scott},
        title = "{Galactic dynamics}",
         year = 1987,
       adsurl = {https://ui.adsabs.harvard.edu/abs/1987gady.book.....B}
}

@ARTICLE{Vorobyov2024,
       author = {{Vorobyov}, Eduard I. and {Skliarevskii}, Aleksandr M. and {Guedel}, Manuel and {Molyarova}, Tamara},
        title = "{Primordial dust rings, hidden dust mass, and the first generation of planetesimals in gravitationally unstable protoplanetary disks}",
      journal = {\aap},
         year = 2024,
        month = jul,
       volume = {687},
          eid = {A192},
        pages = {A192},
          doi = {10.1051/0004-6361/202349104},
archivePrefix = {arXiv},
       eprint = {2404.16151},
 primaryClass = {astro-ph.EP},
       adsurl = {https://ui.adsabs.harvard.edu/abs/2024A&A...687A.192V}
}

@ARTICLE{Draine2001,
       author = {{Weingartner}, Joseph C. and {Draine}, B.~T.},
        title = "{Dust Grain-Size Distributions and Extinction in the Milky Way, Large Magellanic Cloud, and Small Magellanic Cloud}",
      journal = {\apj},
         year = 2001,
        month = feb,
       volume = {548},
       number = {1},
        pages = {296-309},
          doi = {10.1086/318651},
archivePrefix = {arXiv},
       eprint = {astro-ph/0008146},
 primaryClass = {astro-ph},
       adsurl = {https://ui.adsabs.harvard.edu/abs/2001ApJ...548..296W}
}

@ARTICLE{2017A&A...605L...2S,
       author = {{Schoonenberg}, Djoeke and {Okuzumi}, Satoshi and {Ormel}, Chris W.},
        title = "{What pebbles are made of: Interpretation of the V883 Ori disk}",
      journal = {\aap},
         year = 2017,
        month = sep,
       volume = {605},
          eid = {L2},
        pages = {L2},
          doi = {10.1051/0004-6361/201731202},
archivePrefix = {arXiv},
       eprint = {1708.03328},
 primaryClass = {astro-ph.EP},
       adsurl = {https://ui.adsabs.harvard.edu/abs/2017A&A...605L...2S}
}

@ARTICLE{Topchieva2024,
       author = {{Topchieva}, A. and {Molyarova}, T. and {Akimkin}, V. and {Maksimova}, L. and {Vorobyov}, E.},
        title = "{Ices on pebbles in protoplanetary discs}",
      journal = {\mnras},
         year = 2024,
        month = may,
       volume = {530},
       number = {3},
        pages = {2731-2748},
          doi = {10.1093/mnras/stae597},
archivePrefix = {arXiv},
       eprint = {2403.02895},
 primaryClass = {astro-ph.EP},
       adsurl = {https://ui.adsabs.harvard.edu/abs/2024MNRAS.530.2731T}
}

@ARTICLE{DongVorobyov2016,
       author = {{Dong}, Ruobing and {Vorobyov}, Eduard and {Pavlyuchenkov}, Yaroslav and {Chiang}, Eugene and {Liu}, Hauyu Baobab},
        title = "{Signatures of Gravitational Instability in Resolved Images of Protostellar Disks}",
      journal = {\apj},
         year = 2016,
        month = jun,
       volume = {823},
       number = {2},
          eid = {141},
        pages = {141},
          doi = {10.3847/0004-637X/823/2/141},
archivePrefix = {arXiv},
       eprint = {1603.01618},
 primaryClass = {astro-ph.SR},
       adsurl = {https://ui.adsabs.harvard.edu/abs/2016ApJ...823..141D}
}

@ARTICLE{Malygin2014,
       author = {{Malygin}, M.~G. and {Kuiper}, R. and {Klahr}, H. and {Dullemond}, C.~P. and {Henning}, Th.},
        title = "{Mean gas opacity for circumstellar environments and equilibrium temperature degeneracy}",
      journal = {\aap},
         year = 2014,
        month = aug,
       volume = {568},
          eid = {A91},
        pages = {A91},
          doi = {10.1051/0004-6361/201423768},
archivePrefix = {arXiv},
       eprint = {1408.3377},
 primaryClass = {astro-ph.SR},
       adsurl = {https://ui.adsabs.harvard.edu/abs/2014A&A...568A..91M}
}

@ARTICLE{Goldreich1973,
       author = {{Goldreich}, Peter and {Ward}, William R.},
        title = "{The Formation of Planetesimals}",
      journal = {\apj},
         year = 1973,
        month = aug,
       volume = {183},
        pages = {1051-1062},
          doi = {10.1086/152291},
       adsurl = {https://ui.adsabs.harvard.edu/abs/1973ApJ...183.1051G}
}

@ARTICLE{Ryodo2021,
       author = {{Hyodo}, Ryuki and {Guillot}, Tristan and {Ida}, Shigeru and {Okuzumi}, Satoshi and {Youdin}, Andrew N.},
        title = "{Planetesimal formation around the snow line. II. Dust or pebbles?}",
      journal = {\aap},
         year = 2021,
        month = feb,
       volume = {646},
          eid = {A14},
        pages = {A14},
          doi = {10.1051/0004-6361/202039894},
archivePrefix = {arXiv},
       eprint = {2012.06700},
 primaryClass = {astro-ph.EP},
       adsurl = {https://ui.adsabs.harvard.edu/abs/2021A&A...646A..14H}
}

@ARTICLE{Lim2025,
       author = {{Lim}, Jeonghoon and {Simon}, Jacob B. and {Li}, Rixin and {Carrera}, Daniel and {Baronett}, Stanley A. and {Youdin}, Andrew N. and {Lyra}, Wladimir and {Yang}, Chao-Chin},
        title = "{Probing Conditions for Strong Clumping by the Streaming Instability: Small Dust Grains and Low Dust-to-gas Density Ratio}",
      journal = {\apj},
         year = 2025,
        month = mar,
       volume = {981},
       number = {2},
          eid = {160},
        pages = {160},
          doi = {10.3847/1538-4357/adb311},
archivePrefix = {arXiv},
       eprint = {2410.17319},
 primaryClass = {astro-ph.EP},
       adsurl = {https://ui.adsabs.harvard.edu/abs/2025ApJ...981..160L}
}

@ARTICLE{Rice2025,
       author = {{Rice}, Ken and {Baehr}, Hans and {Young}, Alison K. and {Booth}, Richard and {Rowther}, Sahl and {Meru}, Farzana and {Hall}, Cassandra and {Koval}, Adam},
        title = "{Dust density enhancements and the direct formation of planetary cores in gravitationally unstable discs}",
      journal = {\mnras},
         year = 2025,
        month = jun,
       volume = {539},
       number = {4},
        pages = {3421-3435},
          doi = {10.1093/mnras/staf714},
archivePrefix = {arXiv},
       eprint = {2505.00363},
 primaryClass = {astro-ph.EP},
       adsurl = {https://ui.adsabs.harvard.edu/abs/2025MNRAS.539.3421R}
}

@ARTICLE{Cuzzi2004,
       author = {{Cuzzi}, Jeffrey N. and {Zahnle}, Kevin J.},
        title = "{Material Enhancement in Protoplanetary Nebulae by Particle Drift through Evaporation Fronts}",
      journal = {\apj},
         year = 2004,
        month = oct,
       volume = {614},
       number = {1},
        pages = {490-496},
          doi = {10.1086/423611},
archivePrefix = {arXiv},
       eprint = {astro-ph/0409276},
 primaryClass = {astro-ph},
       adsurl = {https://ui.adsabs.harvard.edu/abs/2004ApJ...614..490C}
}

@ARTICLE{Oberg-snow2011,
       author = {{{\"O}berg}, Karin I. and {Murray-Clay}, Ruth and {Bergin}, Edwin A.},
        title = "{The Effects of Snowlines on C/O in Planetary Atmospheres}",
      journal = {\apjl},
         year = 2011,
        month = dec,
       volume = {743},
       number = {1},
          eid = {L16},
        pages = {L16},
          doi = {10.1088/2041-8205/743/1/L16},
archivePrefix = {arXiv},
       eprint = {1110.5567},
 primaryClass = {astro-ph.GA},
       adsurl = {https://ui.adsabs.harvard.edu/abs/2011ApJ...743L..16O}
}

@ARTICLE{Booth2019,
       author = {{Booth}, R.~A. and {Ilee}, J.~D.},
        title = "{Planet-forming material in a protoplanetary disc: the interplay between chemical evolution and pebble drift}",
      journal = {\mnras},
         year = 2019,
        month = aug,
       volume = {487},
       number = {3},
        pages = {3998-4011},
          doi = {10.1093/mnras/stz1488},
archivePrefix = {arXiv},
       eprint = {1905.12639},
 primaryClass = {astro-ph.EP},
       adsurl = {https://ui.adsabs.harvard.edu/abs/2019MNRAS.487.3998B}
}

@ARTICLE{Lebreuilly2020,
       author = {{Lebreuilly}, U. and {Commer{\c{c}}on}, B. and {Laibe}, G.},
        title = "{Protostellar collapse: the conditions to form dust-rich protoplanetary disks}",
      journal = {A\&A},
         year = 2020,
        month = sep,
       volume = {641},
          eid = {A112},
        pages = {A112},
          doi = {10.1051/0004-6361/202038174},
archivePrefix = {arXiv},
       eprint = {2007.06050},
 primaryClass = {astro-ph.SR},
       adsurl = {https://ui.adsabs.harvard.edu/abs/2020A&A...641A.112L}
}

@software{OpTool2021,
       author = {{Dominik}, Carsten and {Min}, Michiel and {Tazaki}, Ryo},
        title = "{OpTool: Command-line driven tool for creating complex dust opacities}",
 howpublished = {Astrophysics Source Code Library, record ascl:2104.010},
         year = 2021,
        month = apr,
          eid = {ascl:2104.010},
archivePrefix = {ascl},
       eprint = {2104.010},
       adsurl = {https://ui.adsabs.harvard.edu/abs/2021ascl.soft04010D}
}

@ARTICLE{Das2025,
       author = {{Das}, Indrani and {Vorobyov}, Eduard and {Basu}, Shantanu},
        title = "{Accretion Bursts in Young Intermediate-mass Stars Make Planet Formation Challenging}",
      journal = {\apj},
         year = 2025,
        month = apr,
       volume = {983},
       number = {2},
          eid = {163},
        pages = {163},
          doi = {10.3847/1538-4357/adb8ee},
archivePrefix = {arXiv},
       eprint = {2502.17114},
 primaryClass = {astro-ph.EP},
       adsurl = {https://ui.adsabs.harvard.edu/abs/2025ApJ...983..163D}
}

@ARTICLE{Woitke2018,
       author = {{Woitke}, P. and {Helling}, Ch. and {Hunter}, G.~H. and {Millard}, J.~D. and {Turner}, G.~E. and {Worters}, M. and {Blecic}, J. and {Stock}, J.~W.},
        title = "{Equilibrium chemistry down to 100 K. Impact of silicates and phyllosilicates on the carbon to oxygen ratio}",
      journal = {\aap},
         year = 2018,
        month = jun,
       volume = {614},
          eid = {A1},
        pages = {A1},
          doi = {10.1051/0004-6361/201732193},
archivePrefix = {arXiv},
       eprint = {1712.01010},
 primaryClass = {astro-ph.EP},
       adsurl = {https://ui.adsabs.harvard.edu/abs/2018A&A...614A...1W}
}

@ARTICLE{Eistrup2016,
       author = {{Eistrup}, Christian and {Walsh}, Catherine and {van Dishoeck}, Ewine F.},
        title = "{Setting the volatile composition of (exo)planet-building material. Does chemical evolution in disk midplanes matter?}",
      journal = {\aap},
         year = 2016,
        month = nov,
       volume = {595},
          eid = {A83},
        pages = {A83},
          doi = {10.1051/0004-6361/201628509},
archivePrefix = {arXiv},
       eprint = {1607.06710},
 primaryClass = {astro-ph.EP},
       adsurl = {https://ui.adsabs.harvard.edu/abs/2016A&A...595A..83E}
}

@ARTICLE{Semenov2003,
       author = {{Semenov}, D. and {Henning}, Th. and {Helling}, Ch. and {Ilgner}, M. and {Sedlmayr}, E.},
        title = "{Rosseland and Planck mean opacities for protoplanetary discs}",
      journal = {\aap},
         year = 2003,
        month = nov,
       volume = {410},
        pages = {611-621},
          doi = {10.1051/0004-6361:20031279},
archivePrefix = {arXiv},
       eprint = {astro-ph/0308344},
 primaryClass = {astro-ph},
       adsurl = {https://ui.adsabs.harvard.edu/abs/2003A&A...410..611S}
}

@ARTICLE{Ormel2007,
       author = {{Ormel}, C.~W. and {Cuzzi}, J.~N.},
        title = "{Closed-form expressions for particle relative velocities induced by turbulence}",
      journal = {\aap},
         year = 2007,
        month = may,
       volume = {466},
       number = {2},
        pages = {413-420},
          doi = {10.1051/0004-6361:20066899},
archivePrefix = {arXiv},
       eprint = {astro-ph/0702303},
 primaryClass = {astro-ph},
       adsurl = {https://ui.adsabs.harvard.edu/abs/2007A&A...466..413O}
}

@ARTICLE{1988ClarkePringle,
       author = {{Clarke}, C.~J. and {Pringle}, J.~E.},
        title = "{The diffusion of contaminant through an accretion disc}",
      journal = {\mnras},
         year = 1988,
        month = nov,
       volume = {235},
        pages = {365-373},
          doi = {10.1093/mnras/235.2.365},
       adsurl = {https://ui.adsabs.harvard.edu/abs/1988MNRAS.235..365C}
}

@ARTICLE{2016ARep...60..879E,
       author = {{Elbakyan}, V.~G. and {Vorobyov}, E.~I. and {Glebova}, G.~M.},
        title = "{Variations in the accretion rate and luminosity in gravitationally unstable protostellar disks}",
      journal = {Astronomy Reports},
         year = 2016,
        month = oct,
       volume = {60},
       number = {10},
        pages = {879-893},
          doi = {10.1134/S1063772916100012},
       adsurl = {https://ui.adsabs.harvard.edu/abs/2016ARep...60..879E}
}

@ARTICLE{Aumatell2011,
       author = {{Aumatell}, Guillem and {Wurm}, Gerhard},
        title = "{Breaking the ice: planetesimal formation at the snowline}",
      journal = {\mnras},
         year = 2011,
        month = nov,
       volume = {418},
       number = {1},
        pages = {L1-L5},
          doi = {10.1111/j.1745-3933.2011.01126.x},
archivePrefix = {arXiv},
       eprint = {1108.0805},
 primaryClass = {astro-ph.EP},
       adsurl = {https://ui.adsabs.harvard.edu/abs/2011MNRAS.418L...1A}
}

@ARTICLE{Saito2011,
       author = {{Saito}, Etsuko and {Sirono}, Sin-iti},
        title = "{Planetesimal Formation by Sublimation}",
      journal = {\apj},
         year = 2011,
        month = feb,
       volume = {728},
       number = {1},
          eid = {20},
        pages = {20},
          doi = {10.1088/0004-637X/728/1/20},
       adsurl = {https://ui.adsabs.harvard.edu/abs/2011ApJ...728...20S}
}

@ARTICLE{Vorobyov2025,
       author = {{Vorobyov}, Eduard I. and {Elbakyan}, Vardan G. and {Skliarevskii}, Alexandr and {Akimkin}, Vitaly and {Kulikov}, Igor},
        title = "{Dust enrichment and growth in the earliest stages of protoplanetary disk formation}",
      journal = {\aap},
         year = 2025,
        month = jul,
       volume = {699},
          eid = {A27},
        pages = {A27},
          doi = {10.1051/0004-6361/202553718},
archivePrefix = {arXiv},
       eprint = {2505.04233},
 primaryClass = {astro-ph.EP},
       adsurl = {https://ui.adsabs.harvard.edu/abs/2025A&A...699A..27V}
}

@ARTICLE{Lyra2026,
       author = {{Baronett}, Stanley A. and {Lyra}, Wladimir and {Aly}, Hossam and {Brouillette}, Olivia and {Carrera}, Daniel and {De Cun}, Victoria I. and {Eriksson}, Linn E.~J. and {Flock}, Mario and {Huang}, Pinghui and {Krapp}, Leonardo and {Lesur}, Geoffroy and {Li}, Rixin and {Li}, Shengtai and {Lim}, Jeonghoon and {Paardekooper}, Sijme-Jan and {Rea}, David G. and {Sengupta}, Debanjan and {Simon}, Jacob B. and {Sudarshan}, Prakruti and {Umurhan}, Orkan M. and {Yang}, Chao-Chin and {Youdin}, Andrew N.},
        title = "{A Comparative Study of the Streaming Instability: Unstratified Models with Marginally Coupled Grains}",
      journal = {arXiv e-prints},
         year = 2026,
        month = mar,
          eid = {arXiv:2603.04558},
        pages = {arXiv:2603.04558},
          doi = {10.48550/arXiv.2603.04558},
archivePrefix = {arXiv},
       eprint = {2603.04558},
 primaryClass = {astro-ph.EP},
       adsurl = {https://ui.adsabs.harvard.edu/abs/2026arXiv260304558B}
}

@ARTICLE{Bitsch2020,
       author = {{Savvidou}, Sofia and {Bitsch}, Bertram and {Lambrechts}, Michiel},
        title = "{Influence of grain growth on the thermal structure of protoplanetary discs}",
      journal = {\aap},
         year = 2020,
        month = aug,
       volume = {640},
          eid = {A63},
        pages = {A63},
          doi = {10.1051/0004-6361/201936576},
archivePrefix = {arXiv},
       eprint = {2005.14097},
 primaryClass = {astro-ph.EP},
       adsurl = {https://ui.adsabs.harvard.edu/abs/2020A&A...640A..63S}
}

@ARTICLE{Mori2021,
       author = {{Mori}, Shoji and {Okuzumi}, Satoshi and {Kunitomo}, Masanobu and {Bai}, Xue-Ning},
        title = "{Evolution of the Water Snow Line in Magnetically Accreting Protoplanetary Disks}",
      journal = {\apj},
         year = 2021,
        month = aug,
       volume = {916},
       number = {2},
          eid = {72},
        pages = {72},
          doi = {10.3847/1538-4357/ac06a9},
archivePrefix = {arXiv},
       eprint = {2105.13101},
 primaryClass = {astro-ph.EP},
       adsurl = {https://ui.adsabs.harvard.edu/abs/2021ApJ...916...72M}
}

@ARTICLE{Nayakshin2025,
       author = {{Lee}, Hans and {Nayakshin}, Sergei and {Booth}, Richard A.},
        title = "{Dust growth and planet formation by disc fragmentation}",
      journal = {\mnras},
         year = 2025,
        month = nov,
       volume = {544},
       number = {1},
        pages = {L18-L23},
          doi = {10.1093/mnrasl/slaf096},
archivePrefix = {arXiv},
       eprint = {2509.09305},
 primaryClass = {astro-ph.EP},
       adsurl = {https://ui.adsabs.harvard.edu/abs/2025MNRAS.544L..18L}
}

@ARTICLE{Yamamuro2023,
       author = {{Yamamuro}, Ryota and {Tanaka}, Kei E.~I. and {Okuzumi}, Satoshi},
        title = "{Massive Protostellar Disks as a Hot Laboratory of Silicate Grain Evolution}",
      journal = {\apj},
         year = 2023,
        month = may,
       volume = {949},
       number = {1},
          eid = {29},
        pages = {29},
          doi = {10.3847/1538-4357/acc52f},
archivePrefix = {arXiv},
       eprint = {2303.09148},
 primaryClass = {astro-ph.EP},
       adsurl = {https://ui.adsabs.harvard.edu/abs/2023ApJ...949...29Y}
}

@ARTICLE{Pavlyuchenkov2023,
       author = {{Pavlyuchenkov}, Ya. N. and {Akimkin}, V.~V. and {Topchieva}, A.~P. and {Vorobyov}, E.~I.},
        title = "{Effect of Dust Evaporation and Thermal Instability on Temperature Distribution in a Protoplanetary Disk}",
      journal = {Astronomy Reports},
         year = 2023,
        month = may,
       volume = {67},
       number = {5},
        pages = {470-482},
          doi = {10.1134/S1063772923050086},
archivePrefix = {arXiv},
       eprint = {2307.15544},
 primaryClass = {astro-ph.EP},
       adsurl = {https://ui.adsabs.harvard.edu/abs/2023ARep...67..470P}
}

@ARTICLE{Okuzumi2019,
       author = {{Okuzumi}, Satoshi and {Tazaki}, Ryo},
        title = "{Nonsticky Ice at the Origin of the Uniformly Polarized Submillimeter Emission from the HL Tau Disk}",
      journal = {\apj},
         year = 2019,
        month = jun,
       volume = {878},
       number = {2},
          eid = {132},
        pages = {132},
          doi = {10.3847/1538-4357/ab204d},
archivePrefix = {arXiv},
       eprint = {1904.03869},
 primaryClass = {astro-ph.EP},
       adsurl = {https://ui.adsabs.harvard.edu/abs/2019ApJ...878..132O}
}

@INPROCEEDINGS{Bergin2019,
       author = {{Zhang}, K. and {Bergin}, E.~A. and {Williams}, J.~P. and {Pinilla}, P. and {Andrews}, S.~M.},
        title = "{Tracing the Water Snowline in Protoplanetary Disks with the ngVLA}",
    booktitle = {Science with a Next Generation Very Large Array},
         year = 2018,
       editor = {{Murphy}, Eric},
       series = {Astronomical Society of the Pacific Conference Series},
       volume = {517},
        month = dec,
        pages = {209},
          doi = {10.48550/arXiv.1810.06604},
archivePrefix = {arXiv},
       eprint = {1810.06604},
 primaryClass = {astro-ph.EP},
       adsurl = {https://ui.adsabs.harvard.edu/abs/2018ASPC..517..209Z}
}

@ARTICLE{Tobin2023,
       author = {{Tobin}, John J. and {van't Hoff}, Merel L.~R. and {Leemker}, Margot and {van Dishoeck}, Ewine F. and {Paneque-Carre{\~n}o}, Teresa and {Furuya}, Kenji and {Harsono}, Daniel and {Persson}, Magnus V. and {Cleeves}, L. Ilsedore and {Sheehan}, Patrick D. and {Cieza}, Lucas},
        title = "{Deuterium-enriched water ties planet-forming disks to comets and protostars}",
      journal = {\nat},
         year = 2023,
        month = mar,
       volume = {615},
       number = {7951},
        pages = {227-230},
          doi = {10.1038/s41586-022-05676-z},
       adsurl = {https://ui.adsabs.harvard.edu/abs/2023Natur.615..227T}
}

@ARTICLE{Cieza2016,
       author = {{Cieza}, Lucas A. and {Casassus}, Simon and {Tobin}, John and {Bos}, Steven P. and {Williams}, Jonathan P. and {Perez}, Sebastian and {Zhu}, Zhaohuan and {Caceres}, Claudio and {Canovas}, Hector and {Dunham}, Michael M. and {Hales}, Antonio and {Prieto}, Jose L. and {Principe}, David A. and {Schreiber}, Matthias R. and {Ruiz-Rodriguez}, Dary and {Zurlo}, Alice},
        title = "{Imaging the water snow-line during a protostellar outburst}",
      journal = {\nat},
         year = 2016,
        month = jul,
       volume = {535},
       number = {7611},
        pages = {258-261},
          doi = {10.1038/nature18612},
archivePrefix = {arXiv},
       eprint = {1607.03757},
 primaryClass = {astro-ph.SR},
       adsurl = {https://ui.adsabs.harvard.edu/abs/2016Natur.535..258C}
}

@ARTICLE{vanHoff2018,
       author = {{van 't Hoff}, Merel L.~R. and {Tobin}, John J. and {Trapman}, Leon and {Harsono}, Daniel and {Sheehan}, Patrick D. and {Fischer}, William J. and {Megeath}, S. Thomas and {van Dishoeck}, Ewine F.},
        title = "{Methanol and its Relation to the Water Snowline in the Disk around the Young Outbursting Star V883 Ori}",
      journal = {\apjl},
         year = 2018,
        month = sep,
       volume = {864},
       number = {1},
          eid = {L23},
        pages = {L23},
          doi = {10.3847/2041-8213/aadb8a},
archivePrefix = {arXiv},
       eprint = {1808.08258},
 primaryClass = {astro-ph.SR},
       adsurl = {https://ui.adsabs.harvard.edu/abs/2018ApJ...864L..23V}
}

@ARTICLE{Leemker2021,
       author = {{Leemker}, M. and {van't Hoff}, M.~L.~R. and {Trapman}, L. and {van Gelder}, M.~L. and {Hogerheijde}, M.~R. and {Ru{\'\i}z-Rodr{\'\i}guez}, D. and {van Dishoeck}, E.~F.},
        title = "{Chemically tracing the water snowline in protoplanetary disks with HCO$^{+}$}",
      journal = {\aap},
         year = 2021,
        month = feb,
       volume = {646},
          eid = {A3},
        pages = {A3},
          doi = {10.1051/0004-6361/202039387},
archivePrefix = {arXiv},
       eprint = {2011.12319},
 primaryClass = {astro-ph.EP},
       adsurl = {https://ui.adsabs.harvard.edu/abs/2021A&A...646A...3L}
}

@ARTICLE{Facchini2024,
       author = {{Facchini}, Stefano and {Testi}, Leonardo and {Humphreys}, Elizabeth and {Vander Donckt}, Mathieu and {Isella}, Andrea and {Wrzosek}, Ramon and {Baudry}, Alain and {Gray}, Malcom D. and {Richards}, Anita M.~S. and {Vlemmings}, Wouter},
        title = "{Resolved ALMA observations of water in the inner astronomical units of the HL Tau disk}",
      journal = {Nature Astronomy},
         year = 2024,
        month = may,
       volume = {8},
        pages = {587-595},
          doi = {10.1038/s41550-024-02207-w},
archivePrefix = {arXiv},
       eprint = {2403.00647},
 primaryClass = {astro-ph.EP},
       adsurl = {https://ui.adsabs.harvard.edu/abs/2024NatAs...8..587F}
}

@ARTICLE{Nakasone2026,
       author = {{Nakasone}, Hiroto and {Notsu}, Shota and {Yoshida}, Tomohiro C. and {Nomura}, Hideko and {Tsukagoshi}, Takashi and {Hirota}, Tomoya and {Honda}, Mitsuhiko and {Akiyama}, Eiji and {Booth}, Alice S. and {Lee}, Jeong-Eun and {Lee}, Seokho},
        title = "{ALMA Band 7 Observations of Water Lines in the Protoplanetary Disk of V883 Ori}",
      journal = {\apj},
         year = 2026,
        month = feb,
       volume = {998},
       number = {1},
          eid = {53},
        pages = {53},
          doi = {10.3847/1538-4357/ae2c82},
archivePrefix = {arXiv},
       eprint = {2512.15108},
 primaryClass = {astro-ph.EP},
       adsurl = {https://ui.adsabs.harvard.edu/abs/2026ApJ...998...53N}
}

@ARTICLE{Sai2026,
       author = {{Sai}, Jinshi and {Vorobyov}, Eduard I. and {Skliarevskii}, Alexandr and {Takami}, Michihiro},
        title = "{Morphological and Kinematic Diagnostic of FU Orionis-type Outburst Mechanisms}",
      journal = {\apj},
         year = 2025,
        month = dec,
       volume = {995},
       number = {1},
          eid = {129},
        pages = {129},
          doi = {10.3847/1538-4357/ae17c3},
archivePrefix = {arXiv},
       eprint = {2510.22927},
 primaryClass = {astro-ph.SR},
       adsurl = {https://ui.adsabs.harvard.edu/abs/2025ApJ...995..129S}
}

@ARTICLE{Pavlyuchenkov2019,
       author = {{Pavlyuchenkov}, Yaroslav and {Akimkin}, Vitaly and {Wiebe}, Dmitri and {Vorobyov}, Eduard},
        title = "{Revealing dust segregation in protoplanetary discs with the help of multifrequency spectral index maps}",
      journal = {\mnras},
         year = 2019,
        month = jul,
       volume = {486},
       number = {3},
        pages = {3907-3914},
          doi = {10.1093/mnras/stz1046},
archivePrefix = {arXiv},
       eprint = {1904.05251},
 primaryClass = {astro-ph.IM},
       adsurl = {https://ui.adsabs.harvard.edu/abs/2019MNRAS.486.3907P}
}

@ARTICLE{2016A&A...586A.103W,
       author = {{Woitke}, P. and {Min}, M. and {Pinte}, C. and {Thi}, W.-F. and {Kamp}, I. and {Rab}, C. and {Anthonioz}, F. and {Antonellini}, S. and {Baldovin-Saavedra}, C. and {Carmona}, A. and {Dominik}, C. and {Dionatos}, O. and {Greaves}, J. and {G{\"u}del}, M. and {Ilee}, J.~D. and {Liebhart}, A. and {M{\'e}nard}, F. and {Rigon}, L. and {Waters}, L.~B.~F.~M. and {Aresu}, G. and {Meijerink}, R. and {Spaans}, M.},
        title = "{Consistent dust and gas models for protoplanetary disks. I. Disk shape, dust settling, opacities, and PAHs}",
      journal = {\aap},
         year = 2016,
        month = feb,
       volume = {586},
          eid = {A103},
        pages = {A103},
          doi = {10.1051/0004-6361/201526538},
archivePrefix = {arXiv},
       eprint = {1511.03431},
 primaryClass = {astro-ph.EP},
       adsurl = {https://ui.adsabs.harvard.edu/abs/2016A&A...586A.103W}
}

\begin{appendix}

\section{Turbulent viscosity, cooling, and heating}
\label{App:visc-cool-heat}
Turbulent viscosity was parameterized using the  \citet{1973ShakuraSunyaev} approach as $\nu = \alpha_{\rm visc} c_{\mathrm{s}} H_{\mathrm{g}}$, where $\nu$ is the kinematic viscosity that enters the viscous stress tensor $\bl{\Pi}$, $c_{\mathrm{s}}$ is the sound speed, and $H_{\mathrm{g}}$ is the gas vertical scale height, calculated using an assumption of local hydrostatic balance in the gravitational field of the star and disk \citep{VorobyovBasu2009}.  We assumed that turbulence is  isotropic, and $\alpha_{\rm visc}$ represents not only the efficiency of mass and angular transport in the disk plane due to turbulence, but also influences dust growth and settling in the dust growth model described in Sect.~\ref{sect:dustgrowth}.  

The radiative cooling  and  heating rates are denoted by $\Lambda$ and $\Gamma$ and are written as \citep{DongVorobyov2016,Pavlyuchenkov2023}
\begin{equation}
\Lambda=\frac{8\tau_{\rm P} \sigma T_{\rm mp}^4 }{1+2\tau_{\rm P} + 
{3 \over 2}\tau_{\rm R}\tau_{\rm P}}, \,\,\, 
\Gamma=\frac{8\tau_{\rm P} \sigma T_{\rm irr}^4 }{1+2\tau_{\rm P} + {3 \over 2}\tau_{\rm R}\tau_{\rm
P}}.
\label{eq:cool-heat}
\end{equation}
Here, $\tau_{\rm R}$ and $\tau_{\rm P}$ are the Rosseland and Planck mean optical depths to the disk midplane (see Sect.~\ref{Sect:opacity} for details), $T_{\rm mp}$ is the temperature in the disk midplane and $T_{\rm irr}$ is the irradiation temperature at the disk surface 
determined from the stellar and background black-body irradiation as
\begin{equation}
T_{\rm irr}^4=T_{\rm bg}^4+\frac{F_{\rm irr}(r)}{\sigma},
\label{fluxCS}
\end{equation}
where $\sigma$ is the Stefan-Boltzmann constant and  $F_{\rm
irr}(r)$ is the radiation flux  absorbed by the disk surface at radial distance  $r$ from the
central star. The latter quantity is calculated as 
\begin{equation}
F_{\rm irr}(r)= \frac{L_\ast}{4\pi r^2} \cos{\gamma_{\rm irr}},
\label{fluxF}
\end{equation}
 where  $\gamma_{\rm{irr}}$ is the incidence angle of radiation arriving at the disk surface at radial distance $r$ \citep[see for details][]{2010VorobyovBasu}.
The background temperature is set equal to $T_{\rm bg}=15$~K. The stellar luminosity $L_{\ast}$ is the sum of the accretion and photospheric luminosities.

\section{Dust diffusion, friction, and small-to-grown conversion}
\label{App:diff-fric}

Diffusion of grown dust was realized according to the model of \citet{1988ClarkePringle}, where $D$ is the turbulent diffusivity of dust, which is related to the turbulent viscosity of gas as $D=\nu / (\mathrm{Sc}+\mathrm{St}^2)$. The Stokes number $\mathrm{St}$ is the product of the Keplerian angular velocity and the stopping time, the latter written as
\begin{equation}
t_{\rm stop} = {8 \over 3} {a_{\rm max} \rho_{\rm s} \over \rho_{\rm g} C_{\rm D} |{\bl v} - {\bl u}|  }.
\label{Eq:stop-time}
\end{equation}
Here, $\rho_{\rm g}$ is the volume density of gas, defined as $\rho_{\rm g}=\Sigma_{\rm g} / (\sqrt{2 \pi} H_{\rm g})$, and $C_{\rm D}$ is the friction parameter (see below).
The Schmidt number $\mathrm{Sc}$ is taken to be unity in this study.
For the justification of the hydrodynamic approach to describing dust dynamics and the coupled dynamics of small dust and gas, we refer to \citet{Vorobyov2022}.

The drag force (per unit mass) links grown dust with gas and can be written as \citep{Weidenschilling1977}
\begin{equation}
    {{\bl f}} = \dfrac{1} {2 m_{\rm d}} C_{\rm D} \, \sigma_{\rm d} \rho_{\rm g} ({{\bl v}} - {{\bl u}}) |{{\bl v}} - {{\bl u}}|,
\label{eq:friction}
\end{equation}
where $\sigma_{\rm d}$ is the dust grain cross section and $m_{\rm d}$ the mass of a dust grain. The dimensionless friction parameter $C_{\rm D}$  is described in details in \citet{Vorobyov2023a} and is based on the works of \citet{Henderson1976} and \citet{Stoyanovskaya2020}.
The Henderson friction coefficient implements a smooth transition in the drag force between the Epstein and Stokes regimes, depending on the local conditions and dust properties.

We are interested in the dynamics of dust grains that are the main mass carriers. Therefore, we use the maximum size of dust grains $a_{\rm max}$ when calculating the values of $\sigma$ and $m_{\rm d}$ in Eq.~(\ref{eq:friction}). This choice may slightly underestimate the friction force ${\bl f}$. A more accurate calculation of ${\bl f}$ involves weighting $m_{\rm d}$ in Eq.~(\ref{eq:friction}) over this dust size bin of [$0.2 a_{\rm max}:a_{\rm max}$], in which $\ge 50\%$ of the total dust mass is concentrated for our choice of $p=3.5$, an update planned for our future works.
To account for the back-reaction of grown dust on dust, the term $\Sigma_{\rm d,gr} f$ is symmetrically included in both the gas and dust momentum equations.

The term $S(a_{\rm max})$ that enters Eqs.~(\ref{eq:contDsmall})--(\ref{eq:momDlarge}) for the dust component is the conversion rate between small and grown dust populations.  For a mathematical and graphical description of small-to-grown dust conversion in the process of dust growth we refer to~\citet{Molyarova2021}. Here, we note that $S(a_{\rm max})$ depends only on the local maximal size of dust grains $a_{\rm max}$, which in turn is determined by the rate of collisional growth and fragmentation, and also by advection (and drift) of dust through the disk.

\section{Dust growth scheme}
\label{App:dust-growth}
The equation describing the dynamical evolution of the maximum dust size $a_{\rm max}$ is as follows:
\begin{equation}
{\partial a_{\rm max} \over \partial t} + ({\bl u} \cdot {\bl \nabla} ) a_{\rm max} = \cal{D},
\label{eq:dustA}
\end{equation}
where the rate of dust growth due to collisions and coagulation is computed in the monodisperse approximation as \citep{2012Birnstiel}
\begin{equation}
\cal{D} = {\rho_{\rm d} \mathit{u}_{\rm rel} \over \rho_{\rm s}}.
\label{eq:dust-growth}
\end{equation}
This rate includes the total volume density of dust, defined as $\rho_{\rm d}=(\Sigma_{\rm d,gr} + \Sigma_{\rm d,sm}) / (\sqrt{2 \pi} H_{\rm d})$, where $H_{\rm d}$ is the dust vertical scale height (see Eq.~\ref{eq:dust-scale-height}), and the relative velocity of particle-to-particle collisions defined as $\mathit{u}_{\rm rel} = (\mathit{u}_{\rm th}^2 + \mathit{u}_{\rm turb}^2)^{1/2}$, where $\mathit{u}_{\rm th}$ and $\mathit{u}_{\rm turb}$ account for the Brownian and turbulence-induced local motion, respectively.  The dust-to-dust collision velocity owing to turbulence is computed following the model of turbulent eddies proposed in \citet{Ormel2007}: 
\begin{equation}
    u_{\rm{turb}} = \sqrt{{\frac{3 \alpha_{\rm visc}}{\mathrm{St}+\mathrm{St}^{-1}}}} c_{\rm s},
    \label{turb_vel}
\end{equation}
where the Stokes number corresponds to dust grains of maximum size $a_{\rm max}$.
When calculating $\rho_{\rm d}$ in Eq.~(\ref{eq:dust-growth}), we take into account dust settling by calculating the effective scale height of grown dust  $H_{\rm d}$ as
\begin{equation}
    H_{\rm d} = H_{\rm g} \sqrt{ {\alpha_{\rm visc} \over \alpha_{\rm visc}+\mathrm{St} } }.
    \label{eq:dust-scale-height}
\end{equation}

\section{Evolution equations for volatiles}
\label{App:volatiles}
The chemical evolution of the surface densities of volatile species is described by the following system of equations:
 \begin{eqnarray}
 \frac{{\partial \Sigma_{s}^{\rm gas} }}{{\partial t}}  + {\bl \nabla}  \cdot 
 \left(\Sigma_{s}^{\rm gas} \bl{v} \right)  &=&  -\lambda_s\Sigma_{s}^{\rm gas}+\eta_s^{\rm sm}+\eta_s^{\rm gr},\label{eq:sig1}\\
 \frac{{\partial \Sigma_{s}^{\rm sm} }}{{\partial t}} + {\bl \nabla}  \cdot 
 \left(\Sigma_{s}^{\rm sm} \bl{v} \right)  &=&  \lambda_s^{\rm sm}\Sigma_{s}^{\rm gas}-\eta_s^{\rm sm} ,\label{eq:sig2}\\
 \frac{{\partial \Sigma_{s}^{\rm gr} }}{{\partial t}}  + {\bl \nabla}  \cdot 
 \left( \Sigma_{s}^{\rm gr} \bl{u} \right)  &=& \lambda_s^{\rm gr}\Sigma_{s}^{\rm gas}-\eta_s^{\rm gr},\label{eq:sig3}
 \end{eqnarray}
where the mass rate coefficients of adsorption $\lambda_{s}$ and desorption $\eta_{s}$  per unit surface area for the species $s$ are calculated for local conditions at every hydrodynamic step, separately for small and grown dust populations. We refer to \citet{Molyarova2021} and \citet{Topchieva2024} for the mathematical description of the adsorption and desorption rates. Here, we note that in our model ice mantles provide feedback on dust dynamics via the fragmentation velocity, which depends on whether dust grains are covered with ice or not.  Dust grains are considered icy if the local total surface density of all ices on grown dust divided by $\Sigma_{\rm d,gr}$ is greater than the threshold value $K$, which is calculated as 
\begin{equation}
 K = \frac{3 a_{\rm ml}\rho_{\rm ice}}{\sqrt{a_*a_{\rm max}} \rho_{\rm s}},
 \label{eq:threshold}
 \end{equation}
i.e., an icy grain must have at least one monolayer of ice. Here, $a_{\rm ml}$ is the thickness of the ice monolayer estimated as the size of a water molecule $3 \times 10^{-8}$\,cm. The material density of ice $\rho_{\rm ice}=1.0$\,g\,cm$^{-3}$ and the mean radius of a grown grain is calculated as $\sqrt{a_* a_{\rm max}}$ for the power-law distribution with $p=3.5$. 
We further note that by computing the evolution of four volatiles (H$_2$O, CO$_2$, CO, and CH$_4$), more complex choices of the fragmentation velocity depending on the composition of ices on dust grains (and not just presence vs. absence of ices) are possible \citep[e.g.,][]{Okuzumi2019}, but these were not considered in this study.

\section{Radial profiles at later stages of disk evolution}
\label{App:later-evolution}
Figures~\ref{fig:radial-add} and \ref{fig:radial-zoom-add} present the azimuthally averaged radial profiles of the main disk properties at $t=482$~kyr and 602~kyr. The main features of the gas, dust, and water distribution in the vicinity of the water snow line remain similar to those at earlier evolution times, described in the context of Figs.~\ref{fig:radial} and \ref{fig:radial-zoom}.  
Local accumulations of grown dust and, in particular, water ice are clearly visible behind the water snow line. A sharp increase in the maximum dust size across the snow line is also present. The position of the water snow line is firmly within the viscous-heating-dominated disk region. 

A notable difference with the earlier evolution times is only seen the optical depth of the disk -- it becomes optically thin in the Rosseland and Planck sense over a larger disk extent, already beyond 20~au after 600~kyr of evolution. A strong contrast in the optical depth and, as a consequence, in the gas temperature, which develops in the vicinity of the water snow line, influences the gas density profile via the elevated viscous mass transport in the disk regions interior to the water snow line. As a result, a characteristic bend in $\Sigma_{\rm g}$ behind the water snow line occurs at these late times of disk evolution, when the contribution of gravitational torques to mass transport through the disk becomes negligible.

\begin{figure}   
    \centering
    \includegraphics[width=1\columnwidth]{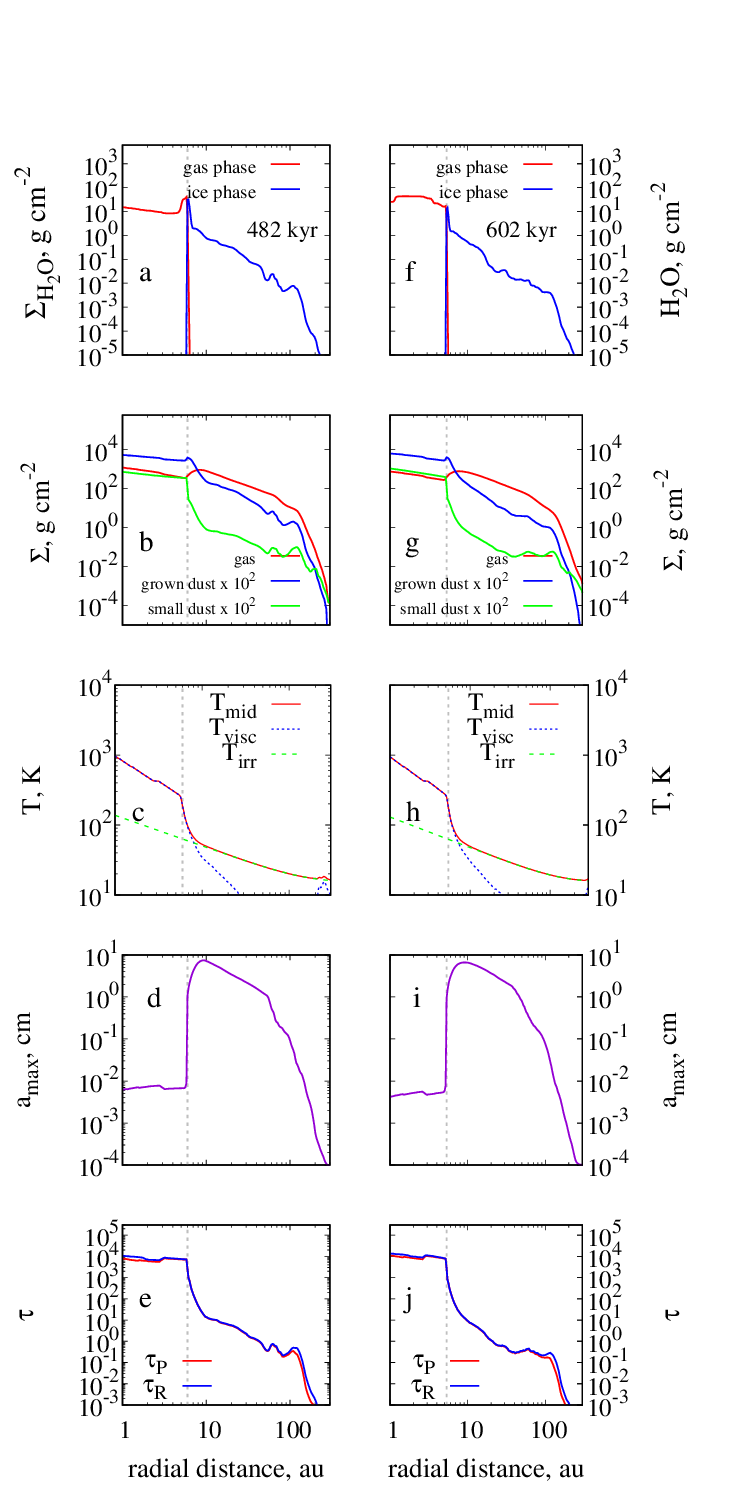}
    \caption{Similar to Fig.~\ref{fig:radial} but at later evolution times of $t=482$~kyr (\textit{left column}) and $t=602$~kyr (\textit{right column}).} 
    \label{fig:radial-add}
\end{figure}

\begin{figure}   
    \centering
    \includegraphics[width=1\columnwidth]{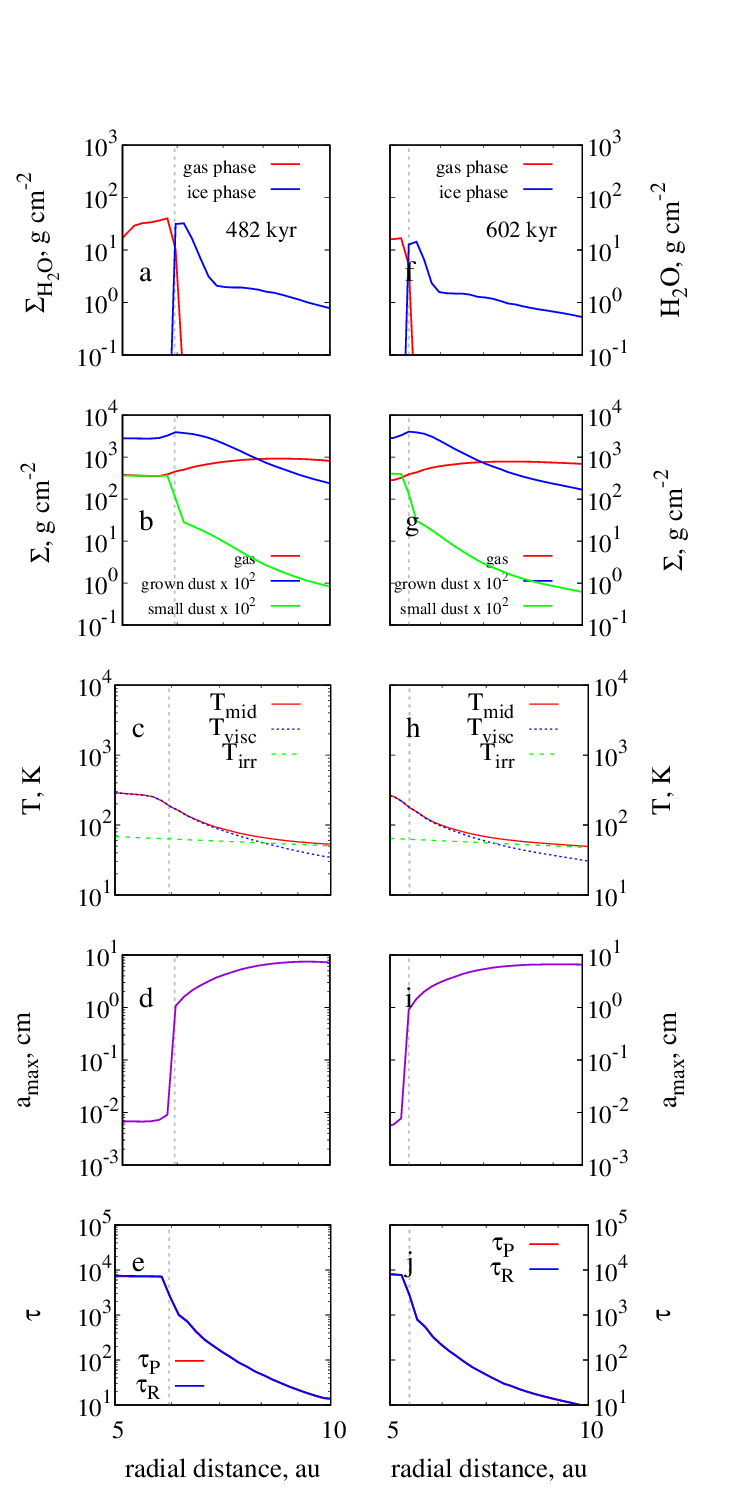}
    \caption{Similar to Fig.~\ref{fig:radial-zoom} but  at later evolution times of $t=482$~kyr (\textit{left column}) and $t=602$~kyr (\textit{right column}).}
    \label{fig:radial-zoom-add}
\end{figure}

\section{Dependence of opacities on their chemical composition}
\label{App:opacities}

In Fig.~\ref{fig:opacities-compositions} we compare dust opacities for grains with different chemical compositions. In particular, we considered dust grains consisting of 
40\% astrosilicates, 10\% troilite and 50\% refractory organics, as in Table~\ref{tab:abundances} for a temperature range of $T_{[10-150]}$. We note that for convenience we neglected changes in grain's composition with temperature. This case is compared to the default OpTool composition (87\% amorphous laboratory silicates and 13\% amorphous carbon) adopted in the DIANA standard dust model \citep{2016A&A...586A.103W}. 
In both considered models, dust grains are assumed to be solid spheres without porosity. 

Figure~\ref{fig:opacities-compositions} demonstrates that the Rosseland and Planck mean opacities for the DIANA standard composition are slightly higher than those used in our work, but the general trends with increasing temperature and maximum dust size remain similar. A notable decrease in opacity with increasing maximum dust size (at moderate and high temperatures) is observed in both of the models. 
Transitions between increase and decrease of opacity with dust growth at lower temperatures is also preserved regardless of grain composition. 

The major causes of the considered effects arise fundamentally due to the character of dust distribution over size and their optical properties, as was mentioned in the main text. The grain composition has small impact on the overall picture.

\begin{figure}   
    \centering
    \includegraphics[width=1\columnwidth]{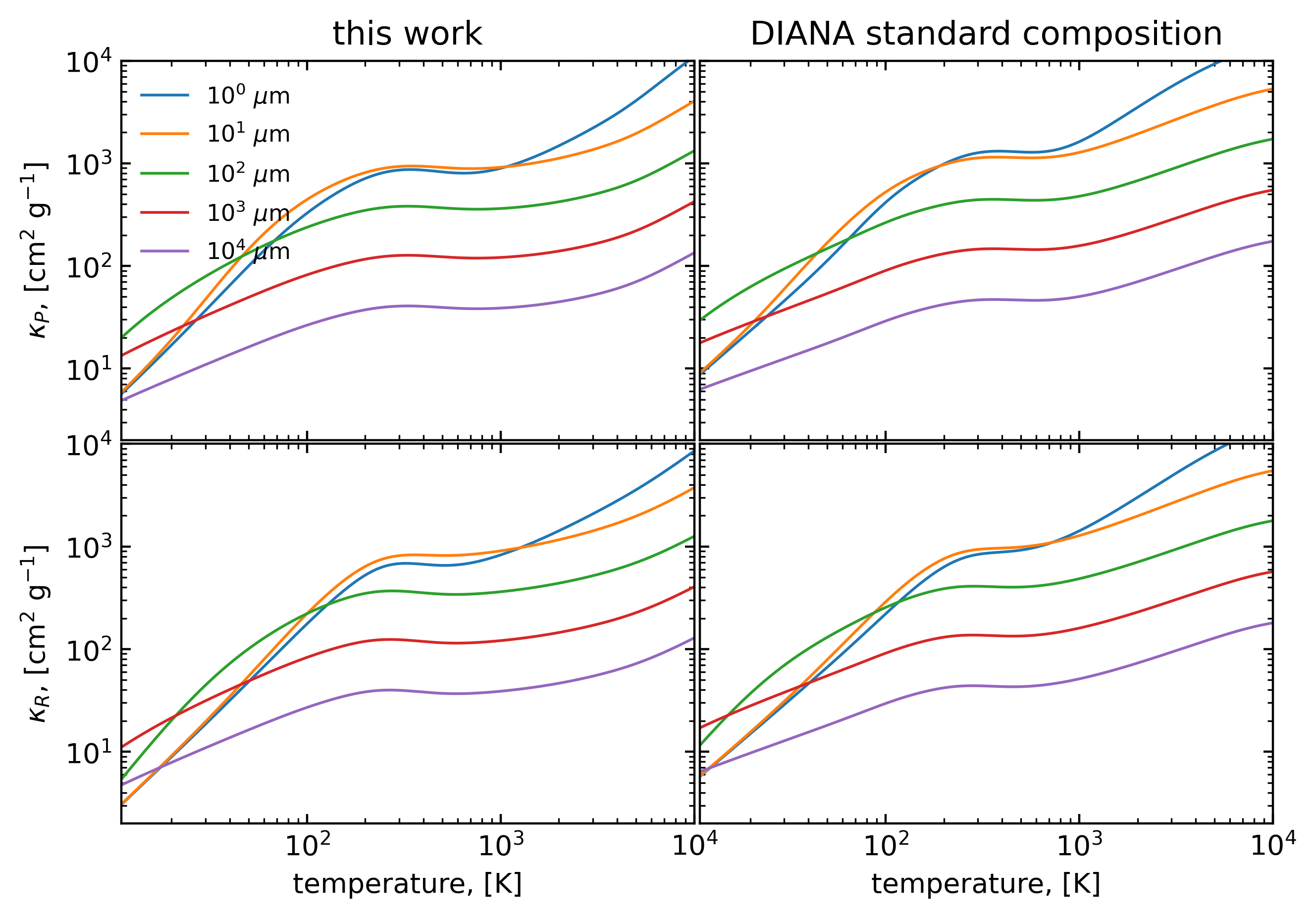}
    \caption{Comparison of opacities calculated assuming different chemical compositions of dust grains. \textit{Left column}: Composition considered in this work (40\% astrosilicates, 10\% troilite, 50\% refractory organics). \textit{Right column}: Opacities for the DIANA standard composition. The Planck and Rosseland mean opacities are presented in the top and bottom rows, respectively. The blue, orange, green, red, and violet lines correspond to different assumed maximum sizes of dust grains, from 1.0~$\mu$m to 1.0~cm in one-decade increments.}
    \label{fig:opacities-compositions}
\end{figure}

\end{appendix}

\end{document}